%% file: Thesis.tex
\documentclass[a4paper, 12pt]{report}
\pdfoutput=1
\usepackage[utf8]{inputenc}
\usepackage{hyperref}
\usepackage{amsmath}
\usepackage{amsthm}
\usepackage{amssymb}
\usepackage[english]{babel}
\usepackage{multirow}
\usepackage{graphicx}
\usepackage{mathrsfs}
\usepackage{textcomp}
\usepackage{color}
\usepackage{cite} 
\usepackage{braket}
\usepackage{siunitx}
\usepackage{mathtools}
\usepackage{verbatim}
\usepackage{fancyhdr}
\usepackage{wrapfig}
\usepackage{float}	
\usepackage[]{frontespizio}
\newcommand{\nocontentsline}[3]{}
\newcommand{\tocless}[2]{\bgroup\let\addcontentsline=\nocontentsline#1{#2}\egroup}
\usepackage{caption}
\usepackage{subcaption}
\usepackage[gen]{eurosym}
\usepackage[left=3cm, right=3cm, bottom=3cm]{geometry}
\usepackage{fourier}
\usepackage[margin=1cm,font=small]{caption}
\usepackage[Rejne]{fncychap}
\makeatletter
\renewcommand*{\@makechapterhead}[1]{%
  \vspace*{0\p@}%
  {\parindent \z@ \raggedright \normalfont
    \ifnum \c@secnumdepth >\m@ne
      \if@mainmatter
        \DOCH
      \fi
    \fi
    \interlinepenalty\@M
    \if@mainmatter
      \DOTI{#1}%
    \else%
      \DOTIS{#1}%
    \fi
  }}
\renewcommand*{\@makeschapterhead}[1]{%
  \vspace*{0\p@}%
  {\parindent \z@ \raggedright
    \normalfont
    \interlinepenalty\@M
    \DOTIS{#1}
    \vskip 0\p@
  }}
\makeatother

\begin{document}

\input{Chapters/frontespizio.tex}

\input{Chapters/Thanks.tex}

\tableofcontents

\input{Chapters/Introduction.tex}

\input{Chapters/Interactionmatter.tex}

\input{Chapters/Neutrondetectors.tex}

\input{Chapters/Instrumentation.tex}

\input{Chapters/Plateaumeasurements.tex}

\input{Chapters/Spatialresolution.tex}

\input{Chapters/Gammasensitivity.tex}

\input{Chapters/Pulseshapediscrimination.tex}

\input{Chapters/Conclusions.tex}

\bibliography{IEEEabrv,bib}

\end{document}

%% file: Chapters/frontespizio.tex
\begin{frontespizio}
\begin{Preambolo*}
\usepackage{fourier}
\renewcommand{\frontsmallfont}[1]{\small Student number }
\newcommand{\compring}{anelli compatti}
\Margini{3cm}{3cm}{3cm}{1cm}
\end{Preambolo*}
\Universita{Milano-Bicocca}
\Facolta{Scienze Matematiche, Fisiche e Naturali}
\Scuola{Master Degree in Physics}
\Logo[5cm]{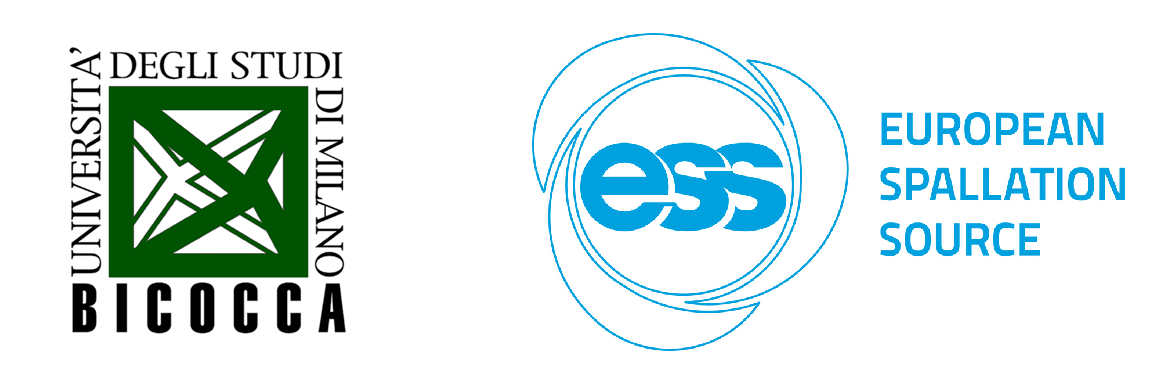}
\Titolo{Characterization of helium-3 tubes}
\Candidato[735658]{Edoardo Rossi}
\NCandidato{Author}
\Relatore {Giuseppe Gorini\\[1ex]\fontsize{10}{12}\normalfont ESS Supervisor:\\ Richard Hall-Wilton}
\NRelatore{Supervisor}{Supervisors}
\Correlatore{Francesco Piscitelli}
\NCorrelatore{Assistant Supervisor}{Assistant Supervisors}
\Piede{Academical Year 2014-2015}
\end{frontespizio}

%% file: Chapters/Thanks.tex

\newpage
\newgeometry{left=4cm, right=4cm, bottom=4cm, top=3.8cm}
\textit{
I would like to thank my family, that has strongly supported me since my first day of university. In particular I want to thank my mother that raised me: without her patience and kindness I would have been completely lost. I thank my little sister Beatrice and I wish her all the best of luck for her studies in the future. I thank my grandparents, Santina and Vittorino, that with their self-denial guaranteed to my family a comfortable life. I thank Roberto that has always provided me good suggestions.
\\
\\
I want to thank Francesco that has been not only an excellent supervisor, but also a perfect role model. He took care of me throughout the development of this thesis and his support has been inestimable. I wish all the best to him and to Federica with the new baby, Ginevra!
\\
I would like to thank all the ESS Detector Group, and now I don't know where to start...
\\
I thank Richard that welcomed me warmly in the Group and that cared about me and motivated me. I thank Anton for the valuable discussions about the gamma sensitivity and the detectors in general. I thank Judith and Maddi that helped me countless times. Maddi made me feel comfortable at the ESS since my first day and Judith is simply outstanding. I thank Irina for the pleasant chats when it was late and the ESS was almost empty. I thank Kelly for the very interesting lessons on LoKI. And I thank Tomek, Anna, Michail, Marita, Rick, Eszter, Hanno and so on. 
\\
I am sorry not to say a few words for everyone...
\\
\\
I want to thank prof. Gorini that gave me this huge opportunity and that provided me prompt assistance from Italy. I thank also all the neutron group of my university, with a special mention for Giorgia, that explained me the basics of the bandGEM detectors before I left for Sweden.
\\
\\
I've been really lucky to have as neighbors the 'Klosterg\aa rden people': Alex, Aitor, Iker and Cris. Moreover, I've been lucky to have as friends Federico, Mateusz, Rodrigo and Jimmy. All of them made my staying in Lund particularly enjoyable with movies every evening, parties, barbecues and so on. I really hope to remain in touch with them, because they are among the most friendly and pleasant people I have ever met. I wish them all the best of luck in their future work. 
\\
I want to thank particularly Alex and Aitor that were present throughout the 6 months of my internship. The beach soccer match between us is arguably the most memorable moment of my staying in Lund!
\\
I thank all my other friends in Lund for the pleasant time spent together.
\\
\\
I would like also to thank my friends in Milan, that I left for 6 months.
}

\restoregeometry
\newpage

%% file: Chapters/Introduction.tex

\chapter{Introduction}

\section{Outline}

Neutrons are electrically neutral particle with a totally negligible electric dipole moment. Thanks to this property they interacts only with nuclei through the strong interaction, without interference due to the electron clouds that surrounds nuclei in matter, for this reason they can penetrate deeper in matter than electrically charged particles and photons. Moreover, while X-rays interacts with the electrons, the neutrons are sensitive to the isotopic composition of the sample instead.
\\
Neutrons of thermal energies ($\approx$ 25 $meV$) have a wavelength of about 1.8 \AA, they correspond to the typical atomic excitations and the interatomic distance in matter. Neutrons are then an important probe for studying the microscopic structure of matter, but neutron scattering experiments are performed only when they provide information that can not be obtained in a simpler, less expensive way, such as X-ray scattering.
\\
\\
Nowadays helium-3 detectors are by far the most common choice for neutron detection, thanks in particular to
their high efficiency (high neutron absorption cross section). It is an inert gas which means that can be safely employed in those environments where users, people using the instruments for their experiments, are present.
\\
Since 2001 the world is experiencing a severe helium-3 shortage. Its price is increasing rapidly and its availability is very limited. This makes the construction of large area detectors ($\approx$ 50 $m^{2}$) not realistic anymore. On the other hand, the neutrons scattering science, in particular the new European Spallation Source (ESS) in Lund (Sweden), is pushing the development toward high fluxes and thus the detector requirements are becoming more and more challenging to fulfill. Although the construction of small area detectors ($\le$ 1 $m^{2}$) would still be possible according to the moderate quantity of helium-3 available, this technology can not cope with the requirements arising today. 
\\
Because of the helium-3 shortage and the increasing instrumental power the detector research in the entire world is focused on the development of detectors based on alternative technologies, such as boron-10, lithium-6 and gadolinium. The ESS will be the first large scale facility based mainly on boron-10 detectors. In order to validate these new technologies, a detailed knowledge of the properties of helium-3 detectors is needed. Although helium-3 detectors have been used for decades and this technology is well known, some of its properties can not be found in the literature.
\\
\\
The work presented in this thesis was carried out at the European Spallation Source (ESS) in Lund (Sweden) in the Detector Group, which is in charge of the research, the design and the maintainin of the neutron detectors that will be used on the instruments. 
\\
We discuss some crucial properties of several helium-3 tubes widely used in the most important facilities in the world. Our goal is to quantify these properties in order to provide a standard for comparing the performance of the helium-3 detectors and the performance of the arising alternative technologies.
\\
Most of the measurements were performed at the Source Facility laboratory, located in the Physics Faculty of the Lund University, where there are a neutron source and several gamma-ray sources. The source facility is a collaboration between the ESS Detector Group and the Department of Nuclear Physics of the Lund University. Some measurements were performed at the beamline R2D2 in the nuclear reactor JEEP II, at the Institute for Energy Technology (IFE), located in Lillestr\o m (Norway). 
\\
\\
In this manuscript we will discuss over the energy resolution we measured for several tubes. The energy resolution of a helium-3 detector is strictly related to the discrimination between gamma-rays and neutrons that can be achieved. In fact, the gamma-rays are the most common background in neutron instruments and thus the detector sensitivity to gamma-rays determines the maximum signal-to-noise achievable. We have also discussed the different contributions to the energy resolution; these provide information about how the helium-3 counters operate.
\\
\\
We have measured the spatial resolution of position sensitive helium-3 tubes in order to quantify and understand the limit that can be achieved with this technology. We performed these measurements at IFE with a collimated neutron beam. The linearity of the spatial resolution has been also investigated.
\\
\\
In the last decade several measurements of the gamma-ray sensitivity have been described in the literature. The gamma-ray sensitivity is the efficiency for miscounting a gamma-ray pulse as a neutron pulse. This quantity is crucial to determine the performance of the detectors.
\\
None of the gamma-ray sensitivity measurements described in the literature involves helium-3 detectors. We have measured the gamma-ray sensitivity for several helium-3 detectors, providing a useful term of comparison for the alternative technologies.
\\
We investigate the different contributions that give rise to the gamma-ray sensitivity in a detector. In order to better understand these contributions we have performed several Monte Carlo simulations.
\\
\\
In order to improve the gamma-ray sensitivity, we tested the feasibility of gamma-ray pulse rejection via Pulse Shape Discrimination (PSD). This technique is commonly used in lithium-6 loaded scintillators, but a comprehensive discussion about its use in helium-3 counters does not exist. While proving its feasibility, we quantify the magnitude of the gamma-ray rejection in helium-3 tubes. The implementation of the PSD technique on helium-3 detectors was also investigated in order to set a possible starting point for the PSD implementation on boron-10 detectors.
\\
\\
In this thesis we describe and quantify several parameters that characterize the performance of helium-3 based counters, in particular helium-3 tubes. These parameters set a standard to systematically compare the different detector technologies.

\section{European Spallation Source}

\begin{minipage}{\textwidth}
\centering
\begin{minipage}[t]{0.49\textwidth}
\begin{figure}[H]
\includegraphics[width=\textwidth]{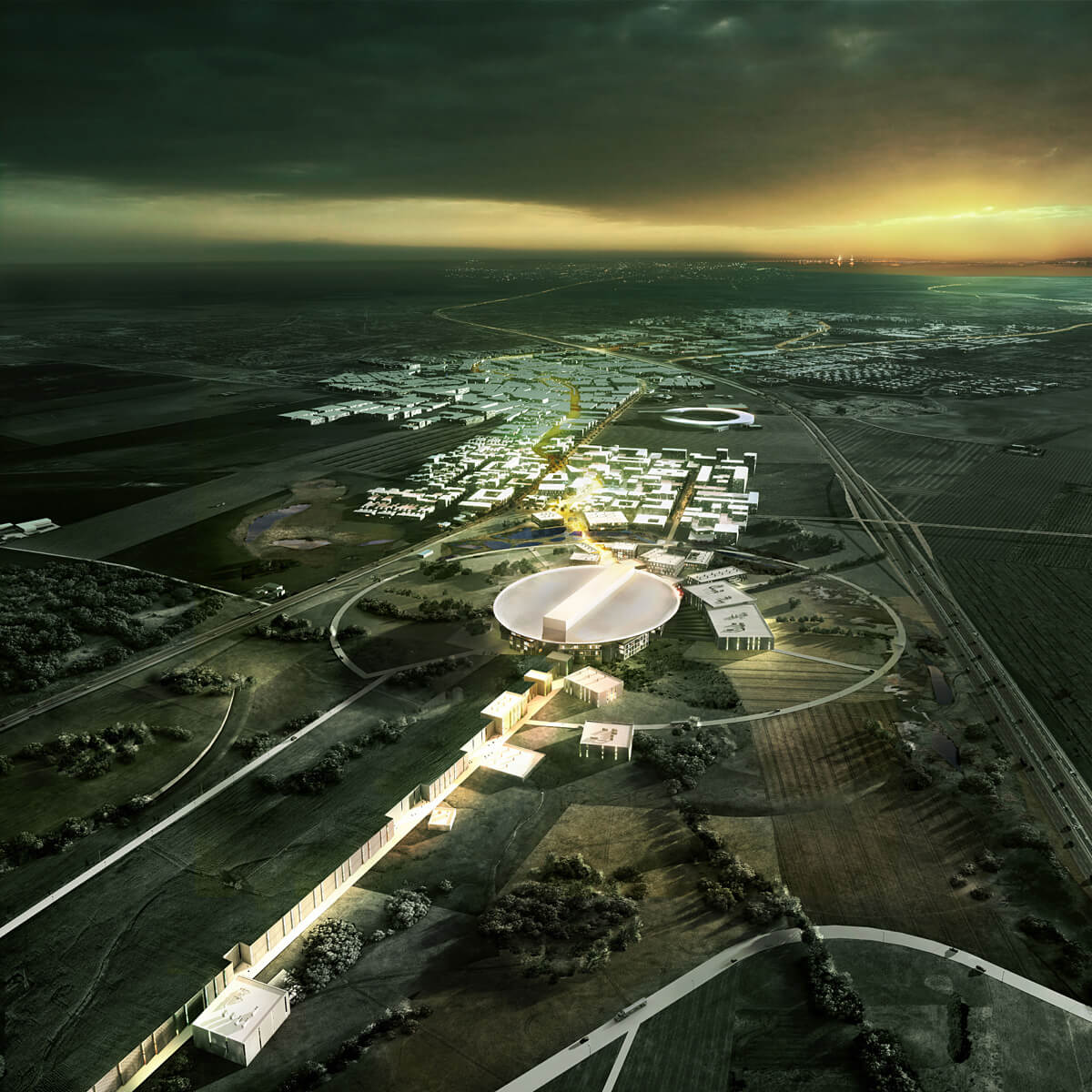}
\caption{Rendering of the ESS construction project.}
\end{figure}
\end{minipage}
\begin{minipage}[t]{0.49\textwidth}
\begin{figure}[H]
\vspace{0.8cm}
\includegraphics[width=\textwidth]{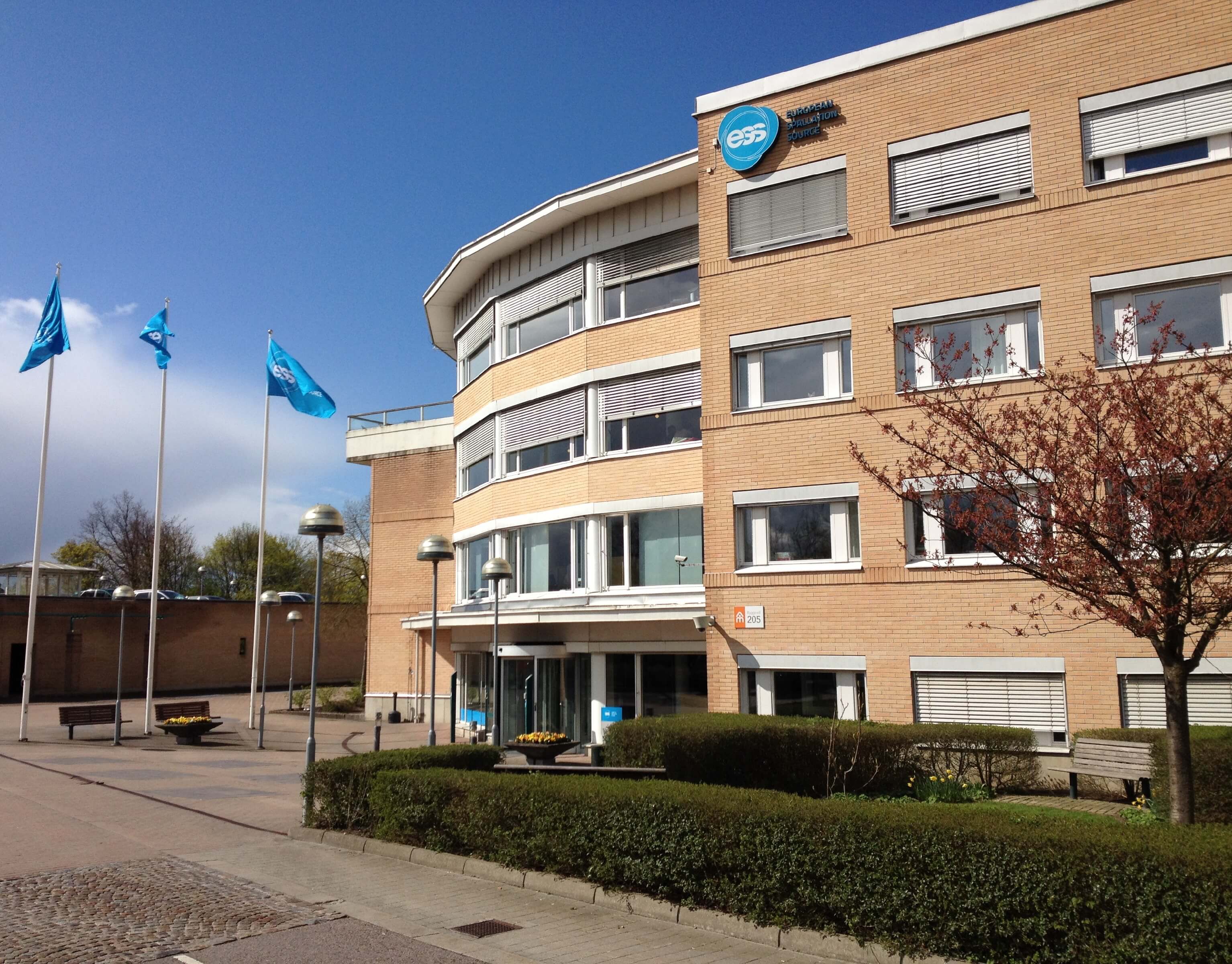}
\caption{Temporary ESS headquarter.}
\end{figure}
\end{minipage}
\end{minipage}
\vspace{1mm}

\begin{figure}[h]
\centering
\includegraphics[width=\columnwidth]{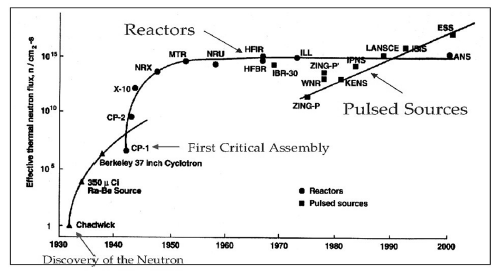}
\caption{The historical trend of the effective thermal neutron flux in the main facilities in the world.}
\label{imm:sources}
\end{figure}

In 1998 the OECD research ministers recommended that a megawatt-class spallation neutron source be built in each of the three developed regions of the world. In 2003 a new concept was put forward for the European Spallation Source (ESS) comprising a 5 $MW$ proton linear accelerator delivering a 2 to 3 millisecond-long pulse to a single target station surrounded by a suite of 20 to 25 neutron instruments.
\\
\\
 The decision to locate the facility in Lund, with the Data Management and Software Centre in Copenhagen, was taken in 2009 and the construction phase started in 2013. The ESS construction cost is calculated to \euro{}1.8 billion, of which nearly half will come from the host countries, the remaining will be provided by the other 15 partner countries (Norway, Spain, Iceland, Netherlands, Germany, France, Estonia, Italy, Czech Republic, Poland, Lithuania, Switzerland, Hungary, Latvia, United Kingdom). The annual operation cost is estimated at about \euro{}140 million. The first neutrons are expected in 2019, with 7 instruments operational from day 1. The facility will be fully operational in 2025 with 22 instruments.
\\
\\
The ESS will produce neutron through spallation. A proton beam is accelerated by a linear accelerator and their collision with the nuclei of a tungsten target. The large energy of the protons is transferred to the nuclei that are split. This reaction generates ions and fast neutrons. The latter are moderated using a water pre-moderator and a $\mathrm{H_2}$ moderator. In this way cold neutrons are obtained that are directed toward the beamlines through the neutron guides.
\\
The spallation source will deliver neutron beams to a suite of research instruments, each devised to extract different kinds of information from the samples studied. ESS will offer unparallelled cold neutron beam brightness, delivering a peak flux 30 times higher than the world's brightest neutron source, and 5 times more power than any spallation source.
\\
\\
Spallation is a more expensive way of producing neutron beams compared with neutrons generated by fission in a nuclear reactor, but it permits to have with relative ease a pulsed beam. This is exploited in Time of Flight instruments and, furthermore, it allows a better temporal characterisation of the sample. Moreover, Figure \ref{imm:sources} shows that the nuclear reactor technology has reached its maximum neutron flux at an effective thermal flux of $\approx$ 10$^{15}$ $n/cm^2\cdot s$, while the pulsed sources can reach larger effective thermal flux and their limit has not been reached yet.

ESS will have long neutron pulses (2.86 $ms$) at a rate of 14 $Hz$ with an average proton power of 5 $MW$. This structure is very well matched to the requirements of low energy neutrons. To fully exploit the long-pulse structure, ESS will rely on components such as guides and choppers to adapt resolution and dynamic range to each individual experiments, rather than hard-wiring these parameters into the design of the source and instruments, as has been done at current spallation and reactor sources. The result will be an unmatched instrument flexibility and the high flux and unique time structure will make possible many investigations that are out of range today, significantly expanding the scientific possibilities.

\begin{figure}[h]
\centering
\includegraphics[width=\columnwidth]{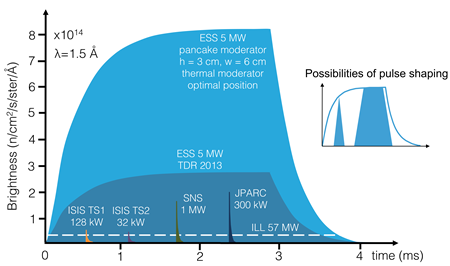}
\caption{Comparison between the ESS pulse and pulses in other neutron facilities.}
\label{imm:comparison}
\end{figure}

The science drivers for ESS are soft condensed matter, life science, magnetic and electric phenomena, chemistry of materials,  energy research, engineering materials and geoscience, archaeology and heritage conservation, fundamental and particle physics. Many of these areas have long traditions using neutron techniques, while in others, the use of neutrons is on the rise. 
\\
The material of this Section is taken from \cite{Peggs2013}.

\section{Helium-3 Shortage}
\label{sec:shortage}

$\mathrm{^{3}He}$ historically has been the uncontested main isotope for neutron detectors. This is a really rare isotope of He (whose main isotope is $\mathrm{^{3}He}$) and it constitutes only the 0.000137\% of the natural helium on Earth. For all practical purpose it is absent in nature.
\\
$\mathrm{^{3}He}$ can be produced using a neutron flux of a nuclear reactor breeding Li, in order to obtain tritium ($\mathrm{^{3}He}$), taking advantage of one of the following reactions:

\begin{equation*}
^{1}n+^{6}Li \longrightarrow ^{3}H + ^{4}He
\end{equation*}
\begin{equation*}
^{1}n+^{7}Li \longrightarrow ^{3}H + ^{4}He+^{1}n
\end{equation*}

$\mathrm{^{3}He}$ is then obtained as a product of tritium decay:

\begin{equation*}
^{3}H \longrightarrow ^{3}He+e^{-}+\overline{\nu}_{e}
\end{equation*}

\begin{figure}[H]
\centering
\includegraphics[width=0.8\textwidth]{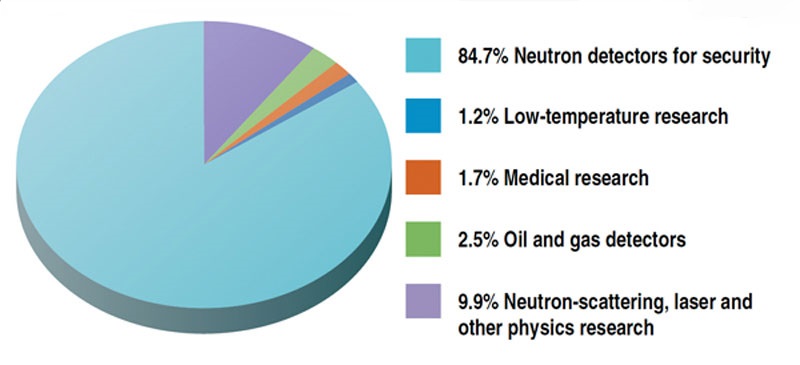}
\caption{Estimate of the $\mathrm{^{3}He}$ usage in the USA \cite{Brown2010}.}
\label{imm:usage}
\end{figure}

Moreover, $\mathrm{^{3}He}$ accumulates in nuclear weapons, because of the tritium decay. It decreases the weapon power and for this reason it is removed and then used for other purposes. Most of the $\mathrm{^{3}He}$ production is from the nuclear weapon stockpiles. 
\\
\\
Until 2001 the $\mathrm{^{3}He}$ production by nuclear weapons exceeded the global demand and the USA accumulated a stockpile. After the terrorist attacks of September 11, 2001 the US federal government began using it for the National Security Programme\footnote{Neutron detectors are used for detecting nuclear and radiological materials.} and hence the annual supply, and thus its price, increased abruptly. Figure \ref{imm:usage} shows that now the majority of the $\mathrm{^{3}He}$ is used for National Security program.

\begin{figure}[H]
\centering
\includegraphics[width=0.8\textwidth]{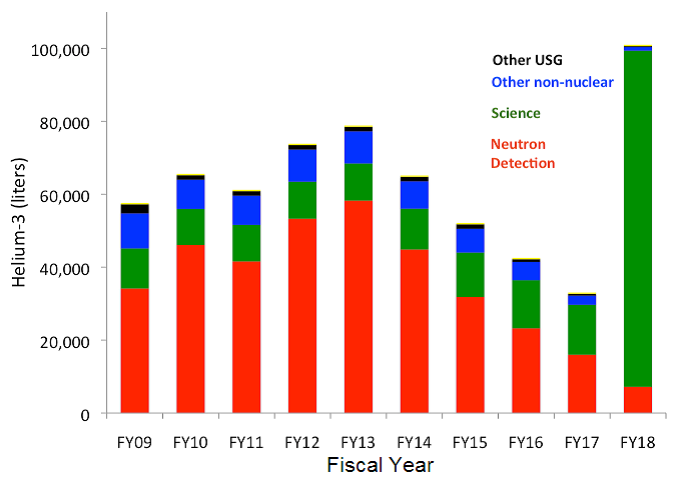}
\caption{The demand of $\mathrm{^{3}He}$ in the last years and the projection for the future \cite{Shea2010}.}
\label{imm:demand}
\end{figure}

The shortage was aggravated by the reduction of the nuclear weapon stockpiles that started in the second half of the '80s with the end of the Cold War. In fact, the USA had 17 519 nuclear weapons in 1985, 10 577 in 1995, 7 700 in 2005 and it is expected to have approximately 3 500 of them in 2022. Russia had 40 000 nuclear weapons in 1986, 25 000 in 1993, 10 000 in 2004 and now slightly more than 5 000.
\\
\\
Other uses of the $\mathrm{^{3}He}$ are cryogenic applications (it has an extremely low boiling point at 4 $K$, that is -269 °$C$) and medical imaging. Both these fields are open research fields and their are developing rapidly, increasing in this way also the $\mathrm{^{3}He}$ demand. Figure \ref{imm:demand} shows that the demand for neutron detection is decreasing, but the demand due to other science applications is strongly increasing. This further reduces the availability of $\mathrm{^{3}He}$ for neutron scattering applications.

\begin{figure}[H]
\centering
\includegraphics[width=0.7\textwidth]{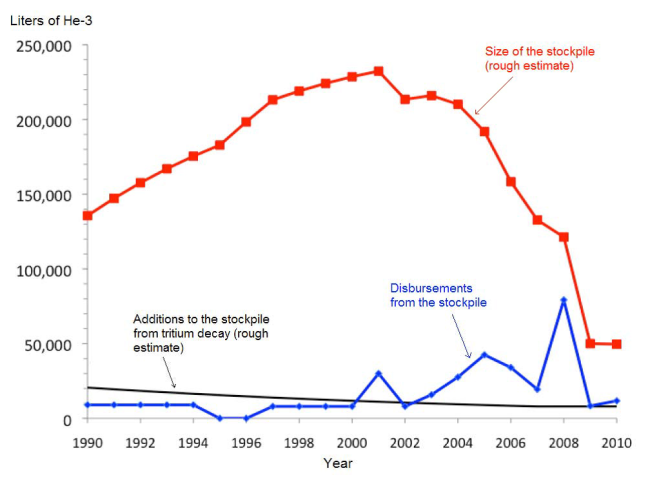}
\caption{The USA helium-3 stockpile from 1990 to 2010 \cite{Shea2010}.}
\label{imm:stockpile}
\end{figure}

After 2001 the demand for $\mathrm{^{3}He}$ exceeded the production and so the USA stockpile began to shrink (Figure \ref{imm:stockpile}).

As a consequence of the shortage the price of the $\mathrm{^{3}He}$ skyrocketed in 2001 from the historical price (between \$100 and \$200 per liter) up to \$2000 per liter. Now, in 2015, the price can be estimate to be between \$2500 and \$5000 per liter.
\\
\\
 $\mathrm{^{3}He}$ is not in the open market and its price is fixed by the producers, mainly the USA and Russian governments. Furthermore, the supply to foreign countries is heavily controlled. The drivers behind the price choice and the supply are not only economical reason, but mainly political. 
\\
In Europe the purchase of big quantities of $\mathrm{^{3}He}$ is particularly problematic, because none of the European countries is a major producer. In the future this circumstance will probably worsen.
\\
\\
The $\mathrm{^{3}He}$ shortage has opened new lines of research in neutron detection. New detectors are studied that exploit the properties of other neutron converter isotopes. The ESS is heavily inserted in this framework of new detectors development as the first large scale facility that will use mainly $\mathrm{^{10}B}$ technologies in neutron scattering instruments. 
\\
\\
The ESS Detector Group R\&D program focuses on alternative isotopes as a neutron converter, such as $\mathrm{^{10}B}$, $\mathrm{^{7}Li}$ and Gd. The choice of new detector technologies is a big challenge and a crucial aspect of the ESS project. For this reason the ESS Detector Group was the first neutron technology group established in Lund in July 2010.

%% file: Chapters/Interactionmatter.tex

\chapter{Interaction of Radiation with Matter}

In this chapter are explained the basic notions of interaction of radiation with matter needed for the comprehension of this thesis. I illustrate the neutron interaction and the photon interaction with matter, the latter is necessary to understand the main characteristics of gamma sensitivity and pulse shape discrimination. Several topics that are closely related to the neutron scattering science, e.g. the coherent and incoherent scattering lengths, are not discussed here, because they are not directly related to the detection mechanisms. The material of this Chapter is taken from \cite{Knoll2010}, \cite{Leo1994}, and \cite{Piscitelli}.

\section{Definition of Cross-Section}

\begin{figure}[H]
\centering
\includegraphics[width=0.7\textwidth]{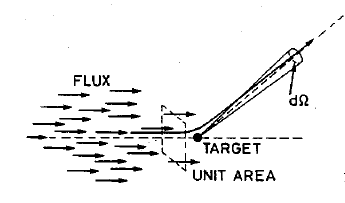}
\caption{The geometry involved in the definition of the cross section.}
\label{imm:cross}
\end{figure}

The collision or interaction of two particles is generally described in terms if the \textit{cross-section}. This quantity essentially gives a measure of the probability for a reaction to occur. Consider a beam of particles incident on target particle and assume the beam to be much broader than the target, as shown in Figure \ref{imm:cross}. Suppose that the particles in the beam are randomly distributed in space and time. The \textit{flux} $\phi$ is given by the number of incident particles per unit area per unit time. Now look at the number of particles scattered into the solid angle d$\Omega$ per unit time. By \textit{scattering} we mean any reaction in which the outgoing particle is emitted in the solid angle $\Omega$.  If we average, the number of particles scattered in a solid angle d$\Omega$  per unit time will tend toward a fixed value dN$_s$. The \textit{differential cross-section} is then defined as:

\begin{equation}
\frac{d\sigma}{d\Omega}(E,\Omega)=\frac{1}{\phi}\cdot \frac{dN_s}{d\Omega}
\label{eq:diffcross}
\end{equation}

that is, $\frac{d\sigma}{d\Omega}$ is the average fraction of particles scattered into d$\Omega$ per unit time per unit incident flux $\phi$ for d$\Omega$ infinitesimal. Note that d$\sigma$ has the dimension of an area. We can interpret d$\sigma$ as the geometrical cross sectional area of the target intercepting the beam. In other words, the fraction of flux incident on this area will interact with the target and scatter into the solid angle d$\Omega$ while all those missing d$\sigma$ will not.

In general the differential cross-section of a process varies with the energy of the reaction and with the angle at which the particle is scattered. We can calculate a \textit{total cross-section}, for any scattering whatsoever at an energy E, as the integral of the differential cross-section over all solid angles as follows:

\begin{equation}
\sigma(E)=\int d\Omega\cdot \frac{d\sigma}{d\Omega}(E,\Omega)
\label{eq:totalcross}
\end{equation}

Consider now a real target, which is usually a slab of material containing many scattering centers. We want to know how many interactions occur on average when that target is exposed to a beam of incident particles. Assuming that the slab is not too thick so that the likelihood of interaction is low, the number of centers per unit perpendicular area which will be seen by the beam is then $n\cdot \delta x$ where $n$ is the volume density of centers and $\delta x$ the thickness of the material along the direction of the beam. If A is the perpendicular area of the target and the beam is broader that the target, the number of incident particles which are eligible for an interaction per unit time is $\phi \cdot A$. The average number of scattered particles into $d \Omega$ per unit time is:

\begin{equation}
N_s(\Omega)=\phi \cdot A \cdot n \cdot \delta x \cdot \frac{d\sigma}{d\Omega}\cdot d\Omega
\label{eq:scatpart}
\end{equation}

The total number of scattered particles into all angles is similarly:

\begin{equation}
N_{tot}=\phi \cdot A \cdot n \cdot \delta x \cdot \sigma
\label{eq:totalscatpart}
\end{equation}

In the case the beam is smaller than the target, we need only to set A equal to the area covered by the beam. We can take another point of view; that is the probability of an incident particle of the beam to be scattered. If we divide Equation \ref{eq:totalscatpart} by the total number of incident particles per unit time ($\phi \cdot A$), we have the probability for the scattering of a single particle in a thickness $\delta x$:

\begin{equation}
P_{\delta x}=n \cdot \sigma \cdot \delta x
\label{eq:probscattering}
\end{equation}

Note that the probability for interaction is proportional to the distance traveled, $dx$, in the first order. 
\\
Let us consider now a more general case of a macroscopic thickness $x$. In particular we search what is the probability for a particle not to suffer an interaction over a distance $x$ traveled in the target. This is known as the \textit{survival} probability. Let's denote with $P(x)$ the probability of not having an interaction after a distance $x$ and with $w dx$  the probability of having an interaction between $x$ and $x+dx$. Using these definitions we find:

\begin{equation}
P(x+dx)=P(x)(1-wdx)
\end{equation}

Now, using the power series expansion of $P(x+dx)$ to the first order in dx:

\begin{equation}
P(x)+\frac{dP}{dx}dx=P(x)-Pwdx \implies dP=-wPdx
\end{equation}

We find now, resolving this differential equation we obtain:

\begin{equation}
P(x)=Ce^{-wx}=e^{-wx}
\end{equation}

where $C=1$ because of the condition $P(0)=1$. Now, using the definition of $w$ and Equation \ref{eq:probscattering} we find:

\begin{equation}
w=n \cdot \sigma \equiv \Sigma
\end{equation}

where $\Sigma$ is called \textit{macroscopic cross-section} and has the dimension of $\frac{1}{\text{length}}$. Finally, the equation for the survival probability can be written:

\begin{equation}
P(x)=e^{-\Sigma x}
\end{equation}

From this equation it's easy to find the equations that gives the probability that a particle interacts after a distance $x$ traveled in the target:

\begin{equation}
\label{eq:probinteraction}
P_{int}=1-e^{-\Sigma x}
\end{equation}

Now, using the survival probability, we can define the \textit{mean free path} as the mean distance $\lambda$ traveled by the particle without suffering a collision, that is

\begin{equation}
\lambda=\frac{\int x P(x) dx}{\int P(x) dx} = \frac{1}{\Sigma} = \frac{1}{N\sigma}
\end{equation}

We point out that in the case of low $\Sigma x$ Equation \ref{eq:probinteraction} can be approximated to:

\begin{equation}
P_{int} \simeq \Sigma \cdot x = n \sigma x
\end{equation}

\section{Interaction of Heavy Charged Particles}

Heavy charged particles interact with matter primarily through coulomb forces between their positive charge and the negative charge of the orbital electrons within the absorber atoms. The interactions of these particles with nuclei are also possible, but such encounters occur only rarely and they are not normally significant in the response of radiation detectors. 
\\
Upon entering any absorbing medium, the charged particle immediately interacts simultaneously with many electrons. In any of such encounter, the electron feels an impulse from the attractive coulomb force as the particle passes its vicinity. Depending on the proximity of the encounter, this impulse may be sufficient either to raise the electron to a higher-lying shell within the absorber atom (excitation) or to remove completely the electron from the atom (ionization). The maximum energy that can be transferred from a charged particle of mass $m$ with kinetic energy $E$ to an electron of mass $m_e$ in a single collision is $4Em_e / m$, that is about 1/2000 of the particle energy for an alpha particle. Because this is a small fraction of the total energy, the primary particle must lose its energy in many such interactions during its passage through an absorber. At any given time, the particle is interacting with many electrons, so the net effect is to decrease its velocity continuously until the particle is stopped. Except at the very end, the tracks tend to be quite straight because the particle is not greatly deflected by any one encounter, and interactions occur in all directions simultaneously.
\\
The product of the interaction of heavy charged particles with a atoms are either excited atoms or ion pairs, which are made up of a free electron and the corresponding positive ion of an absorber atom from which an electron has been totally removed. The ion pairs have a natural tendency to recombine to form neutral atoms, but in ionization detectors (as the helium-3 detectors discussed in this thesis) this recombination is suppresed so that the ion pairs can be used as the basis of the detector response.

\tocless\subsection{Specific Energy Loss}

A critical quantity for analyzing heavy charged particles interaction in matter is the \textit{linear stopping power} for charged particles in a given absorber. This is simply defined as the differential energy loss for that particle within the material, divided by the corresponding differential path length:

\begin{equation}
S=-\frac{dE}{dx}
\end{equation}

For particles with a given charge state, S increases as the particle velocity is decreased. The classical expression that describes the specific energy loss is known as the \textit{Bethe formula} and is written

\begin{equation}
-\frac{dE}{dx}=\frac{4 \pi e^{4}z^{2}}{m_{e} \nu^{2}}nB
\end{equation}

where

\begin{equation}
B \equiv Z \left[ log\left( \frac{2 m_{e} \nu^{2}}{I}\right) - log\left( 1-\frac{\nu^2}{c^2}\right) -\frac{\nu^2}{c^2}\right]
\end{equation}

Where $\nu$ and $ze$ are the velocity and charge of the primary particle and $n$ and $Z$ are respectively the density and the atomic number of the absorber atoms in the target. $I$ represents the average excitation and ionization potential of the absorber and is normally treated as an experimentally determined parameter for each element. For non relativistic charged particles only the first term in $B$ is significant. 
\\
We can notice that B varies slowly with the particle energy, so, for nonrelativistic particles, dE/dx varies approximately as $1/\nu^2$, or inversely with particle energy. This behavior can be heuristically explained by noting that because the charged particle spends a greater time in the vicinity of any given electron when its velocity is low, the impulse felt by the electron, and hence the energy transfer, is largest. Moreover particles with the greatest charge will have the largest specific energy loss. In comparing different materials as absorbers, $dE/dx$ depends primarily on the product $NZ$, which is outside the logarithmic term and represents the electron density of the absorber. The stopping power for many different types of charged particles approaches a near-costant broad minimum value at relativistic energies, that is several hundred MeV. This specific energy loss corresponds to about 2 MeV per g/cm$^2$ in light materials.

\tocless\subsection{Bragg Curve}

A plot of the specific energy loss along the track of a charged particle is known as a \textit{Bragg curve}. For high energies this quantity increases with the distance of penetration, because the particle is slowing down and so its specific energy loss is increasing. At a certain depth the velocity of the particle becomes low enough and the ion starts recombination with the electrons he collides with. In this case the charge of the ion is reduced and so the specific energy loss, that is the Bragg curve starts decreasing fast with the distance of penetration after a certain depth. Charged particles with the greatest number of nuclear charges begin to pick up electrons early in their slowing down process.

\tocless\subsection{Particle range}

\begin{figure}[H]
\centering
\includegraphics[width=0.6\textwidth]{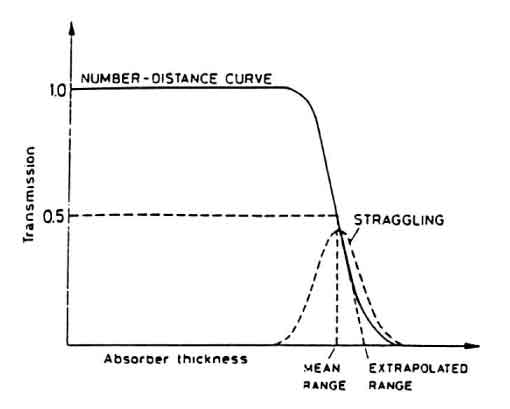}
\caption{Number-distance curve and straggling. In a first approximation, distribution of ranges is approximately Gaussian in form.}
\label{imm:range}
\end{figure}

A very important property for charged particles interaction in matter is how far these particles will penetrate before they lose all of their energy. In a first approximation, the energy loss is continuous, this distance must be a well defined number, the same for all identical particles with the same initial energy in the same type of material. This quantity is called the \textit{range} of the particle, and depends on the type of the material, the particle type and its energy.
\\
Actually the energy loss is not continuous, but statistical in nature. Indeed, two identical particles with the same initial energy will not in general suffer the same number of collisions with the same energy loss. In fact a measurements with an ensemble of identical particles will show a statistical distribution centered about some mean value. This phenomenon is known as \textit{range straggling}. The mean value of the distribution of ranges is known as \textit{mean range} and corresponds to the midpoint of the descending point of the number-distance curve, as shown in Figure \ref{imm:range}. This is the thickness at which roughly half the particles are absorbed. More commonly, however, what is desired is the thickness at which all the particles are absorbed, in which case the point at which the curve drops to the background level should be taken. This point is usually found by taking the tangent to the curve at the midpoint and extrapolating to the zero-level. This value is known as the \textit{extrapolated range}.

From a theoretical point of view, we can calculate the mean range of a particle of a given energy, $E_0$, by integrating the $dE/dx$ formula:

\begin{equation}
R(E_{0})=\int^{E_0}_{0} \left( \frac{dE}{dx} \right)^{-1} dE
\end{equation} 

This yields the approximate pathlength travelled. However, this Equation ignores the effect of multiple Coulomb scattering, which causes the particle to follow a zigzag path through the absorber. Thus, the range, defined as the straight-line thickness, will generally be smaller than the total zigzag pathlength.
\\
As it turns out, however, the effect of multiple scattering is generally small for heavy charged particles, so that total path length is, in fact, a relatively good approximation to the straight-line range. 

\tocless\subsection{Energy Loss of Electrons and Positrons}

Electrons and positrons while passing through matter are influenced by really high accelerations, because of their low mass. This brings deviation from the straight-line path and causes the emission of electromagnetic radiation, called in this case \textit{bremsstrahlung} ("braking radiation"). At energies of a few $MeV$ or less, this process is still a relatively small factor. However, as the energy is increased, the probability of bremsstrahlung quickly shoots up so that at a few 10's of $MeV$, loss of energy by radiation is comparable to or greater than the collision-ionization loss. 
\\
\\
In the measurements discussed in this thesis only electrons with less than 1.332$MeV$ are present. For this reason bremsstrahlung has been neglected.
\\
\\
There is another effect that is a consequence of the fact that electrons undergo large-angle deflections along their tracks, that is the phenomenon of backscattering. In fact an electron entering one surface of an absorber may undergo sufficient deflection so that it re-emerges from the surface through which it entered. Backscattering is most pronounced for electrons with low incident energy and absorbers with high atomic number.

\section{Interaction of Gamma-Rays}
\label{sec:interactiongamma}

Although a large number of possible interaction mechanisms are known for gamma-rays in matter, only three major types play an important role in radiation measurements: \textit{photoelectric electric absorption}, \textit{Compton scattering} and \textit{pair production}. All these processes lead to the partial or complete transfer of the gamma-ray photon energy to electron energy. They result in sudden and abrupt changes in the gamma-ray photon history, in that the photon either disappears entirely or is scattered through a significant angle. 

\tocless\subsection{Photoelectric Absorption}

In the photoelectric absorption a photon undergoes an interaction with an absorber atom in which the photon completely disappears. In its place, an energetic photoelectron is ejected by the atom from one of its bound shells. For gamma-rays of sufficient energy, the most probable origin of the photoelectron is the most tightly bound or K shell of the atom. The photoelectron appears with an energy given by:

\begin{equation}
E_{e^{-}}= h \nu - E_{b}
\end{equation}

where $E_{b}$ represents the binding energy of the photoelectron in its original shell.
\\
In addition to the photoelectron, the interaction also creates an ionized absorber atom with a vacancy in one of its bound shells. This vacancy is quickly filled through capture of a free electron from the medium and/or rearrangement of electrons from other shells of the atom. Therefore, one or more characteristic X-ray photons may also be generated. In some fraction of the cases, the emission of an Auger electron may substitute for the characteristic X-ray in carrying away the atomic excitation energy.
\\
\\
The photoelectric process is the predominant mode of interaction for gamma-rays (or X-rays) of relatively low energy. The process is also enhanced for absorber materials of high atomic number $Z$. No single analytic expression is valid for the probability of photoelectric absorption per atom over all ranges of $E_{\gamma}$ and $Z$, but a rough approximation is:

\begin{equation}
\tau \sim \frac{Z^{n}}{E_{\gamma}^{3.5}}
\end{equation}

where the exponent $n$ varies between 4 and 5 over the gamma-ray energy region of interest.

\tocless\subsection{Compton Scattering}

Compton scattering is the scattering of photons on free electrons. In matter the electrons are bound; however, if the photon energy is high with respect to the binding energy, this latter energy can be ignored and the electrons can be considered as essentially free. In this process the incoming gamma-ray is deflected through an angle $\theta$ with respect to its original direction. The photon transfers a portion of its energy to the electron that leaves the atom, so then it is known an a \textit{recoil electron}. Because all angles of scattering are possible, the energy transferred to the electron can vary from zero to a large fraction of the gamma-ray energy.
\\
Applying energy and momentum conservation to the scattering process we can find a relation between the scattering angle $\theta$ of the photon and the final energy of the electron, assumed to be initially at rest:

\begin{equation}
\label{eq:compton}
E_{e^{-}}= h \nu \frac{A \cdot (1-cos\theta)}{1+A\cdot (1-cos\theta)}
\end{equation}

where $A=h\nu /m_{e}c^2$. For small scattering angles $\theta$, very little energy is transferred. Some of the original energy is always retained by the incident photons, even in the extreme case of $\theta=\pi$. 
\\
Using Equation \ref{eq:compton} one finds that the maximum recoil energy allowed by kinematics for the recoil electron is given by

\begin{equation}
E_{max}=h \nu \cdot \frac{2\cdot A}{1+2\cdot A}
\end{equation}

The probability of Compton scattering depends on the number of electrons available as scattering targets and therefore increases linearly with $Z$.

Finally, the differential cross section for Compton scattering can be calculated using quantum electrodynamics and is known as the Klein-Nishina formula, which we won't recall here. From this formula is shown the strong tendency for forward scattering at high values of the gamma-ray energy, that is $\gtrsim$100 $keV$.

\tocless\subsection{Thomson and Rayleigh Scattering}

Two classical processes related to Compton scattering are \textit{Thomson} and \textit{Rayleigh} scattering.
\\
Thomson scattering is  the scattering of free electrons in the classical limit. This process happens in the low-energy limit, when the electric field of the incident photon accelerates the electron, causing it, in turn, to emit radiation at the same frequency as the incident photon. This is a diffusion process typical of waves.
\\
Rayleigh scattering, on the other hand, is the scattering of photons by atoms as a whole. This is also called \textit{coherent} scattering.
\\
In both processes, the scattering is characterized by the fact that no energy is transferred to the medium, but only the direction of the photon is changed. At the relatively high energies of x-rays and gamma-rays, Thomson and Rayleigh scattering are very small and for most purposes can be neglected.

\tocless\subsection{Pair Production}

if the gamma-ray energy exceeds twice the rest-mass energy of an electron (1.02 $MeV$), the process of pair production is energetically possible. As a practical manner, the probability of this interaction remains very low until the gamma-ray energy approaches several $MeV$ and therefore pair production is predominantly confined to high-energy gamma rays. In the interaction (which must take place in the coulomb field of a nucleus), the gamma-ray photon disappears and is replaces by an electron-positron pair. All the excess energy carried in by the photon above the 1.02 $MeV$ required to create the pair goes into kinetic energy shared by the positron and the electron. Because the positron will subsequently annihilate after slowing down in the absorbing medium, two annihilation photons are normally produced as secondary products of the interaction.
\\
\\
No simple expression exists for the probability of pair production, but its magnitude varies approximately as the square of the absorber atomic number. The importance of pair production rises sharply with energy.

\tocless\subsection{Total Cross Section}

The total cross section for interaction of a photon is given by the sum of the cross sections of these processes:

\begin{equation}
\sigma_{tot} = \sigma_{photoelectric} + \sigma_{Compton} + \sigma_{pair} + \sigma_{coherent}
\end{equation}

\begin{figure}[H]
\centering
\includegraphics[width=0.8\textwidth]{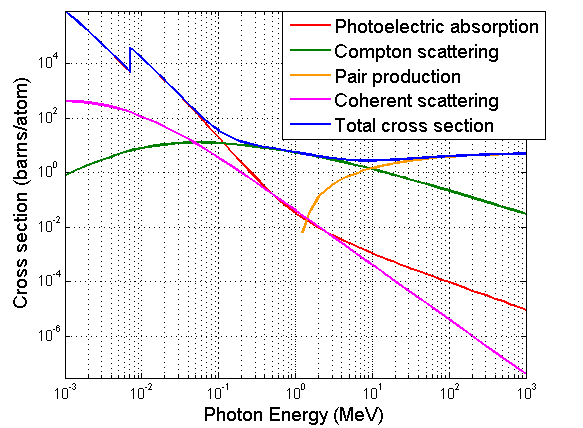}
\caption{Total cross section in stainless steel taken from \cite{Berger} .}
\label{imm:crosssteel}
\end{figure}

Figure \ref{imm:crosssteel} shows the total cross section of interaction in stainless steel.

A few characteristics of the total cross section are valid for most of the elements. A peak is present at about 10 $keV$ and corresponds to the K edge peak for the photoelectric effect. The photoelectric effect is the dominant process up to $\approx$100 $keV$, while between $\approx$100 $keV$ and $\approx$10 $MeV$ is the Compton scattering the dominant one. Pair production begins at 1.02 $MeV$, but only at over 10 $MeV$ it gives an important contribution to the cross section. Coherent scattering, on the other hand, is always about two order of magnitude lower than the dominating process and so it is usually neglected.

\section{Interaction of Neutrons}

In common with gamma-rays neutrons carry no charge and therefore cannot interact in matter by means of the Coulomb force. Instead, its principal means of interaction is through the strong force with nuclei. These reactions are, of course, much rarer in comparison because of the short range of this force. As a result of the interaction, the neutron may either totally disappear and be replaced by one or more secondary radiations, or else the energy or direction of the neutron is changed significantly. The important process in the energy region of our interest ($\approx$ 10$^{-2}$-10$^{5}$ $eV$) are: \textit{elastic scattering}, \textit{inelastic scattering}, \textit{neutron capture} and \textit{fission}.
\\
Often in neutron scattering applications, wavelength is used instead of energy. Wavelength is related to the typical distances analyzable of the sample. We recall that wavelength and energy are related by the deBroglie equation:

\begin{equation}
\lambda [\AA]= \frac{h}{\sqrt{2Em}} = \frac{0.285}{\sqrt{E[eV]}}
\end{equation}
\\
\\
The relative probabilities of the various types of neutron interactions change dramatically with neutron energy. We will divide neutrons in two categories on the basis of their energy, either \textit{fast neutrons} or \textit{slow neutrons}. We define the dividing line to be about 0.5 $eV$ (0.4 $\AA$), the energy of the abrupt drop in absorption cross section in cadmium (the cadmium cutoff energy). Much of the population in the slow energy range , because of the thermalisation process with the elements surrounding the source, will be \textit{thermal neutrons} which, at room temperature (300 $K$), have an average energy of about 25 $meV$ (1.8 $\AA$).

\tocless\subsection{Elastic Scattering}

Neutrons scatters from the nuclei in matter. This is the principal mechanism of energy loss for fast neutrons and this often serve to bring fast neutrons into thermal equilibrium with the absorber medium, creating thermal neutrons. This process is called \textit{moderation}. At energies of several $MeV$ the collision may be treated nonrelativistically and very simply with conservation laws. We find the relation between the final kinetic energy of the neutron in the laboratory system and the scattering angle in the center-of-mass system:

\begin{equation}
E= E_0 \cdot \frac{A^{2}+1+2 A\cdot \cos (\theta_{cm})}{(A+1)^2}
\end{equation}

where $E_{0}$ is the initial energy of the neutron, $A$ is the atomic mass number of the absorber material and $\theta_{cm}$ is the scattering angle in the center-of-mass system. From this equation it is easy to see that the energy of the scattered neutron is limited to the range:

\begin{equation}
\left(\frac{A-1}{A+1}\right)^2 \cdot E_0 \le E \le E_0
\end{equation}

and the limits correspond to scattering at $cos(\theta_{cm})=\pm1$. The lighter the nucleus and the more recoil energy is it absorbs from the neutron. In particular a proton can absorb all the neutron energy in a single collision. This implies that the slowing down of neutrons is most efficient when protons or light nuclei are used. This explains the use of hydrogenous materials such as water, paraffin ($\mathrm{CH_2}$) or polyethylene ($\mathrm{(C_2 H_4)_n}$) in connection with neutron moderators and shielding.

\tocless\subsection{Inelastic Scattering}

Inelastic scattering is a scattering process in which the nucleus is left in an excited state which may later decay by gamma-ray or some other form of radiative emission and the neutron loses a greater fraction of its energy than it would in an equivalent elastic collision. In order for the inelastic reaction to occur, the neutron must have sufficient energy to excite the nucleus, usually on the order of 1 $MeV$ or more. Below this energy threshold, only elastic scattering may occur. 

\tocless\subsection{Neutron Capture}

In neutron capture reactions the neutron is captured by a nucleus and either a gamma or charged particles are emitted. Some common reactions are $(n,\gamma)$, $(n, p)$, $(n, t)$ and $(n, \alpha)$. The slower the neutron is and the more time it will spend in the proximity of a nucleus, increasing the probability of being absorbed. For this reason the cross-section for neutron capture goes approximately as $\simeq 1/v$, where $v$ is the velocity of the neutron. Depending on the element, there may also be resonance peaks superimposed upon this $1/v$ dependence. Neutron capture generally occurs in the $eV$ to $keV$ region.
\\
Neutron capture reactions often are two-body reactions. The initial kinetic energy of the slow neutron (<1 $eV$) can be neglected in comparison to the much higher energy released by the reaction, i.e. the Q-value (usually >100 $keV$). In this approximation conservation laws impose that the neutron capture fragments are emitted back-to-back. The emission angles therefore do not depend on the initial velocity of the neutrons, but can be considered randomly distributed.
\\
\\
The slow neutron interactions of real importance are neutron-induced reactions that can create secondary radiations of sufficient energy to be detected directly. Because the incoming neutron energy is so low, all such reactions must have a positive Q-value to be energetically possible.

\tocless\subsection{Nuclear Fission}

Nuclear fission is a process in which one nucleus splits in two smaller nuclei. This can occur as a radioactive decay (\textit{spontaneous fission}), but this phenomenon is very rare except in a few very heavy isotopes, such as $^{240}Pu$ and $^{252}Cf$. More commonly nuclear fission occurs as the result of the nuclear excitation energy produced when a nucleus captures a neutron. The de-excitation of the nucleus creates two smaller fragments, neutrons (usually 2 or 3) and some secondary radiation, such as gamma-rays. Isotopes that can sustain this reaction are called fissile nuclei and they they are heavy isotopes. The most commonly used are $^{235}U$ and $^{239}Pu$.
\\
\\
Typical fission events have a Q-value of about 200 $MeV$ of energy. This energy is hundreds of times higher than the typical Q-value for neutron capture reactions. As for neutron capture, fission is most likely at thermal energies and its cross-section goes approximately as $\simeq 1/v$.

\section{Radioactive Sources}

\tocless\subsection{Activity}

The activity $A$ of a radioisotope source is defined as its rate of decay:

\begin{equation}
A(t)=\left| \frac{dN(t)}{dt} \right|= \lambda N(t) \qquad \Longrightarrow \qquad N(t)=N_{0} e^{-\lambda t}
\end{equation}

where $N(t)$ and $N_0$ are the number of radioactive nuclei at the time $t$ and $t=0$ respectively. $\lambda$ is called \textit{decay constant} and is the probability per unit time for a nucleus to decay. In the SI the activity is measured in $Bq$, defined as 1 decay per second, but historically has been widely used the Curie $Ci$, defined as 3.7$\cdot$10$^{10}$ decays per second (the activity of one gram of $^{226}Ra$). The activity can be equivalently expressed as the average lifetime of radioisotope $\tau=\ / \lambda$ or the half-time $t_{1 / 2}= \tau \cdot \log 2$, the amount of time required for the nuclei population to fall to half its initial value.
\\
By knowing the activity at time $t=0$ it is possible to calculate the activity at time $t$:

\begin{equation}
A(t)=\lambda \cdot N(t) = \lambda \cdot N_{0} e^{-\lambda t} = A_{0} \cdot e^{-\lambda t}
\end{equation}

It should be emphasized that activity measures the source decay rate, which is not necessarily the emission rate of radiation produced in its decay. Frequently, a given radiation will be emitted in only a fraction of all the decays, thus a knowledge of the decay scheme of the particular isotope is necessary to infer a radiation emission rate from its activity. Moreover, the decay of a radioisotope may lead to a daughter product whose activity also contributes to the radiation yield from the source. The \textit{branching ratio} for a decay is the fraction of particles which decay by an individual decay mode with respect to the total number of particles which decay. With the branching ratio is it possible to infer a radiation emission rate from the activity of a given source.

\tocless\subsection{$\alpha$-n Neutron Sources}

The most common portable source of neutrons is obtained by the bombardment of $Be$ with $\alpha$-particles emitted by the decay of an other isotopes, e.g. $^{241}Am$ and $^{239}Pu$. The $\alpha$-particles must have an energy greater than a few $MeV$ in order to overcome the Coulomb repulsion between the particle and the nucleus ($Be$ is used because of its low Coulomb force). The reaction is the following:

\begin{equation}
^{4}_{2}\alpha + ^{9}_{4}Be \quad \longrightarrow \quad ^{12}_{6}C+^{1}_{0}n
\end{equation}

which has a Q-value of 5.71 $MeV$. The resulting neutrons are fast and they must be moderated if slow neutrons are needed, therefore often the source is surrounded by moderating materials. Most of the $\alpha$-particles are simply stopped in the target, and only 1 in about $10^{4}$ reacts with a $Be$ nucleus. About 70 neutrons are produced per $MBq$ of $^{239}Pu$.

%% file: Chapters/Neutrondetectors.tex

\chapter{Neutron Detectors}

In this Chapter the main property of gas detectors are discussed, focusing on the characteristics of proportional counters and Position Sensitive Detectors (PSD). After a general introduction, a more detailed description of $^{3}He$ proportional tubes and the analysis of their Pulse Height Spectrum (PHS) are given. Finally, a brief overview of the most important alternative technologies to $^{3}He$ for neutron detection is given. Some of the material of this Chapter is taken from \cite{Knoll2010}, \cite{Leo1994}, and \cite{Piscitelli}.

\section{Gaseous Detectors}
\label{sec:gasdetectors}

Gaseous ionization detectors were the first electrical devices developed for radiation detectors. These instruments are based on the direct collection of the ionization electrons and ions produced in a gas by passing electrons. Because of the greater mobility of electrons and ions, a gas is the obvious medium to use for the collection of ionization from radiation. 
\\
The basic configuration configuration consists of a container with conducting walls and a wire in it. The container is filled with a suitable gas, usually a noble gas such as argon or helium. The wire is kept at a positive voltage while the walls are grounded. We consider a cylindrical geometry, as in the case of the $^{3}He$ detectors we used. In this case a radial electric field is established:

\begin{equation}
E=\frac{1}{r}\frac{V_0}{\log (b/a)}
\end{equation}

where $V_0$ is the voltage applied to the wire, $b$ is the inside radius of the cylinder and $a$ is the radius of the wire. If a particle now penetrates the cylinder, a certain number of electron-ion pairs will be created, either directly, is the radiation is a charged particle, or indirectly through secondary radiation if the particle is neutral. The mean number of pairs created is proportional to the energy deposited in the counter. Under the action of the electric field, the ions will be accelerated towards the cathode, i.e. the walls, and the electrons towards the anode wire, where they are collected.
\\
The current signal observed depends on the field intensity. At zero voltage no charge is collected as the ion-electron pairs recombine under their own electrical attraction. As the voltage is raised, however, the recombination forces are overcome and the current begins to increase as more and more of the electron-ion pairs are collected before they recombine. At some point all created pairs will be collected and further increases in voltage show no effect. A detector working in this region is called an \textit{ionization chamber} since it collects the ionization produced directly by the passing radiation. The signal current, however, usually it is too small for being of practical use.
\\
If we know increase the voltage at a certain point the electric field is strong enough to accelerate freed electrons to an energy where they are also capable of ionizing gas molecules in the cylinder. The electrons liberated in these secondary ionizations then accelerate to produce still more ionization and so on. This is called \textit{Townsend avalanche} and the fractional increase in the number of electrons per unit path length is governed by the Townsend equation:

\begin{equation}
\frac{dn}{n}=\alpha dx
\end{equation}

$\alpha$ is called first Townsend coefficient for the gas. Generally its value increases with increasing field strength. Several direct measurements of the first Townsend coefficient in helium can be found in the literature \cite{Cavalleri1969, Chanin1964}. Even analytical equations for its calculation have been theoretically found \cite{Abdelnabi1953}. For a spatially constant field (as in parallel plate geometry), $\alpha$ is a constant in the Townsend equation. Its solution then predicts that the density of electrons grows exponentially with distance as the avalanche progresses:

\begin{equation}
n(x)=n_0 e^{\alpha x}
\end{equation}

\begin{figure}[H]
\begin{center} 
\includegraphics[width=0.6\textwidth]{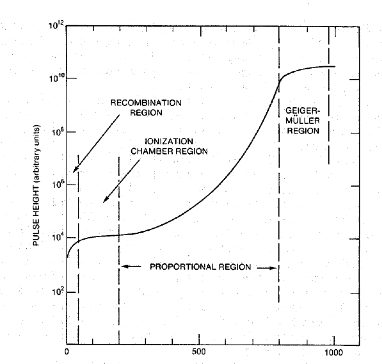} 
\end{center} 
\caption{Gain vs. applied HV curves. The ionization chamber, proportional and Geiger M{\"u}ller regions of operation are also shown. \cite{Crane1991}.} 
\label{imm:gainvoltage}
\end{figure}

For the cylindrical geometry the electric field increases in the direction that the avalanche progresses, and the growth with the distance is even steeper. Since the electric field is strongest near the anode, the avalanche occurs very quickly and almost entirely within a few radii of the wire. \\
The number of electron-ion pairs in the avalanche is directly proportional to the number of primary electrons. The result is a proportional amplification of the current, with a multiplication factor depending on the working voltage. A detector operating in this domain is known as \textit{proportional chamber}.
\\
\\
If the voltage is further increased, the total amount of ionization created through multiplication becomes sufficiently large that the space charge created distorts the electric field about the anode. Thus proportionality begins to be lost. Increasing $V$ still higher we reach the \textit{Geiger-M{\"u}ller region}. In this region the photons emitted by molecules in the gas, previously excited by the avalanche, play an important role. In fact if the gas multiplication is relatively low, as in a proportional tube, the number of excited molecules formed in a typical avalanche is not very large. Also, because most gases are relatively transparent in the visible and UV regions, the probability of photoelectric absorption of any given photon is also relatively low. However, in the Geiger-M{\"u}ller region the multiplication represented by a single avalanche is high enough for creating at least a photon that moves in the gas and is absorbed elsewhere, triggering new avalanches. The process is terminated when a large concentration of positive ions is created near the anode wire. Their presence begins to reduce the magnitude of the electric field in the vicinity of the wire. In this way the gas multiplication is diminished and so the Geiger discharge is terminated.
\\
For a fixed applied voltage, the point at which the Geiger discharge is terminated will always be the same, in the sense that a given density of positive ions will be needed to reduce the electric field below the minimum value required for further multiplication. Thus, each Geiger discharge is terminated after developing about the same total charge and so all the output pulses are about the same size, and their amplitude can provide no information about the properties of the incident radiation.
\\
\\
Figure \ref{imm:gainvoltage} shows an example of a gain-voltage plot.

\tocless\subsection{Charge division readout}
\label{sec:chargedivision}

\begin{figure}[H]
\centering
\includegraphics[width=1\textwidth]{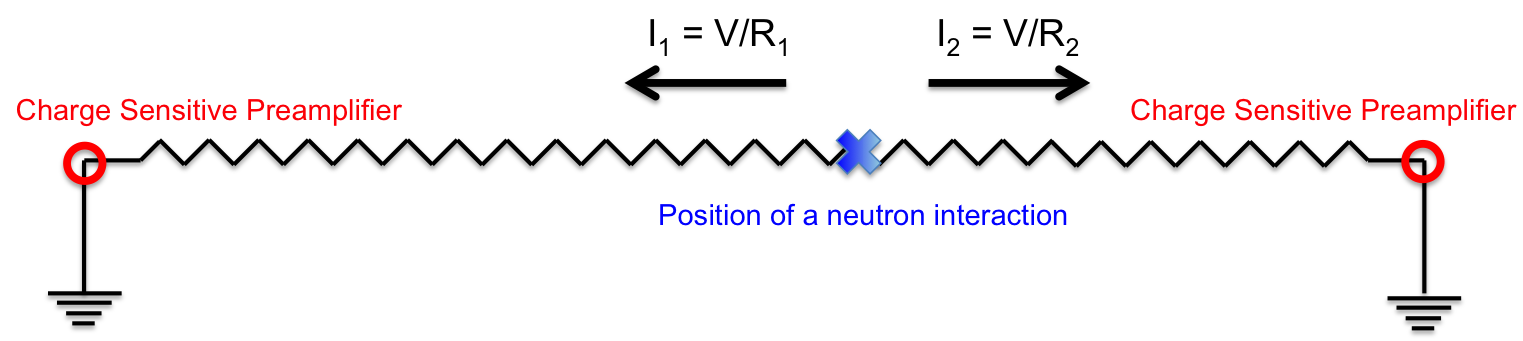}
\caption{Explanation of the charge division readout. The charge generated by the primary particle is collected in the wire. An electric potential difference appears and thus there are two currents flowing in the two directions along the wire. The intensity of the currents is given by the Ohm's law.}
\label{imm:resistanceexp}
\end{figure}

Position Sensitive Detectors (PSD) can reconstruct the position of interaction of a particle. A common technique is the charge division readout, that reconstructs the position of interaction along one direction. This technique can be used with detectors with one or more resistive wires. A charge division detector must have connectors at both ends of the wire in order to be able to collect both the signals.
\\
\\
To a first approximation, the wire can be modeled as a simple resistance with both the ends grounded. When a particle interacts in the detector, it creates electrons that are collected in the wire. This charge is localized in a precise point of the wire and it generates an electric potential difference between that point and the grounded edges of the wire. This creates two currents flowing in the two directions along the wire. The current intensities are given by the Ohm's law (Figure \ref{imm:resistanceexp}).

The currents are then collected at the two connectors with charge sensitive preamplifiers.  The charge sensitive preamplifiers convert the currents in voltage signals, that are then analyzed. The position of interaction of the particle $x$ can be reconstructed using the amplitudes of the two currents:

\begin{equation}
x= \frac{I_1}{I_1+I_2}\cdot L
\label{eq:cd}
\end{equation}

where $I_1$ and $I_2$ are the amplitudes of the signals and $L$ is the length of the tube.
\\
\\
When a particle interacts near the center of the wire, the signals traveling in the two directions encounter the same resistance and thus they have, at a first approximation, the same amplitude (Figure \ref{imm:cdcenter}).
\\
When a particle interacts near one edge of the wire, the situation is asymmetrical. One signal is generated close to the end of the tube and therefore it experiences only a little attenuation, while the other, that travels along almost all the wire, is much more attenuated. In this way its amplitude is lower than the amplitude of the first signal (Figure \ref{imm:cdlateral}).

 \begin{minipage}{\textwidth}
\centering
\begin{minipage}[t]{0.49\textwidth}
\begin{figure}[H]
\includegraphics[width=\textwidth]{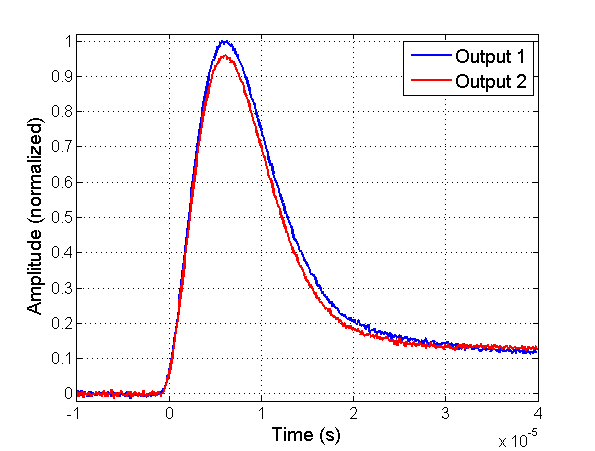}
\caption{Signals generated by a particle interacting near the center of the wire.}
\label{imm:cdcenter}
\end{figure}
\end{minipage}
\begin{minipage}[t]{0.49\textwidth}
\begin{figure}[H]
\includegraphics[width=\textwidth]{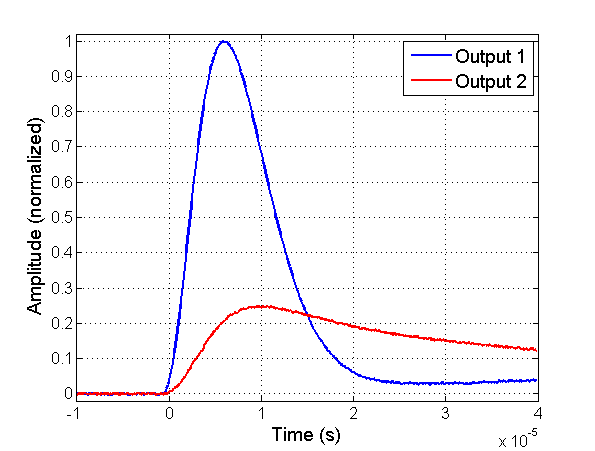}
\caption{Signals generated by a particle interacting near an end of the detector.}
\label{imm:cdlateral}
\end{figure}
\end{minipage}
\end{minipage}
\vspace{2mm}

These examples explain why the ratio between one signal and the the total signal (the sum of the two signals) is related to the position of interaction of the particle.
\\
\\
A better modeling considers also parasitic capacitances present along the wire. The parasitic capacitance is given by the system formed by the grounded walls of the tubes and the positive wire. This capacitance generates non-linear effects that are significant in particular near the edges of the tube. Moreover, this parasitic capaticance inserts a delay between the two signals coming from the tube. These effects do not alter significantly the spatial resolution or the charge division readout technique. For this reason, the discussion of these effects is beyond the aims of this thesis.

\tocless\subsection{Counting Curves}
\label{sec:countingcurves}

\begin{figure}[H]
\begin{center} 
\includegraphics[width=0.6\textwidth]{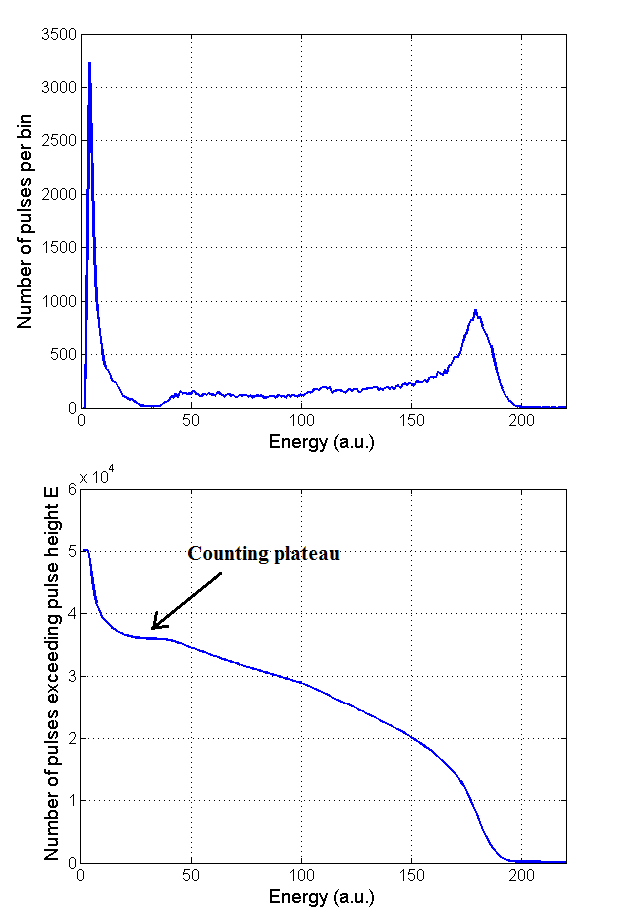} 
\end{center} 
\caption{Differential and integral PHS for a $\mathrm{^{3}He}$ proportional tube.} 
\label{imm:integraldist}
\end{figure}

When radiation detectors are operated in pulse counting mode, a common situation often arises in which the pulses from the detector are fed to a counting device with a fixed discrimination level. Signal pulses must exceed a given threshold in order to be registered by the counting circuit.
\\
In setting up a pulse counting measurement, it is often desirable to establish an operating point that will provide maximum stability over long periods of time. In fact, small drifts of the threshold or of the gain could be expected in any real application. One would like to establish conditions under which these drifts would have minimal influence on the measured counts. Regions of minimum slope on the \textit{integral pulse height distribution} are called \textit{counting plateaus} and represent areas of operation in which minimum sensitivity to drifts in discrimination level are achieved.
\\
We recall that the integral pulse height spectrum (PHS) is given by the number of pulses whose amplitude exceeds that of a given value of the abscissa (see Figure \ref{imm:integraldist}).

Note that where minima appear in the differential PHS, counting plateaus appear in the integral PHS.
\\
\\
Plateaus in counting data can be observed with a different procedure. For a particular radiation detector it is possible to vary the gain by changing the applied voltage to it. An experiment can be carried out in which the number of pulses recorded over a fixed threshold is measured as a function of the gain applied, the result is called \textit{counting curve}. In order to select an operating point of maximum stability, plateaus are sought in the counting curve and the voltage is selected to lie at a point of minimum slope on this counting curve (Figures \ref{imm:countingcurve} and \ref{imm:countingcurve2}). 

\begin{minipage}{\textwidth}
\centering
\begin{minipage}[t]{0.49\textwidth}
\begin{figure}[H]
\includegraphics[width=\textwidth]{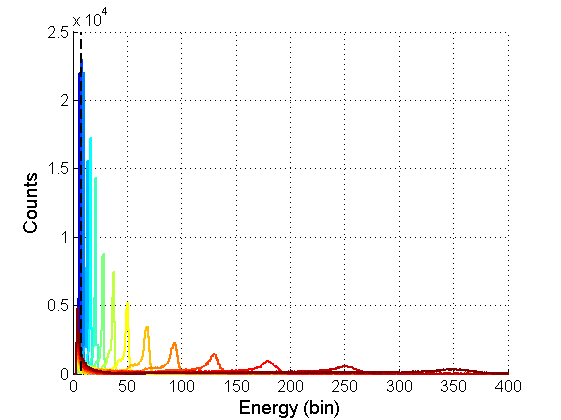}
\caption{PHS at several HV and threshold (dotted line) set for counting pulses for the counting curve.}
\label{imm:countingcurve}
\end{figure}
\end{minipage}
\begin{minipage}[t]{0.49\textwidth}
\begin{figure}[H]
\includegraphics[width=\textwidth]{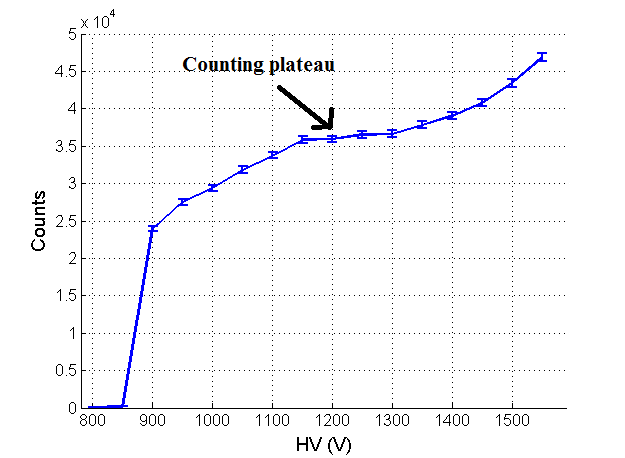}
\caption{Counting curve.}
\label{imm:countingcurve2}
\end{figure}
\end{minipage}
\end{minipage}
\vspace{1mm}

\section{Helium-3 Proportional Tubes}
\label{sec:he3}

Most slow neutron detectors take advantage of neutron capture reactions, whose fragments generate free electrons in the detector gas that give rise to a detectable signal. The most common choice for slow neutron conversion is the $\mathrm{^{3}He}$, that converts neutrons through the reaction:

\begin{equation}
^{1}n + ^{3}He  \quad \longrightarrow \quad ^{1}H + ^{3}H \qquad Q=764 \  keV
\end{equation}

\begin{figure}[H]
\begin{center} 
\includegraphics[width=0.9\textwidth]{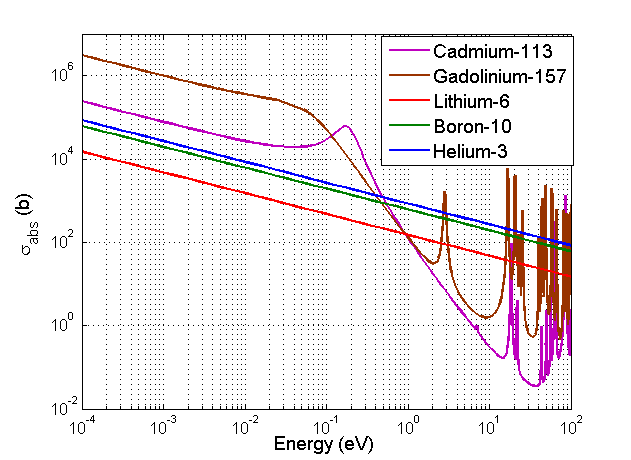} 
\end{center} 
\caption{Absorption cross section as a function of the neutron energy of the most important neutron absorbers.} 
\label{imm:neutroncross}
\end{figure}

The proton ($\mathrm{^{1}H}$) has an energy of 573 $keV$ and the tritium ($\mathrm{^{3}H}$) of 191 $keV$ and they are emitted back-to-back. The cross-section for thermal neutrons is 5333 $b$ and has a $1/v$ dependence on the neutron velocity.
\\
\\
The intrinsic advantage of $\mathrm{^{3}He}$ is its high cross-section of interaction, therefore high efficiency for detectors. Figure \ref{imm:neutroncross} shows the absorption cross section as a function of the neutron energy of the most important neutron absorbers. Although $\mathrm{^{157}Gd}$ and $\mathrm{^{113}Cd}$ have a larger absorption cross section, their use in detector is complicated for technical reason. Moreover, $\mathrm{^{3}He}$ is a noble gas, therefore it operates both as a neutron converter and as a proportional gas. Unlike the $\mathrm{BF_{3}}$, the only other neutron converter that can also operate as a proportional converter, the $\mathrm{^{3}He}$ is non toxic. Usually $\mathrm{^{3}He}$ tubes operates at high pressures (up to 20 $bar$), in order to increase the efficiency.

$\mathrm{^{3}He}$ major drawbacks are its price and availability. In fact, since 2001 the world is experiencing a severe $\mathrm{^{3}He}$ shortage as a consequence of its use in the US National Security Programme and the reduction of the nuclear weapon stockpiles (Section \ref{sec:shortage}). For this reason its price is very high (between \$2500 and \$5000 per liter and the price is increasing over time) and its availability is very limited. 
\\
Moreover, concerning the performances, the $\mathrm{^{3}He}$ are limited in spatial resolution and counting rate capability.
\\
\\
The mean free path for a neutron in the gas depends on the $\mathrm{^{3}He}$ density (pressure) and on the neutron energy (wavelength):

\begin{equation}
\lambda_{mfp}[cm]=\frac{13.88}{p_{^{3}He}[bar]\cdot \lambda_{wavelength} [\AA]}
\end{equation}

Note that the pressure is not the total pressure of the gas in the detector, which includes the stopping and quench gas, but only the $\mathrm{^{3}He}$ pressure.
\\
Referring to Equation \ref{eq:probinteraction}, the efficiency of a $\mathrm{^{3}He}$ detector of a fixed thickness $d$ is given by :

\begin{equation}
\epsilon=1-\exp{(-0.07 \cdot p_{^{3}He}[bar] \cdot \lambda_{wavelength}[\AA] \cdot d[cm])}
\end{equation}

\begin{figure}[H]
\centering 
\includegraphics[width=0.7\textwidth]{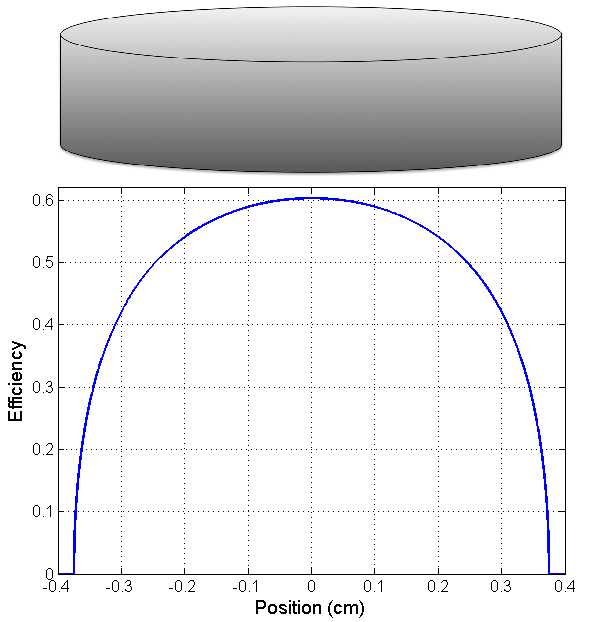} 
\caption{Efficiency for thermal neutrons of a $\mathrm{^{3}He}$ tube as a function of the position of the neutron beam. The $\mathrm{^{3}He}$ pressure is 10 $bar$ and the diameter of the tube is 8 $mm$. The efficiency is calculated analytically.} 
\label{imm:efficiencyposition}
\end{figure}

The efficiency increases as the neutron wavelength increases, because of the cross section of interaction behavior. The $\mathrm{^{3}He}$ pressure is usually between 1 and 20 $bar$. In the case of beam monitors it is of the order of $mbar$, in order to achieve low efficiencies. 
\\
\\
Often $\mathrm{^{3}He}$ detectors have a cylindrical shape, in order to exploit the gas gain in the proportional region. In these tubes the $\mathrm{^{3}He}$ thickness depends on the position and thus the efficiency is not constant across it. Figure \ref{imm:efficiencyposition} shows the efficiency for thermal neutrons as a function of the position of the neutron beam. The efficiency is calculated analytically in the case of a tube with a diameter of 8 $mm$ and a $\mathrm{^{3}He}$ pressure of 10 $bar$. 

\begin{figure}[H]
\centering 
\includegraphics[width=0.7\textwidth]{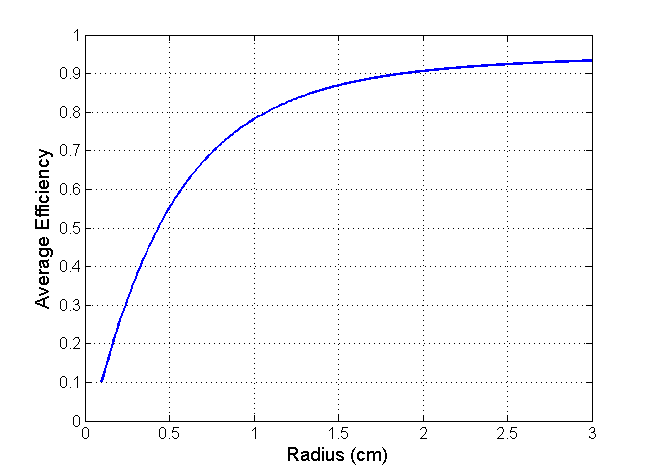} 
\caption{Average efficiency for thermal neutrons as a function of the radius of the tube. The $\mathrm{^{3}He}$ pressure is 10 $bar$.} 
\label{imm:averageefficiency}
\end{figure}

The efficiency of the tube is defined as the average efficiency along its surface. Figure \ref{imm:averageefficiency} shows the average efficiency for thermal neutron as a function of the radius of the tube. The $\mathrm{^{3}He}$ pressure is 10 $bar$.

\begin{figure}[H]
\begin{center} 
\includegraphics[width=0.8\textwidth]{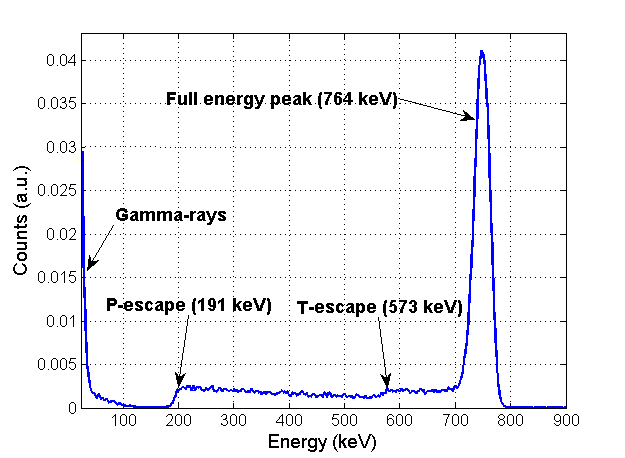} 
\end{center} 
\caption{PHS of a $\mathrm{^{3}He}$ detector.} 
\label{imm:HEspectrum}
\end{figure}

Note that the average efficiency saturates at $\approx$ 93\%. Some neutrons are scattered in the walls of the detector before entering the gas. If a neutron scatters in the wall and is then directed outward from the detector, it will not be detected. Usually a few percent of neutrons are lost in this way and this effect sets a limit to the maximum achievable efficiency.
\\
\\
Usually other gases are added to the $\mathrm{^{3}He}$, up to 50\% of the total gas composition. Stopping gases (usually Ar, Kr and/or $\mathrm{CF_4}$) are elements or molecules with a relatively high number of electrons. These elements interact with the fragments and then they reduce their range, in order to improve the spatial resolution and to decrease the wall effect (explained later in this Section). These gases are chosen because of their good behavior as proportional gas. Quench gas (usually $\mathrm{CO_2}$ or $\mathrm{CH_4}$) are polyatomic gas molecules that can absorb X-rays through rotational excitation. They improve the proportionality of the gas gain, limiting the effects typical of a Geiger-M{\"u}ller counter.
\\
\\
Figure \ref{imm:HEspectrum} shows the PHS of a $\mathrm{^{3}He}$ detector.

\begin{figure}[H]
\begin{center} 
\includegraphics[width=0.8\textwidth]{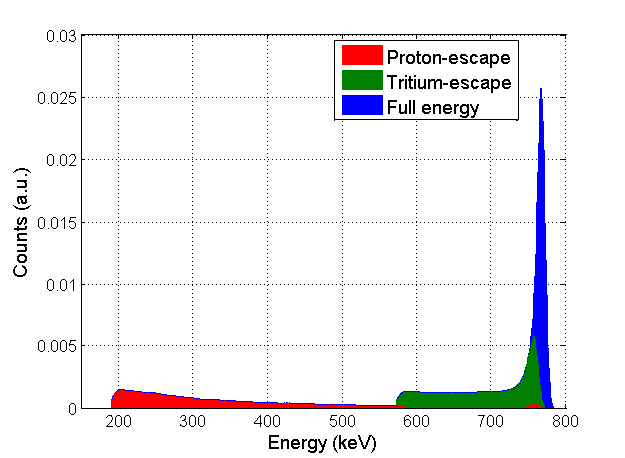} 
\end{center} 
\caption{Full energy peak, proton-escape shoulder and tritium-escape shoulder contributes to the $\mathrm{^{3}He}$ PHS. } 
\label{imm:HEdivided}
\end{figure}

\begin{figure}[H]
\begin{center} 
\includegraphics[width=0.8\textwidth]{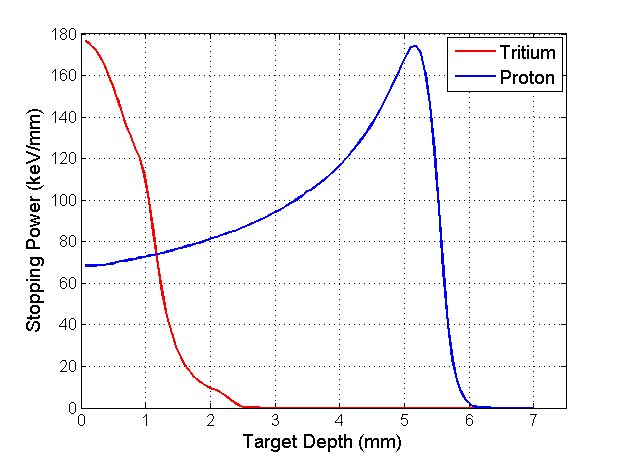} 
\end{center} 
\caption{Tritium (191 $keV$) and proton (573 $keV$) stopping power in $\mathrm{^{3}He}$ at 10 $bar$, obtained with the package SRIM \cite{Ziegler}.} 
\label{imm:stoppingfragments}
\end{figure}

The full energy peak is generated by the neutron fragments that release all their energy (764 $keV$) in the gas. This is not true for all neutron fragments because of the \textit{wall effect}: when a neutron interacts close to the detector walls, one of the fragments can escape from the gas and enter the walls before having released all its energy. Since the fragments are emitted back-to-back only one at a time of them can escape in this way. The escaping fragment can release from very low energies up to its entire energy in the gas, while the other fragment (directed toward the center of the detector) necessarily releases all its energy.
\\
\\
To count the neutron signals an amplitude threshold is set in the minimum region between the neutron PHS and the gamma-ray exponential, i.e. the neutron valley. Using this threshold no neutron signals are discarded and the gamma discrimination is large. This threshold is particularly useful because it is unambiguous and easy to identify.
\\
\\
The PHS can be divided in three contributes: the full energy peak (764 $keV$), the tritium-escape shoulder (573-764 $keV$) and the proton-escape shoulder (191-764 $keV$) (see Figure \ref{imm:HEdivided}). The distances traveled in the gas by the escaping fragments are randomly distributed, therefore the shapes of the shoulder do not depend on the geometry of the detector. Their shapes depend only on the fragment stopping powers (Figure \ref{imm:stoppingfragments}).

Note that a minimum at the beginning of the trace of a fragment corresponds to a peak at low energies in its shoulder and vice versa. For this reason the proton produces a proton-escape peak at 191 $keV$. 
\\
\\
The extent of the wall effect depends on the detector geometry and on the range of the two fragments in the gas (i.e. pressure and gas composition). The larger the detector with respect of the range of the fragments, the more pronunced the full energy peak is. 

\section{Technologies Alternative to Helium-3}

As explained in Section \ref{sec:shortage}, the world is experiencing a severe $\mathrm{^{3}He}$ shortage and in the last decade the development of detectors based on alternative technologies has become particularly crucial. In this Section we briefly the properties of the most promising lines of research, e.g. $\mathrm{^{10}B}$ and $\mathrm{^{6}Li}$ based detectors. The aim of this Section is to provide the basic information needed for evaluate the pros and cons of these detectors. We point out that several other technologies are under development, but a complete discussion is beyond the goal of this thesis.

\subsection{Boron-10 Detectors}

$\mathrm{^{10}B}$ converts neutrons through one of the following reactions:

\begin{align}
^{1}n + ^{10}B \quad &\longrightarrow \quad ^{7}Li + ^{4}He \qquad Q=2.79 \  MeV  \\
^{1}n + ^{10}B \quad &\longrightarrow \quad ^{7}Li + ^{4}He + \gamma \qquad Q=2.31 \  MeV
\end{align}

The second reaction, with a branching ratio of 96\%, is the most probable one. The absorption cross section of $\mathrm{^{10}B}$ for thermal neutron is relatively large (3840 $b$).
\\
Boron is very abundant on Earth and it can be found in several minerals. $\mathrm{^{10}B}$ occurs with a natural isotopic abundance of 20\%, thus it is a relatively cheap and accessible isotope.
\\
\\
There are two classes of $\mathrm{^{10}B}$ detectors:

\begin{itemize}
\item $\mathrm{BF_{3}}$ detectors
\item Solid $\mathrm{^{10}B}$-layers detectors
\end{itemize}

\begin{figure}[H]
\begin{center} 
\includegraphics[width=0.8\textwidth]{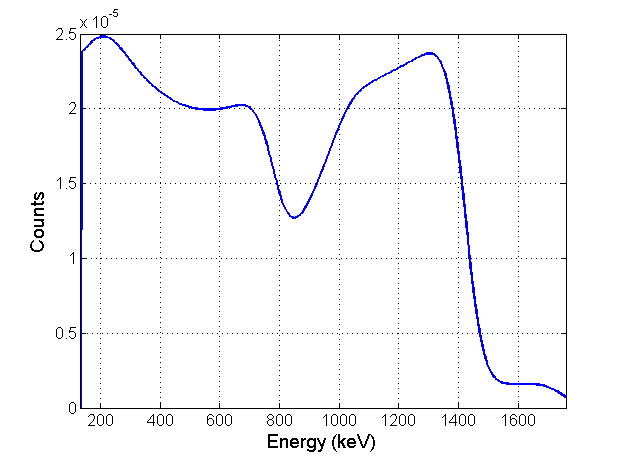} 
\end{center} 
\caption{PHS of a $\mathrm{^{10}B}$-layers based detector.} 
\label{imm:B10spectrum}
\end{figure}

The first class exploit the gaseous state of $\mathrm{BF_{3}}$. This gas acts as both a proportional gas and a neutron detection material, such as $\mathrm{^{3}He}$. $\mathrm{BF_{3}}$ are universally constructed using a cylindrical shape and they have a PHS with the same shape as $\mathrm{^{3}He}$ detectors. Because the performance of $\mathrm{BF_3}$ as a proportional gas is poor when operated at higher pressure, its pressure in typical tubes rarely exceeds 1 $bar$. For this reason, in order to reach a large efficiency is necessary to build tubes with a large radius (up to 20 $cm$).
\\
Although $\mathrm{BF_3}$ provide high efficiency and good performances overall, they are not widely used because of the high toxicity of this gas. Its use could be very dangerous in environments where users, people using the instruments for their experiments, are working. For this reason these detectors are not a common choice for neutron instruments,  although they were widely used at the outset of neutron science in the '50s and '60s. Nowadays the research focuses on the development of new safety systems.
\\
\\
The second class exploits solid boron layers. The ESS will take advantage of mainly this class of detectors. An aluminium substrate is coated on both sides with boron thin layers ($\approx$ 1 $\mu m$). Such blades (composed of the two boron layers and the aluminium substrate) are inserted in a multi wire proportional chamber, usually with a gas mixture of Ar and $\mathrm{CO_2}$. The neutron is converted in a boron layer and one fragment can enter the gas, while the other is absorbed either in the layer or in the aluminium substrate. The fragments that enter the gas generate a signal as in the common proportional chambers. In these detectors therefore the neutron converter does not act as a proportional gas, as in the other neutron gaseous detectors.
\\
\\
Fragments lose their energy while they travel in the boron layers. For this reason the PHS of $\mathrm{^{10}B}$-layers based detectors extends down to low energies (Figure \ref{imm:B10spectrum}).

With a single blade and with the neutron beam orthogonal to the layers the maximum efficiency achievable is $\approx$ 7\%. Such an efficiency can not cope with the efficiency of $\mathrm{^{3}He}$ detectors commonly used in neutron instruments.
\\
To overcome this problem there are two viable choices. The first possibility consists in a configuration of several blades in order to make neutrons cross several boron layers. The second possibility consists in using a low angle incidence for the neutron beams, so neutrons pass through a relatively long distance of boron in a single layer.
\\
These two approaches have different applications and with both it is possible to reach an efficiency for thermal neutrons up to 60\%, still lower than several $\mathrm{^{3}He}$ detectors. On the other hand, these detectors can reach a very high counting rate capability and high spatial resolution.

\subsection{Lithium-6 Detectors}

$\mathrm{^{6}Li}$ neutron conversion reaction is:

\begin{equation}
^{1}n + ^{6}Li \quad \longrightarrow \quad ^{3}H + ^{4}He \qquad Q=4.78 \  MeV
\end{equation}

\begin{figure}[H]
\begin{center} 
\includegraphics[width=0.8\textwidth]{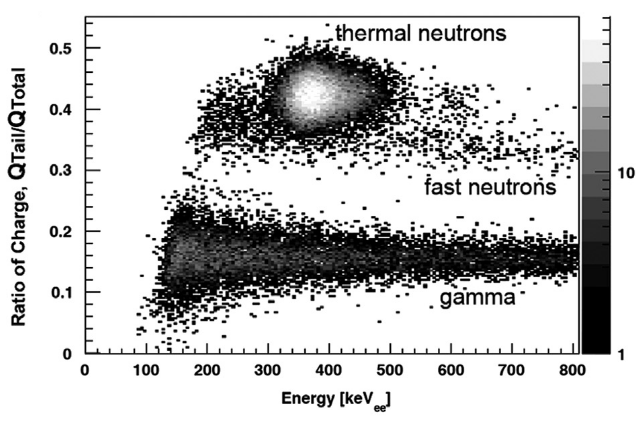} 
\end{center} 
\caption{Neutron and gamma-ray patterns in a $\mathrm{^{6}Li}$ scintillator \cite{Zaitseva2012}. The gamma-ray pulses can be rejected by using a threshold on $Q_{tail}/Q_{total}$} 
\label{imm:lithiumPSD}
\end{figure}

The thermal neutron cross section for this reaction is 940 $b$. The lower cross section is generally a disadvantage but is partially offset by the higher $Q$ value and resulting greater energy given to the reaction products. 
\\
Lithium is widely distributed on Earth, although it does not naturally occur in elemental form due to its high reactivity. $\mathrm{^{6}Li}$ occurs with a natural isotopic abundance of about 7\% and is also widely available in separated form.
\\
\\
Because a stable lithium-containing proportional gas does not exist, a lithium equivalent to the $\mathrm{^{3}He}$ or $\mathrm{BF_3}$ tubes is not available. Li-based detectors are usually scintillators, often based on a lithium compound dispersed in ZnS(Ag) with thickness up to 10 $mm$. All the fragments energy is normally absorbed in the scintillator, thus the PHS consists essentially in a single full energy peak. Lithium scintillators can reach a high efficiency ($\approx$ 60\%) and a good spatial resolution. Moreover they can achieve a very high counting rate capability. They are very sensitive, however, to the gamma-ray background. 
\\
\\
By using the Pulse Shape Discrimination technique (see Chapter \ref{ch:PSD}) a very low gamma-ray sensitivity can be achieved. A scintillator signal has a slow and a fast component and the ratio between these two depends on the linear stopping power of the particle. In this way, it is possible to discriminate between the neutron fragments and the photo-electrons generated by a gamma-ray, since the latter have a stopping power a few order of magnitudes lower than the former. 
\\
Figure \ref{imm:lithiumPSD} shows the neutron and the gamma-ray patterns in a $\mathrm{^{6}Li}$ scintillator. The gamma-ray pulses can be rejected by using a threshold on the ratio between the slow component of the signal and the total signal. The use of the Pulse Shape Discrimination technique, however, requires a more complex data recording system.

%% file: Chapters/Instrumentation.tex

\chapter{Instrumentation and Preliminary Characterization}

In the first part of this Chapter the instrumentation (detectors and electronics) used for the measurements is described. The facilities where these measurements took place are also discussed.
\\
\\
In the second part the counting curve, the gain and the energy resolution of our detectors are described. The counting curve measurements allow us to find the best operating voltage for the detectors (Section \ref{sec:countingcurves}). The analysis of the gain and of the energy resolution measurements gives information about the tubes. Similar analyis can be found in the literature \cite{Ravazzani2006}. The energy resolution provides information about the space charge effect and its magnitude and is related to the gamma discrimination. 

\section{Instrumentation}

\subsection{Detectors}
\label{sec:detector}

\begin{figure}[H]
\includegraphics[width=\textwidth]{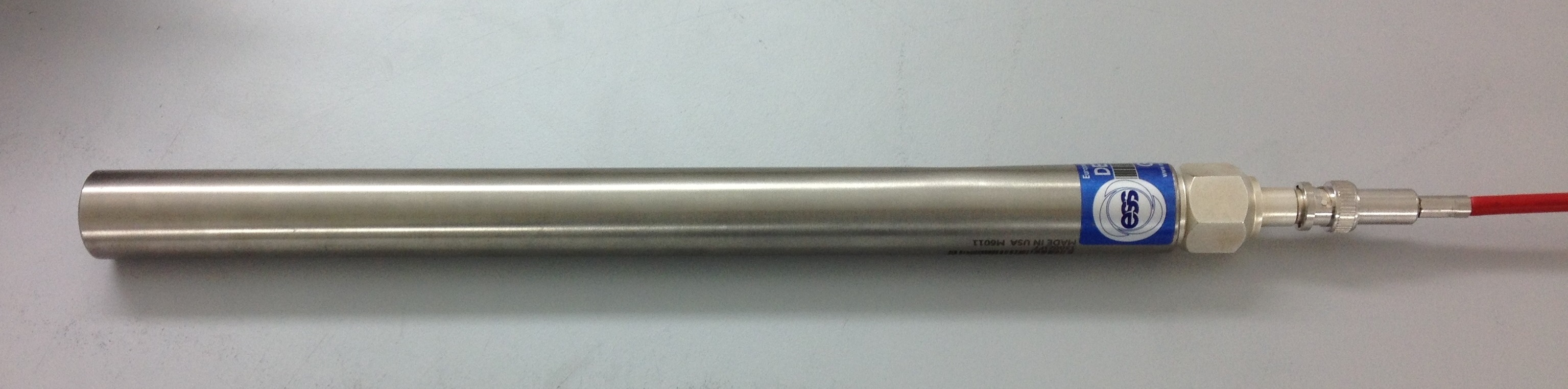}
\caption{Tube RS P4 0810 227 by GE Reuter Stokes.}
\label{imm:tube1}
\end{figure}

\begin{figure}[H]
\hspace{-1.5cm}
\includegraphics[width=1.2\textwidth]{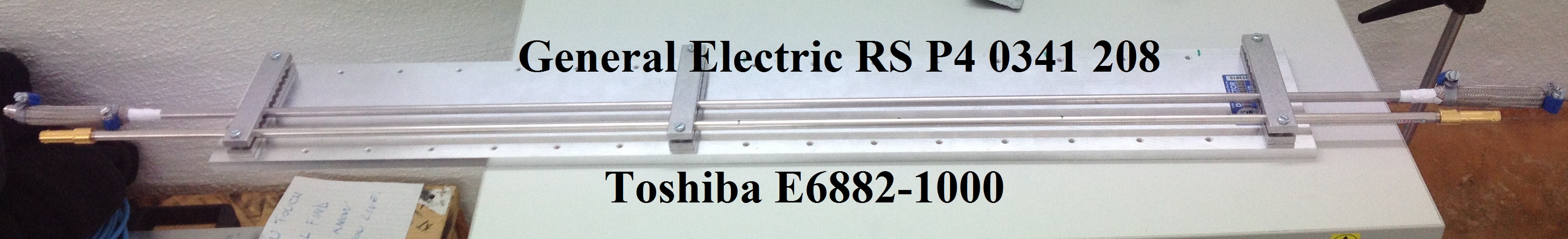}
\caption{PSD E6882-1000 by Toshiba and RS P4 0341-208 by GE Reuter Stokes.}
\label{imm:tube2}
\end{figure}

The following $\mathrm{^{3}He}$-based detectors are available in the laboratory:

\begin{itemize}
\item RS P4 0810 227, manufactured by General Electric (GE) Reuter Stokes (RS 25 $cm$). It is made in stainless steel and it has an active length of 10 $inch$, a radius of 1 $inch$, a gas pressure of 10 $bar$ and an efficiency of 96.4\% (at 2.5\AA)\footnote{The calibration was performed by F. Piscitelli at the Institute Laue-Langevin (ILL)} (see Figure \ref{imm:tube1}).
\item PSD E6882-1000 manufactured by Toshiba (Toshiba 1 $m$). It is made in stainless steel and it has an active length of 1 $m$, a radius of 8 $mm$, and a pressure of 10 $bar$ (see Figure \ref{imm:tube2}).
\item RS P4 0341 208 manufactured by GE Reuter Stokes (RS 1 $m$). It is made in stainless steel and it has an active length of 1 $m$, a radius of 8 $mm$, and a pressure of 10 $bar$, exactly as the previous tube (see Figure \ref{imm:tube2}).
\end{itemize}

\begin{figure}[H]
\centering
\includegraphics[width= 0.7\textwidth]{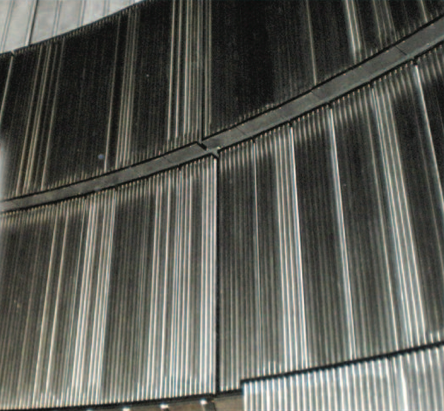}
\caption{Bundle of several $\mathrm{^{3}He}$ tubes in a large-area instrument.}
\label{imm:sanstubes}
\end{figure}

The last two tubes are PSD, thus they can measure the position of interaction of a neutron through the charge division readout (see Section \ref{sec:chargedivision}). 
\\
These detectors are widely used in the most important neutron scattering facilities in the world. For example, detectors with approximately the same design as the Toshiba 1 $m$ tube are used in the Small Angle Neutron Scattering (SANS) instrument \textit{Taikan}, at J-PARC in T$\mathrm{\bar{o}}$kai (Japan). Detectors with a design very similar to the RS $1$ m tube are used in several SANS instruments worldwide: \textit{SANS-1} at FRMII in Garching (Germany), \textit{D22} at ILL in Grenoble (France), \textit{Wish} at ISIS in Harwell (United Kingdom) and \textit{EQ-SANS} at SNS in Oak Ridge (USA).
\\
These tubes are commonly used in large-area instruments in bundles of hundreds of them (see Figure \ref{imm:sanstubes}).

The gas composition in the tubes is unknown. We expect a few $bar$ of Ar (and/or Kr, $\mathrm{CF_4}$), in order to decrease the range of the neutron fragments and to decrease the mean ionization energy of the gas, and $\mathrm{CO_2}$, as quenching gas. Other gases can be present though. We expect a much larger pressure of Ar/$\mathrm{CO_2}$ in the position sensitive tubes, since in them the range of the fragments must be short, in order to improve the spatial resolution.

\subsection{Electronics}

We used 3 different charge sensitive preamplifiers:

\begin{itemize}
\item \textit{ORTEC 142PC}, with gain of 4 $V/pC$, decay time of $\approx$75 $\mu s$ and minimum rise time of 25 $ns$.
\item \textit{CAEN A422A}, with gain of $\approx$0.7 $V/pC$, decay time of 300 $\mu s$ and minimum rise time of 50 $ns$.
\item \textit{FAST ComTec CSP10}, with gain of 1.4 $V/pC$, decay time of 140 $\mu s$ and rise time of 7 $ns$.
\end{itemize}

The ORTEC was our standard choice because of its high gain. At high bias voltages the tube gas gain become too high to prevent the saturation of the shaper or of the MCA. In these cases the CAEN preamplifier has been used, because of its lower gain. A low rise time allows to analyze fast signal changes. For this reason FAST preamplifier has been used when the shape of the pulses was critical, e.g. in the Pulse Shape Discrimination measurements.
\\
\\
After the preamplifiers we used a shaping amplifier: \textit{ORTEC 855 Dual Spec Amplifier}. It shapes the pulses with a semi-gaussian shape with peaking time equal to 6.6 $\mu s$ and 50\% pulse width equal to 9.9 $\mu s$ and a gain range between 5 and 1250. 
\\
We used a \textit{iseg NHQ 226L} as HV power supply. It provides up to 6 $kV$ with an accuracy of 2 $V$, and a current up to 1 $mA$. 
\\
\\
We used two different MCA: \textit{Amptek MCA-8000D} and \textit{CAEN DT5780}. \\
The Amptek MCA-8000D has a single input channel and has a high speed ADC (100 $MHz$, 16 bit) with digital pulse height measurement. The input accepts semigaussian type pulses of peaking time $\ge$500 $ns$ and with amplitude between 0 and 10$V$. 
\\
The CAEN DT5780 has two inputs with high speed ADCs (100 $MHz$, 14 bit). It is equipped with a Field Programmable Gate Array (FPGA) featuring the real-time Digital Pulse Processing for Pulse Height Analysis, i.e. it can handle the signals in real-time, providing a digital shaping and the information we need, such as the energy (i.e. pulse height) or timing information or sample of the waveforms. The CAEN DT5780 has been used to acquire the pulse height from the two channels of a tube, when the charge division readout method is used. These measurements are discussed in Chapter \ref{ch:spatialresolution}.
\\
\\
The oscilloscope \textit{LeCroy HDO4054} has been used to record the signal traces. It digitalize the signal amplitude with a 12-bit resolution and it has a rise time of 700 $ps$.

\subsection{Facilities}
\label{sec:facilities}

\begin{figure}[H]
\centering
\includegraphics[width=0.8\textwidth]{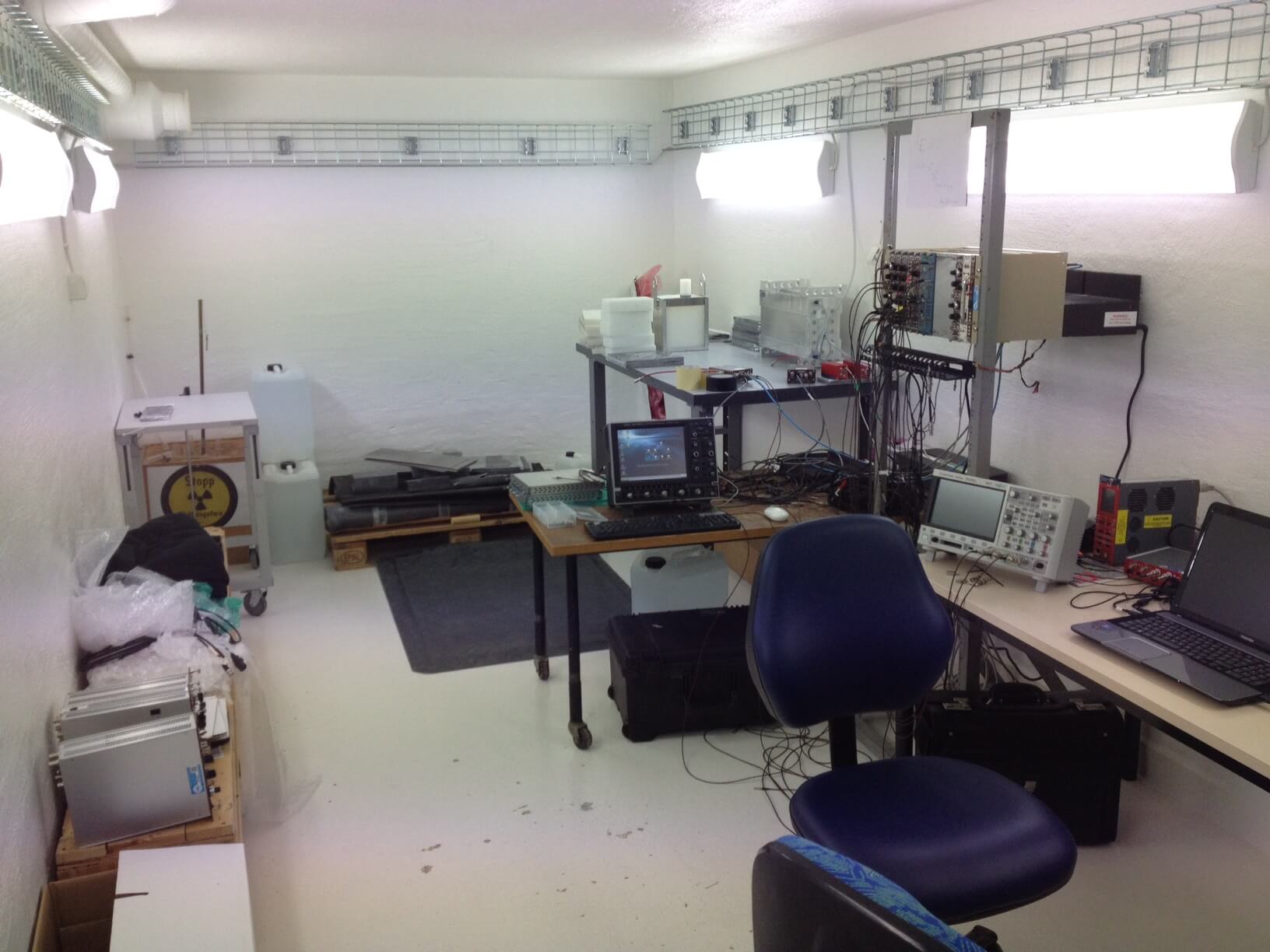}
\caption{Source Facility laboratory.}
\label{imm:sourcefacility}
\end{figure}

Most of the measurements were performed in the \textit{Source Facility} laboratory, located in the Physics Faculty of the Lund University. Figure \ref{imm:sourcefacility} shows a picture of the Source Facility laboratory. A Pu-Be neutron source and two $\gamma$-sources ($\mathrm{^{60}Co}$ and $\mathrm{^{137}Cs}$) are available. Several borated polyethylene and borated aluminium plates are also available to shield the detectors from the background neutrons. In particular, borated polyethylene is a common choice for neutron shielding because it can both moderate the fast neutrons and absorb the slow ones.

\begin{figure}[H]
\centering
\includegraphics[width=0.8\textwidth]{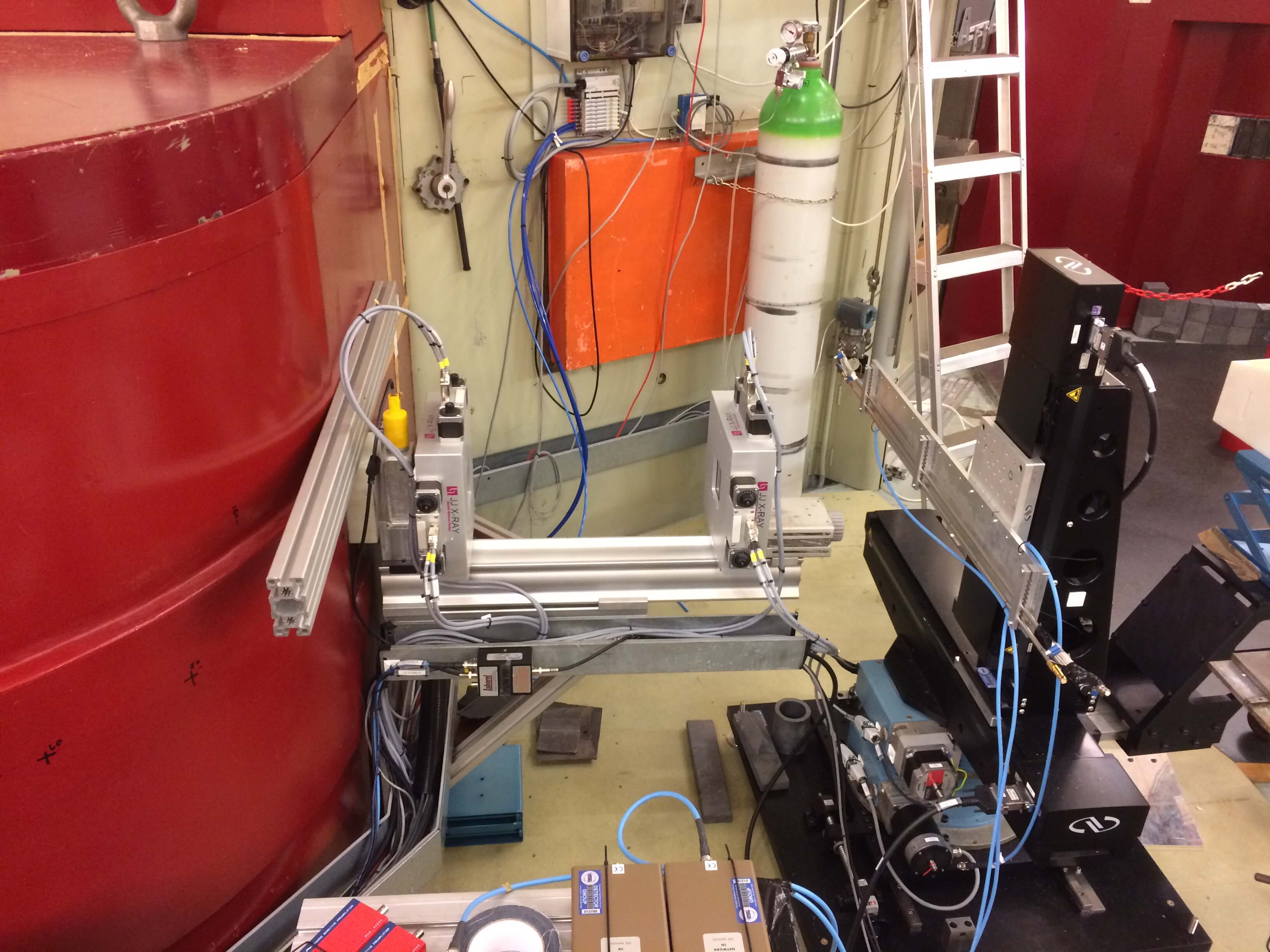}
\caption{Beamline R2D2 at IFE.}
\label{imm:IFE}
\end{figure}

Some measurements were performed at the beamline \textit{R2D2}, at the reactor JEEP II, at the Institute for Energy Technology (IFE), located in Lillestr\o m, Norway. The neutron beam coming out from the reactor is directed on several $Ge$ monochromator crystals which provide neutrons with wavelength of 2 \AA. The flux of the beam is $\approx$10$^{5}$ $\frac{n}{cm^{2}\cdot s}$. Two slits allow to change the size of the beam. The beamline is shown in Figure \ref{imm:IFE}.

%% file: Chapters/Plateaumeasurements.tex

\section{Preliminary characterization}

\subsection{Counting Curves}

\begin{figure} [H]
\centering
\includegraphics[width=0.7\columnwidth]{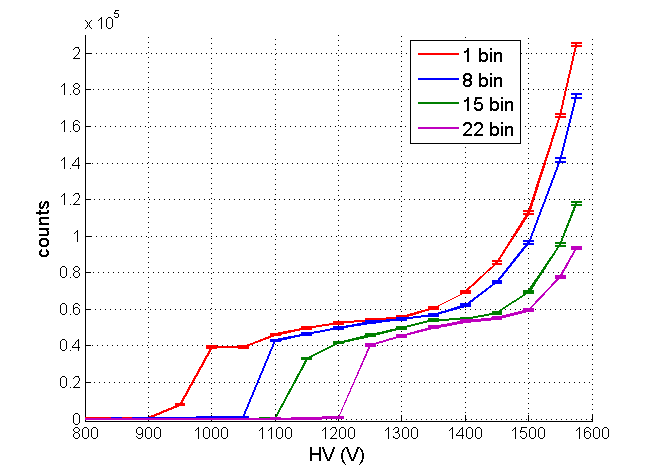}
\caption{Counting curves obtained with several amplitude thresholds (RS 25 $cm$).}
\label{imm:plateau}
\end{figure}

The $\mathrm{^{3}He}$ detector counting curve exhibits a very flat counting plateau because of the valley between the neutron and the gamma-ray PHS (see Figures \ref{imm:countingcurve} and \ref{imm:countingcurve2}). In order to guarantee high counting stability with respect to small voltage fluctuations, the operational HV is usually selected in the counting plateau. 
\\
\\
Figure \ref{imm:plateau} shows the counting curves of the RS 25 $cm$ tube obtained with several amplitude thresholds to count the pulses.

\begin{figure} [H]
\centering
\includegraphics[width=0.7\columnwidth]{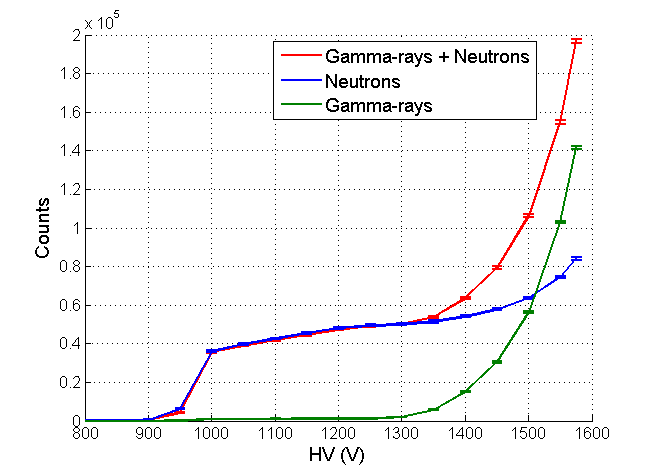}
\caption{Counting curves in presence of a neutron source and/or a $\mathrm{^{60}Co}$ gamma-ray source (RS 25 $cm$).}
\label{imm:comparison}
\end{figure}

Note that the larger the amplitude threshold, the larger the HV at which the counting plateau occurs is. The plateau is wider at lower amplitude thresholds, thus in this case the counting system is more stable. If there are not other requirements on the HV, it is preferable to select an amplitude threshold as small as possible (but large enough to avoid the electronic noise).
\\
\\
Figure \ref{imm:comparison} shows the effect on the counting curves of the neutron pulses and of the gamma-ray pulses.

\begin{figure} [H]
\centering
\includegraphics[width=0.7\columnwidth]{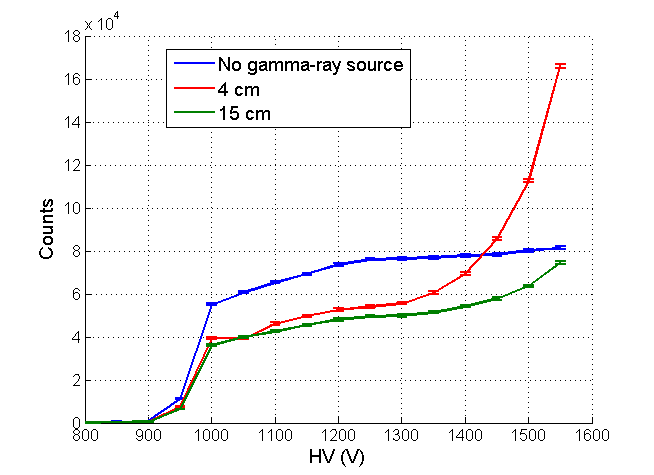}
\caption{Counting curves with a gamma-ray source at several distances from the detectors. Note that the counting plateaus lie at different counts because we changed the experimental setup, i.e. the neutron flux.}
\label{imm:compgammas}
\end{figure}

The gamma-ray signals generate an approximately exponential counting curve at large HV. At low HV they do not affect the counting curve significantly.
\\
\\
Figure \ref{imm:compgammas} shows the counting curves with the detector subject to several gamma-ray irradiations. 

The more intense the gamma-ray flux, the steeper the exponential shape of the counting curve is.

\subsection{Gain}
\label{sec:gainn}

We characterize the gas gain of the $\mathrm{^{3}He}$ tubes. The neutron signals from the tube is amplified by the ORTEC 142PC preamplifier and then shaped by the ORTEC 855 Dual Spec Amplifier. The PHS is then recorded by the Amptek MCA-8000D. We record a PHS with several HV and every PHS measurement lasts 2-3 minutes. The relative gain is given by the position of the full energy peak maximum. We perform a gaussian fit on the full energy peak. In order to not take into account the proton- and tritium-shoulders, the fit involves only the right part of the peak, i.e. energies larger than the position of the maximum. This procedure also provides information about the width of the peak, i.e. the energy resolution.

\begin{figure} [H]
\centering
\includegraphics[width=0.8\columnwidth]{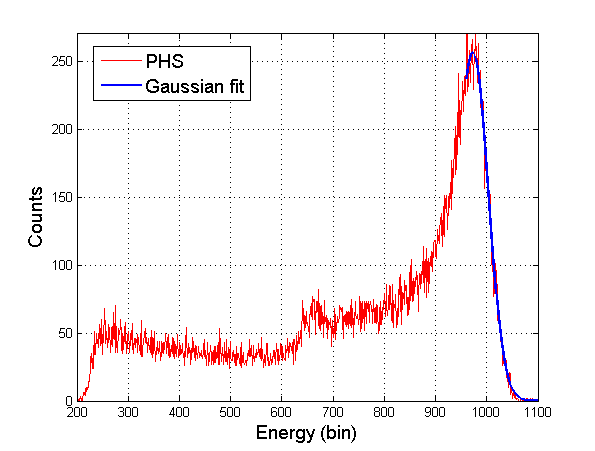}
\caption{The gaussian fit on the full energy peak. In order to avoid the proton- and tritium-shoulders, the fit involves only the right part of the peak, i.e. energies larger than the maximum of the peak.}
\label{imm:fitpeak}
\end{figure}
 
\begin{figure} [H]
\centering
\includegraphics[width=0.8\columnwidth]{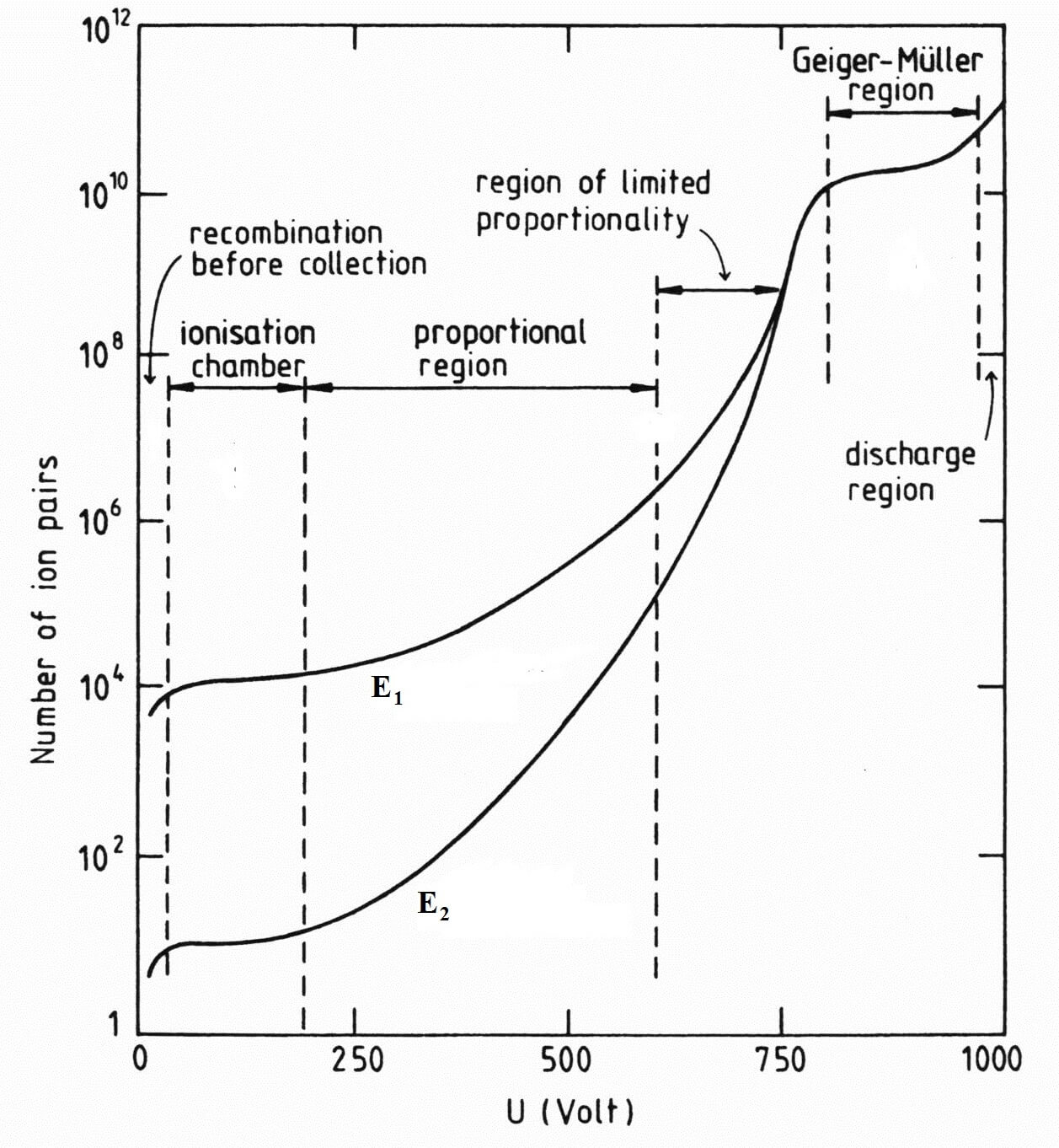}
\caption{Gain of the $\mathrm{^{3}He}$ tubes as a function of the applied HV.}
\label{imm:gain}
\end{figure}

Figure \ref{imm:gain} shows the gain as a function of the HV applied to the tubes.

From about 150 $V$ to 400 $V$ the gain of the RS 1 $m$ detector is constant. In this HV range the tube operates as an ionization chamber. The gain shown in Figure \ref{imm:gain} is normalized to this ionization chamber gain that by definition is 1 (no charge loss or multiplication occurs). Note that the absolute gain of the other tubes can be slightly different from the value shown because of different gas compositions in the tubes.

\subsection{Energy Resolution}

The gaussian fit used to calculate the gain (Section \ref{sec:gainn}) provides also the standard deviation of the full energy peak. This quantity is related to the energy resolution of the acquisition system. 
\\
The fitting process calculates the $\mathrm{\sigma}$ and the position of the full energy peak in arbitrary units, that is in units of bins of the MCA. To obtain the energy resolution we divide the $\sigma$ obtained by the gaussian fit (which has a.u. dimension) by the position of the full energy peak $\mu$ (in a.u. dimension):

\begin{equation}
\sigma = \frac{\sigma [a.u.]}{\mu [a.u.]}
\end{equation}

$\mathrm{\sigma} $ is a noise-to-signal ratio ($\mathrm{\sigma  [a.u.]}$ is the noise amplitude and $\mathrm{\mu  [a.u.]}$ is the signal amplitude). $\mathrm{\sigma }$ depends inversely on the energy resolution: as the energy resolution improves, $\mathrm{\sigma}$ reduces and vice versa. We will refer to $\mathrm{\sigma}$ as energy resolution.
\\
\\
There are 3 contributions to the loss of energy resolution ($\sigma_{tot}$): 

\begin{itemize}
\item The electronic noise ($\sigma_{electronic}$)
\item The poissonian uncertainty on the production of secondary electrons ($\sigma_{poisson}$)
\item The space charge effect ($\sigma_{space \ charge}$)
\end{itemize}

\begin{figure}[H]
\centering
\includegraphics[width=0.6\textwidth]{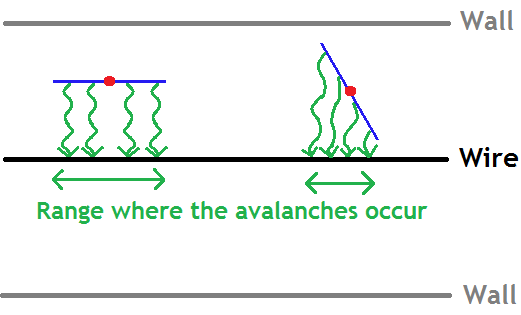}
\caption{The neutron interacts (red dot) emitting the fragments back-to-back (blue lines). The fragments generate secondary electrons that are directed toward the wire (green lines) by the electric field. Two neutron fragments that are emitted perpendicular with respect to the axis wire generate avalanches in a narrower range than two neutron fragments that are emitted parallel to the wire.}
\label{imm:spacechargeexplanation}
\end{figure}

The total energy resolution is:

\begin{equation}
\sigma_{tot}= \sqrt{\sigma_{electronic}^2+\sigma_{poisson}^2+\sigma_{space \ charge}^2}
\end{equation}
\\
The electronic noise does not depend on the HV applied to the tube. As the HV increases, the noise-to-signal decreases as a consequence of the increment of the signal amplitude. For this reason $\sigma_{electronic}$ decreases as the HV increases.
\\
\\
The poissonian uncertainty on the production of secondary electron is given by the ratio mean/(standard deviation) of the Poisson distribution:

\begin{equation}
\sigma_{poisson}=A \cdot \frac{\sqrt{N}}{N}= \frac{A}{\sqrt{N}}
\end{equation}

where $N$ is the number of secondary charge collected and the constant factor and $A$ is a constant that depends on the gain of the acquisition system and on the Fano factor. Note that as the HV applied increases (and the charge multiplication increases as well), $\sigma_{poisson}$ decreases.
\\
\\
The space charge effect arises when the slow ions generated during the Townsend avalanches distort sufficiently the electric field to impede the growth of later avalanches. The space charge effect is relevant only when the gas gain is large enough to create a significant number of ions. 
\\
When the space charge effect is significant, the gas gain depends also on the orientation of the neutron fragments with respect to the wire. If the neutron fragments are emitted perpendicular with respect to the axis of the wire, the avalanches occur in a narrow range, thus the space charge effect can be very significant. The opposite effect occurs when the fragments are emitted parallel with respect to the axis of the wire. 
\\
\\
This phenomenon is explained in Figure \ref{imm:spacechargeexplanation}. The neutron interacts (red dot) emitting the fragments back-to-back (blue lines). The fragments generate secondary electrons that are directed toward the wire (green lines) by the electric field. Two neutron fragments that are emitted perpendicular with respect to the axis wire generate avalanches in a narrower range than two neutron fragments that are emitted parallel to the wire.

\begin{figure} [H]
\centering
\includegraphics[width=0.8\columnwidth]{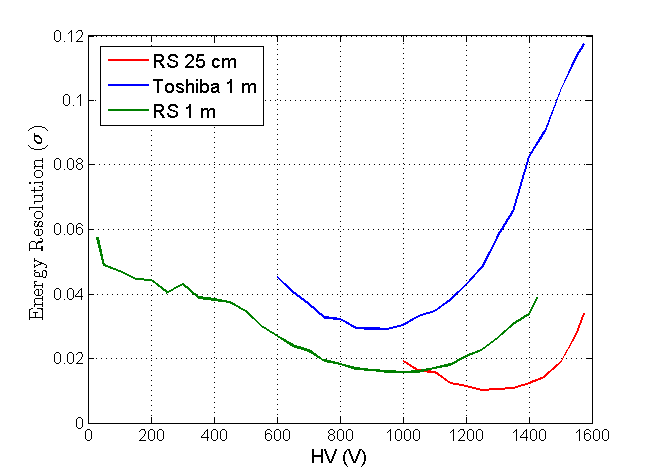}
\caption{The energy resolution as a function of the applied HV.}
\label{imm:enres}
\end{figure}

Because of the space charge effect, the gain depends also on the orientation of the neutron fragments. For this reason, several neutron fragments that release the same energy in the gas, give rise to pulses with a different amplitude and thus the space charge effect deteriorates the energy resolution. This effect is more significant as the gain (HV) increases: $\sigma_{space \ charge}$ increases as the HV increases.
\\
\\
Figure \ref{imm:enres} shows the energy resolution as a function of the HV applied to the tubes. 

\begin{figure} [H]
\centering
\includegraphics[width=0.8\columnwidth]{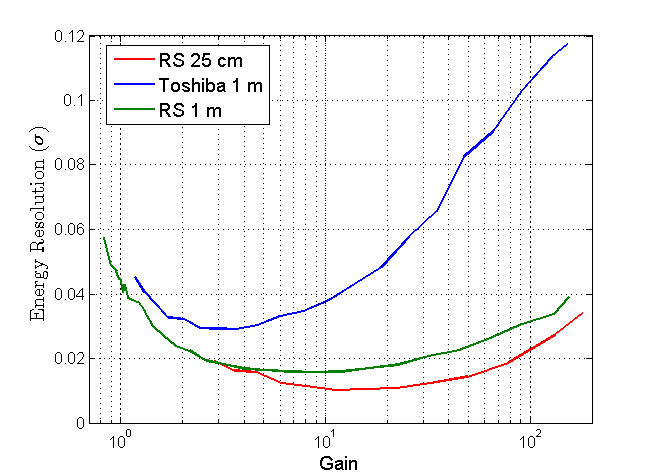}
\caption{The energy resolution as a function of the gain.}
\label{imm:enresgain}
\end{figure}

Since at low HV the space charge effect is negligible, $\sigma$ decreases as the HV increases (as $\sigma_{electronic}$ and $\sigma_{poisson}$). When the gain increases $\sigma_{space \ charge}$ becomes the main contribution to the energy resolution and $\sigma$ increases as the HV increases. The energy resolution has a minimum that occurs at 900-1000 $V$ for the RS and the Toshiba 1 $m$ tubes and at $\approx$1300 $V$ for the RS 25 $cm$ tube. Usually these tubes are operated at relatively large HV (> 1300 $V$), in such configurations the space charge effect is the most important contribution to the energy resolution.
\\
\\
Figure \ref{imm:enresgain} shows the energy resolution as a function of the gas gain. 

The space charge effect becomes the main contribution to the energy resolution at a gain of about 10 for the RS 1 $m$ and the RS 25 $cm$. It happens at a gain of only 2-3 for the Toshiba 1 $m$ tube.
\\
\\
The energy resolution is related to the gamma-ray discrimination of a detector. In fact, to count the neutron signals with $\mathrm{^{3}He}$ proportional counters an amplitude threshold is set in the neutron valley (Section \ref{sec:he3}). The shape and the position of the neutron valley strongly depends on the energy resolution of the detectors. 
\\
Two PHS with different energy resolution are shown in Figures \ref{imm:enresvalley} and \ref{imm:enresvalley2}. The blue PHS has a better energy resolution than the red one.

\begin{minipage}{\textwidth}
\centering
\begin{minipage}[t]{0.49\textwidth}
\begin{figure}[H]
\includegraphics[width=\textwidth]{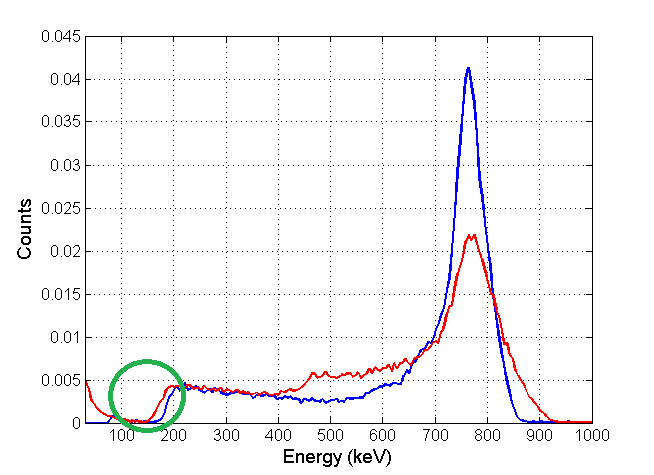}
\caption{PHS with different energy resolution.}
\label{imm:enresvalley}
\end{figure}
\end{minipage}
\begin{minipage}[t]{0.49\textwidth}
\begin{figure}[H]
\includegraphics[width=\textwidth]{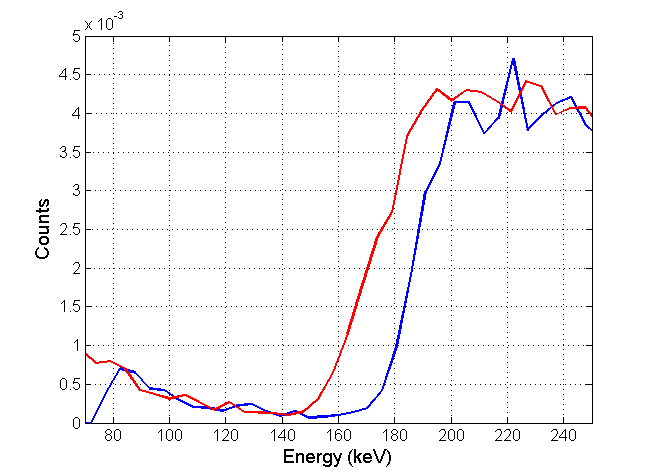}
\caption{Detail of the neutron valley.}
\label{imm:enresvalley2}
\end{figure}
\end{minipage}
\end{minipage}
\vspace{1mm}

As the energy resolution deteriorates, the neutron valley shifts at smaller energies. For this reason the gamma discrimination deteriorates, since a larger portion of the gamma-ray exponential is counted. If there are no other requirements, it is preferable to operate a detector at the HV where the minimum energy resolution occurs in order to maximize the gamma discrimination.

%% file: Chapters/Spatialresolution.tex

\chapter{Spatial Resolution}
\label{ch:spatialresolution}

The results shown in this Chapter were obtained at the beamline R2D2 (Section \ref{sec:facilities}), in order to take advantage of a collimated neutron beam. We discuss the expected reconstructed position distribution of a neutron beam and the effect of the beam width on the measured spatial resolution. The aim of this Chapter is to quantify the spatial resolution of the 1 $m$ long PSD tubes discussed in Section \ref{sec:detector}. The behavior of the spatial resolution as a function of the HV applied to the tubes is discussed here. Some measurements for testing the linearity of the spatial resolution with respect to the position of the beam along and across the wire are also shown.

\section{Introduction}

The spatial resolution is the resolution that is achieved to reconstruct the position of interaction of a neutron in the detector. In a charge division detector the main contribution to the spatial resolution is the amplitude fluctuation due to the electronic noise, e.g. the thermal noise in the resistive wire. These random fluctuations are superimposed to the real signal and they alter the results obtained by using the Equation \ref{eq:cd}. For this reason the reconstructed position distribution of a neutron beam is spread out.
\\
\\
We assume that the reconstructed position distribution of a perfectly punctiform beam has a gaussian shape. We define the spatial resolution as the \textit{Full Width at Half Maximum} (FWHM) of the peak. In the case of gaussian shape, the FWHM is related to the standard deviation $\sigma$ through the formula:

\begin{equation}
FWHM=2\cdot \sqrt{2\cdot \log{(2)}}\cdot \sigma \approx 2.355 \cdot \sigma
\end{equation}

\begin{figure}[H]
\centering
\includegraphics[width=0.8\textwidth]{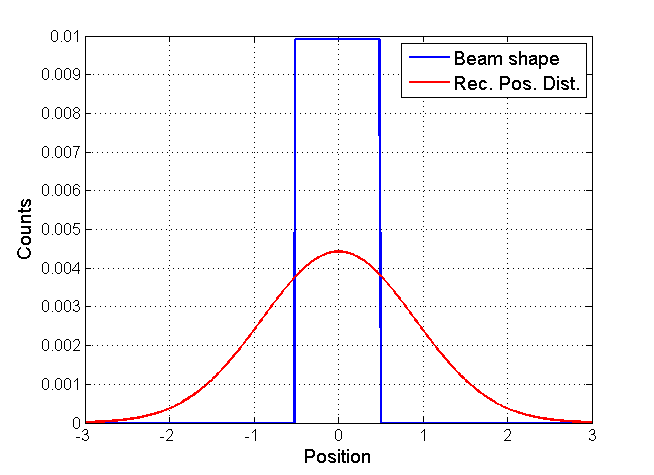}
\caption{Comparison between the neutron beam shape (blue) and the expected reconstructed position distribution (red).}
\label{imm:posresdis}
\end{figure}

A real neutron beam is not punctiform, but has a non-negligible width and a shape. In this case the reconstructed position distribution of every neutron has a gaussian shape and the reconstructed position distribution of the beam is the convolution of the beam shape and the gaussian distribution. At a first approximation the beam can be considered a square beam, i.e. with very precise spatial borders. Figure \ref{imm:posresdis} shows a comparison between the beam shape (blue) and the expected reconstructed position distribution (red).

\begin{figure}[H]
\centering
\includegraphics[width=0.8\textwidth]{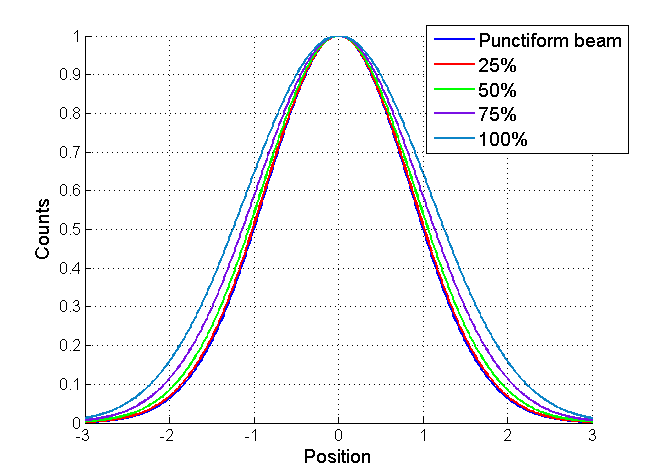}
\caption{Expected reconstructed position distributions with several beam width over spatial resolution ratios. By definition the FWHM of the one obtained with a punctiform beam (blue) is the spatial resolution.}
\label{imm:spres}
\end{figure}

Even though it is really similar, the shape of the reconstructed position distribution is not exactly gaussian. It has a peak in the position of the center of the original neutron beam and its width is related to the spatial resolution and to the beam width as well.
\\
\\
Note that as the beam is sufficiently narrower than the spatial resolution, the beam width does not alter the spatial resolution measured. Figure \ref{imm:spres} shows the expected reconstructed position distribution with several beam width over spatial resolution ratios. By definition the FWHM of the reconstructed position distribution obtained with a punctiform beam (blue) is the spatial resolution. 
\\
\\
The FWHM of the distributions do not depend significantly on the beam width. If the beam width is less than 50\% of the spatial resolution, the difference between the measured spatial resolution and the real one is less than 1\%. This uncertainty is approximately one order of magnitude smaller than the experimental uncertainty we can achieve, therefore we do not take it into account. In the experiments where we quantify the spatial resolution, the beam width is less than 20\% of the spatial resolution. In the case of a beam width equal to the spatial resolution, the difference increases to 25\%.

\section{Spatial Resolution as a Function of the HV}

\begin{figure}[H]
\centering
\includegraphics[width=\textwidth]{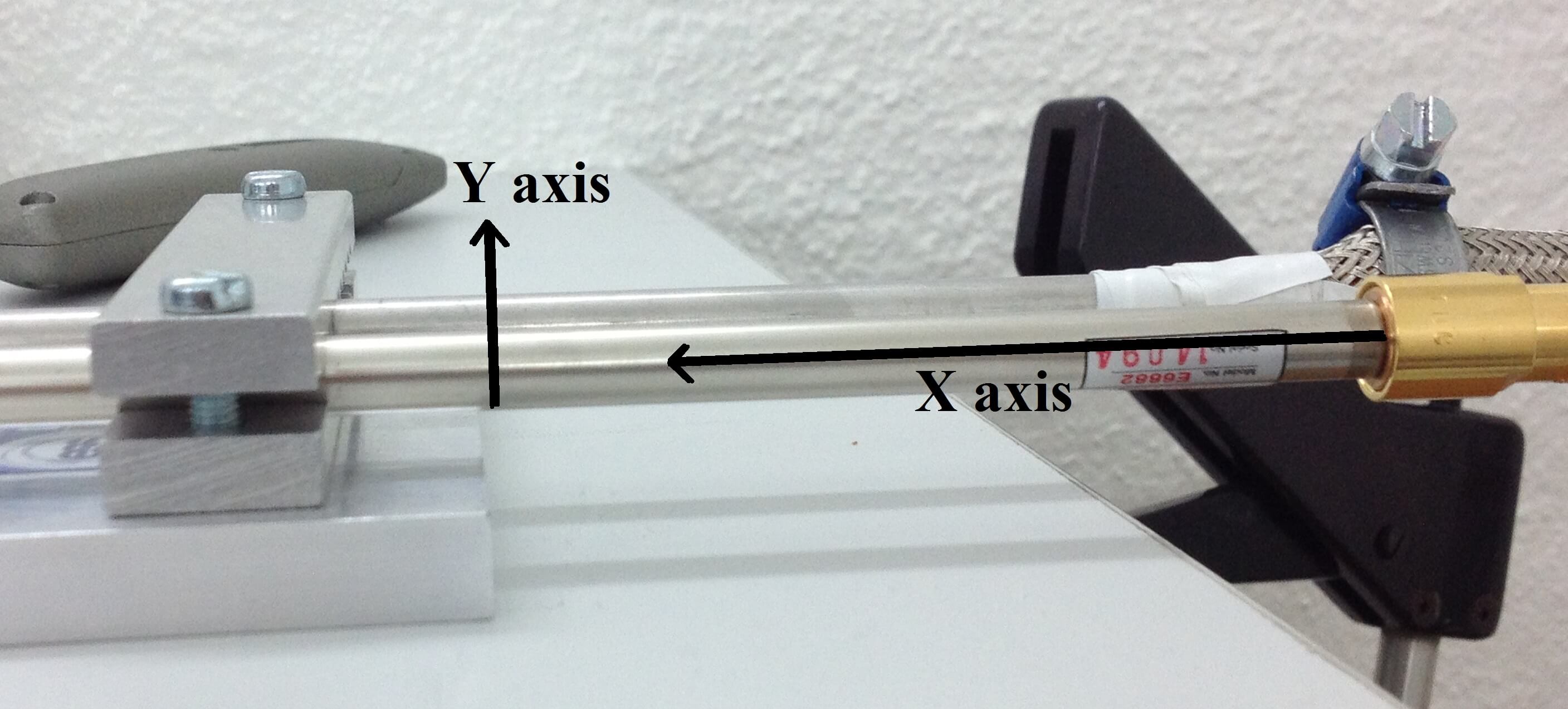}
\caption{X and Y axis.}
\label{imm:axis}
\end{figure}

Figure \ref{imm:axis} shows how we define the X and the Y axis of the tube. The X axis is along the tube axis, while the Y is across it. Note that the charge division method allows to reconstruct the position of interaction of a neutron along the X axis.

The measurements are performed with a 1 $mm$ x 1 $cm$ beam. We choose a beam width along the X axis of 1 $mm$ in order to have a beam smaller than the spatial resolution (that is >5  $mm$). In order to have a sufficiently high beam flux with respect to the background we increase the size of the beam along the Y axis to 1 $cm$. The signals from the two connectors of the tube are amplified by the CAEN 422A preamplifier and then shaped by the ORTEC 855 Dual Spec Amplifier. The outputs are recorded with a CAEN DT5780 MCA, which allows us to record the amplitude of the two signals at the same time. The beam is placed hits the tubes in the center of both axis.
\\
\\

\begin{figure}[H]
\centering
\includegraphics[width=0.8\textwidth]{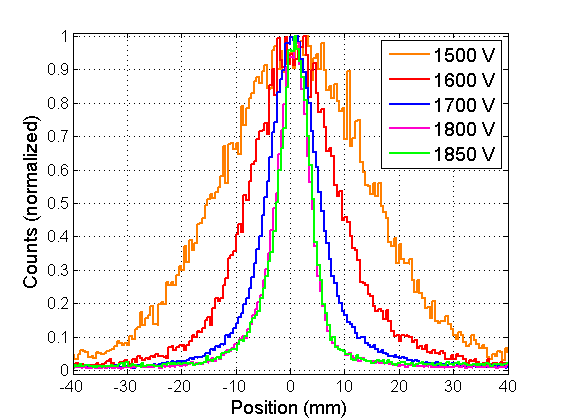}
\caption{Reconstructed position distributions of a 1 $mm$ x 1 $cm$ beam with several HV with the Toshiba tube.}
\label{imm:toshibaspatial}
\end{figure}

The reconstructed position distributions obtained with the Toshiba detector for several HV are shown in Figure \ref{imm:toshibaspatial}.

The spatial resolution improves by increasing the HV as a consequence of the larger signal-to-noise ratio on the outputs. As the gain increases, the electronic noise becomes less significant with respect to the amplitudes of the signals, hence the spatial resolution improves.
\\
\\
Figure \ref{imm:tosnumspat} and Table \ref{tab:tosnumspat} show the spatial resolution for the Toshiba tube as a function of the applied HV.

\begin{minipage}{\textwidth}
\hspace{-1.5cm}
\begin{minipage}[]{0.7\textwidth}
\begin{figure}[H]
\includegraphics[width=\textwidth]{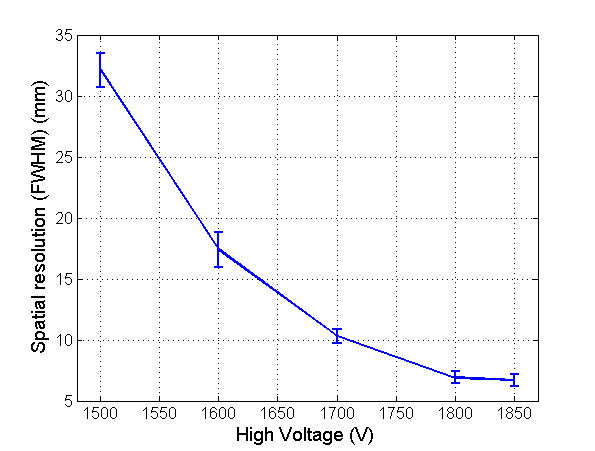}
\caption{Spatial resolution for the Toshiba tube with several HV.}
\label{imm:tosnumspat}
\end{figure}
\end{minipage}
\begin{minipage}[t]{0.5\textwidth}
\begin{tabular}{|c|c|c|}
\hline
HV ($V$) & Spatial Res. (FWHM) ($mm$)\\ \hline \hline
1500 & 32.1 $\pm$ 1.4\\ 
1600 & 17.4 $\pm$ 1.4\\ 
1700 & 10.3 $\pm$ 0.6\\ 
1800 & 6.9 $\pm$ 0.5\\ 
1850 & 6.7 $\pm$ 0.5\\ \hline
\end{tabular}
\captionof{table}{Spatial resolution of the Toshiba tube.}
\label{tab:tosnumspat}
\end{minipage}
\end{minipage}
\vspace{1mm}

\begin{figure}[H]
\centering
\includegraphics[width=0.8\textwidth]{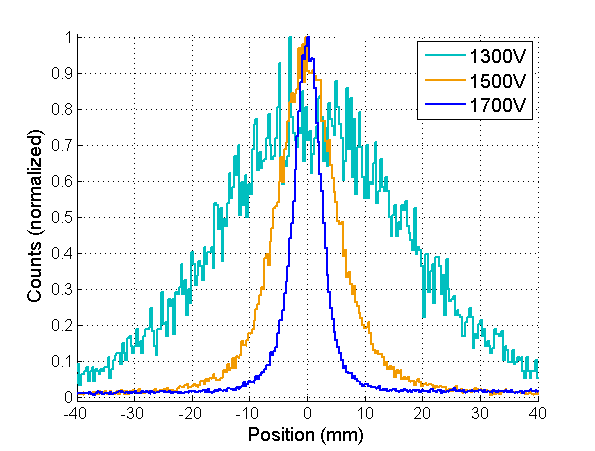}
\caption{Reconstructed position distributions obtained with a  1 $mm$ x 1 $cm$ beam with several HV applied to the RS 1 $m$ tube.}
\label{imm:reuterspatial}
\end{figure}

From Figure \ref{imm:tosnumspat} and Table \ref{tab:tosnumspat} we observe that the spatial resolution does not improve further above 1800 $V$ because the gas gain saturates. The maximum spatial resolution of this tube is (6.7 $\pm$ 0.5) $mm$ that corresponds to about 0.7\% of the total active length of the wire.
\\
\\
The reconstructed position distributions obtained with the RS 1 $m$ tube for several HV are shown in Figure \ref{imm:reuterspatial}. Figure \ref{imm:reuternumspat} and in Table \ref{tab:reuternumspat} show the spatial resolution as a function of the applied HV.

\begin{minipage}{\textwidth}
\hspace{-1.5cm}
\begin{minipage}[]{0.7\textwidth}
\begin{figure}[H]
\includegraphics[width=\textwidth]{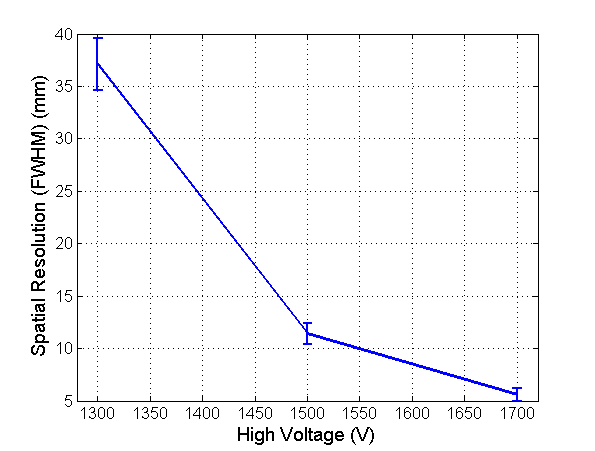}
\caption{The spatial resolution for the RS 1 $m$ tube with several HV.}
\label{imm:reuternumspat}
\end{figure}
\end{minipage}
\begin{minipage}[t]{0.5\textwidth}
\begin{tabular}{|c|c|c|}
\hline
HV ($V$) & Spatial Res. (FWHM) ($mm$)\\ \hline \hline
1300 & 37 $\pm$ 3\\ 
1500 & 11.4 $\pm$ 1.0\\ 
1700 & 5.6 $\pm$ 0.6\\  \hline
\end{tabular}
\captionof{table}{Spatial resolution of the RS 1 $m$ tube.}
\label{tab:reuternumspat}
\end{minipage}
\end{minipage}
\vspace{1mm}

This detector reaches a spatial resolution of (5.6 $\pm$ 0.6) $mm$, that correspond to about 0.6\% of the total length of the wire. The difference in the spatial resolution between the two tubes is a consequence of the different gas composition.

\section{X axis scan}

\begin{figure}[H]
\centering
\includegraphics[width=0.8\textwidth]{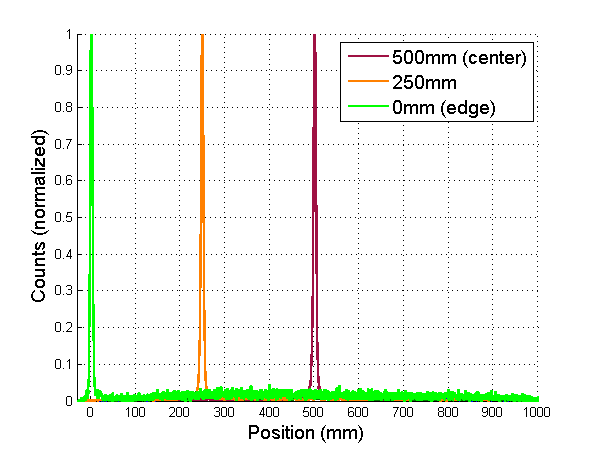}
\caption{Reconstructed positions with the beam hitting the tube in 3 evenly-spaced positions.}
\label{imm:GEXaxis}
\end{figure}

By using the same experimental setup, we measure the linearity along the X axis of the tube (see Figure \ref{imm:axis}). We apply a HV of 1800 $V$ to the Toshiba tube and of 1700 $V$ to the RS tube, in order to have the minimum spatial resolution. The beam scans the X scan, while it is hitting the tube in its center along the Y axis.
\\
Figure \ref{imm:GEXaxis} shows the reconstructed position when the beam hits the RS tube in 3 evenly-spaced positions: one in the center, one at the edge and one in the middle of the two. 

Figures \ref{imm:GEX} and \ref{imm:ToshibaX} show the reconstructed position  in different positions respectively for the tubes. Note that we normalize and we shift the peaks to the center of the axis in order to superimpose and easily compare the responses.

 \begin{minipage}{\textwidth}
\centering
\begin{minipage}[t]{0.49\textwidth}
\begin{figure}[H]
\includegraphics[width=\textwidth]{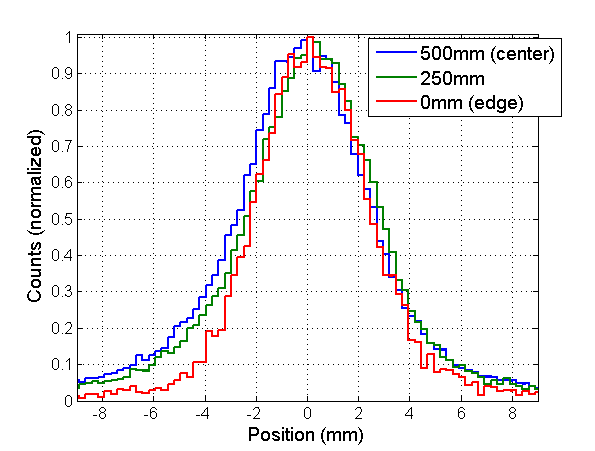}
\caption{Reconstructed positions measured with the RS 1 $m$ tube.}
\label{imm:GEX}
\end{figure}
\end{minipage}
\begin{minipage}[t]{0.5\textwidth}
\begin{figure}[H]
\includegraphics[width=\textwidth]{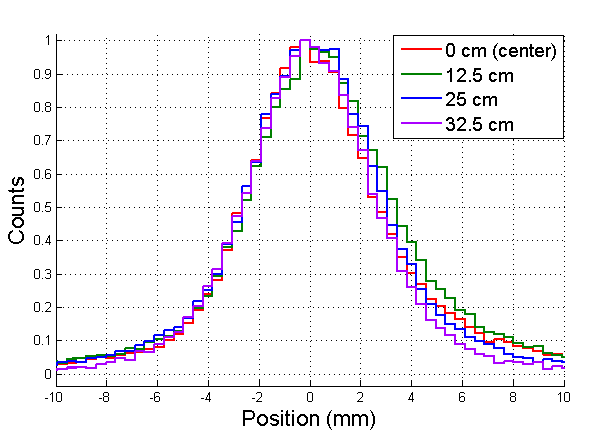}
\caption{Reconstructed positions measured with the Toshiba tube.}
\label{imm:ToshibaX}
\end{figure}
\end{minipage}
\end{minipage}
\vspace{2mm}

Note that spatial resolution does not vary along the length of the tube. A small difference can be noticed when the beam hits the edge of the RS tube. This is due to non-linear effects introduced by parasitic capacitances in the tube, that compress the position reconstruction near the edges. 

\section{Y axis scan}

\begin{figure}[H]
\centering
\includegraphics[width=0.8\textwidth]{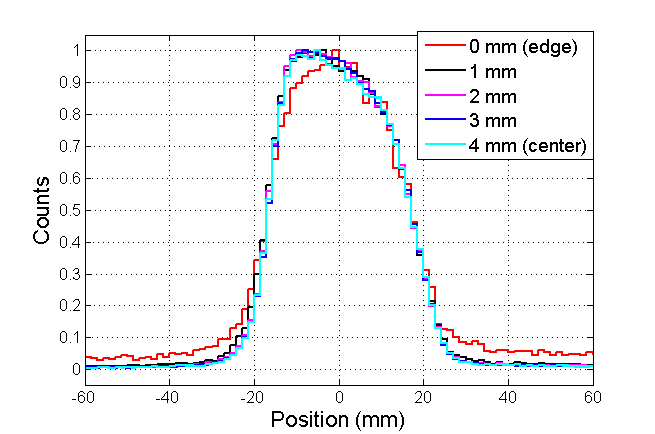}
\caption{Peaks measured varying the Y position of the beam.}
\label{imm:ToshibaYaxis}
\end{figure}

We use a 2.5 $cm$ x 0.3 $mm$ beam to measure the linearity of the spatial resolution of the Toshiba tube while the beam scans the Y axis (see Figure \ref{imm:axis}). We use a width of only 0.3 $mm$ along the Y axis in order to have a good precision on the position along this axis. In order to have enough neutron flux with respect to the background we increase the size of the beam along the X axis to 2.5 $cm$. This Y-scan is then performed in the center of the X axis. The beam is placed in the center of the tube along the X axis. The experimental set-up is the same as in the previous measurements. A HV of 1800 $V$ is applied in order to have the minimum spatial resolution. Figure \ref{imm:ToshibaYaxis} shows the result of this measurement.

Note that the spatial resolution does not vary along the Y axis. The asymmetrical shape of the peaks is due to the non-uniformity of the neutron beam along the X axis. 

%% file: Chapters/Gammasensitivity.tex

\chapter{Gamma Sensitivity}

The most significant source of background in neutron detectors originates from gamma-rays. For this reason the sensitivity of a neutron detector to gamma-rays is a crucial property, as it defines the best achievable signal-to-noise ratio. The gamma sensitivity is the probability for a gamma-ray to give rise to a detectable signal in the detector. Gamma-rays are generated directly from the neutron source (spallation, fission reactor, $\mathrm{\alpha}$-Be etc.), or by interaction of neutrons with any material in their path, e.g. neutron collimators, choppers or beam stops. For example, a common source of gamma-ray background originates from the hydrogen neutron reaction:

\begin{equation}
^{1}n + ^{1}H  \quad \longrightarrow \quad ^{2}H + \gamma \ (\mathrm{2.1} \ MeV)
\end{equation}

Some gamma-rays are also part of the environmental background, such as atmospheric gamma-rays. Gamma-rays generate pulses in the neutron detector that can be misidentified as neutrons. Even though gamma sensitivity is generally low, e.g. 10$\mathrm{^{-6}}$, several neutron scattering applications involve a very low neutron flux. In these cases gamma-rays alter significantly the experimental results. 
\\
\\
The gamma sensitivity has been widely quantified since the rising of $\mathrm{^{3}He}$-alternative technologies in the last decade, even though in literature gamma-ray sensitivity measurements can be traced back to the '80s \cite{Jensen1983}, but only related to $\mathrm{^{6}Li}$-based scintillators. Recent examples of such a measurement involve $\mathrm{^{10}B}$-lined detectors \cite{Khaplanov2013}, $\mathrm{^{6}Li}$ scintillators \cite{Sykora2012}, and hydrogen GEM detectors for fast neutrons \cite{Croci2013}. One can even find some criteria on the acceptable gamma sensitivity for neutron detectors used for national security applications in the United States \cite{Kouzes2011}. In this case a low neutron signal must be detectable in the presence of a natural gamma-ray background, and the system should not create alarms due to a change in the gamma flux.
\\
\\
Despite the importance of gamma sensitivity for neutrons detectors, in the literature it is not possible to find such a measurement performed with $\mathrm{^{3}He}$ detectors, because in the past this was not considered a priority. The aim of this Chapter is to discuss the gamma sensitivity of some $\mathrm{^{3}He}$ tubes, in order to provide a standard to compare the performances of $\mathrm{^{3}He}$ neutron detectors with those of alternative technology detectors. The conversion of the gamma-rays and the transmission of the generated electrons in the detector walls are also discussed. The experimental results establish the relationship between the gamma sensitivity and the space charge effects and the gas composition in the detectors.

\section{Definition of Gamma Sensitivity}
\label{sec:definitiongamma}

We define the gamma sensitivity as the efficiency of recording a gamma-ray of a given energy:

\begin{equation}
GS(E)=\frac{N_{counts}}{N_{incident}}=\frac{N_{counts}}{A \cdot \frac{\Omega}{4\pi} \cdot t}
\end{equation}

Where $N_{counts}$ is the number of gamma-rays counted and $N_{incident}$ is the number of gamma-rays incident on the detector. $A$ is the activity of the gamma-ray source, $\Omega$ is the solid angle subtended by the detector at the source and $t$ is the duration of the measurement. 
\\
We consider the activity of the source $A$ and the duration of the measurements $t$ without uncertainty, and a poissonian error on the gamma counts. The relative error on gamma sensitivity is:

\begin{equation}
\frac{\delta GS}{GS} = \sqrt{\left( \frac{1}{N_{counts}}\right)^2+ \left( \frac{\delta \Omega}{\Omega} \right)^{2}}
\end{equation}

In order to decrease the quantity $1/N_{counts}$ a long measurement is needed and to decrease the quantity $\delta \Omega / \Omega$ it is necessary to measure carefully the distance between the source and the tube.
\\
\\ 
The given definition is ambiguous because it does not take into account the energy required for a gamma-ray pulse to be recorded. In $\mathrm{^{10}B}$-lined detectors an accepted procedure consists in varying the voltage and fix a threshold, and then count the gamma-ray pulses with an energy over that threshold. This is the same procedure as the one used for counting curve measurements. This makes sense for this kind of detectors because their spectrum extends down to low energies and so there is no unambiguous choice for the threshold used for counting neutrons (see Figure \ref{imm:B10PHS}).

\begin{minipage}{\textwidth}
\centering
\begin{minipage}[t]{0.49\textwidth}
\begin{figure}[H]
\includegraphics[width=\textwidth]{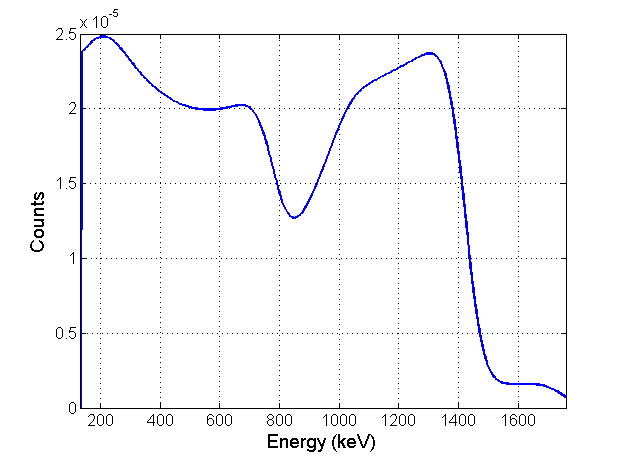}
\caption{$\mathrm{^{10}B}$ PHS.}
\label{imm:B10PHS}
\end{figure}
\end{minipage}
\begin{minipage}[t]{0.49\textwidth}
\begin{figure}[H]
\includegraphics[width=\textwidth]{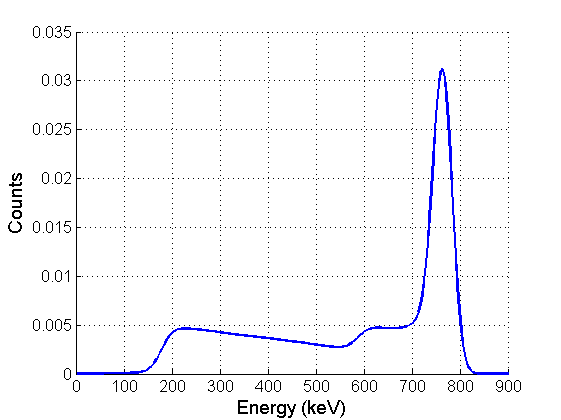}
\caption{$\mathrm{^{3}He}$ PHS.}
\label{imm:He3PHS}
\end{figure}
\end{minipage}
\end{minipage}
\vspace{1mm}

The peculiar shape of the $\mathrm{^{3}He}$ PHS (Figure \ref{imm:He3PHS}) has a reasonable and, at a first approximation, unambiguous threshold choice to count the neutron pulses: the neutron valley between the neutron PHS and the gamma-ray exponential. The most straightforward approach consists in counting the gamma-ray pulses that have an amplitude over this threshold. The Section \ref{sec:sensitivityvalley} shows the experimental results found with this approach. In reality, the neutron valley depends critically on external factors, such as energy resolution (that depends on the HV and on the electronics) and the gamma flux, that should be completely independent from gamma sensitivity.
\\
\\
The dependence on energy resolution is shown in Figures \ref{imm:valleyen} and \ref{imm:valleyenzoom}. These PHS are obtained by a Monte Carlo simulation implemented in MATLAB. While the energy resolution gets worse, the neutron valley is shifted toward lower energies. The valley position depends on energy resolution, that is directly related to the high voltage applied to the tube and to the electronics used.

\begin{minipage}{\textwidth}
\centering
\begin{minipage}[t]{0.49\textwidth}
\begin{figure}[H]
\includegraphics[width=\textwidth]{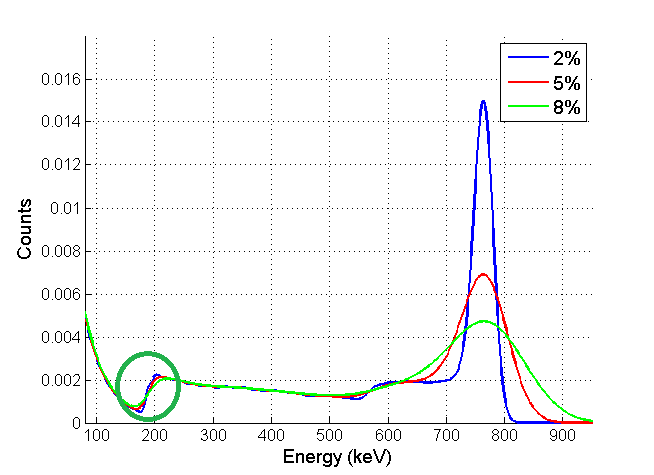}
\caption{PHS with different energy resolution.}
\label{imm:valleyen}
\end{figure}
\end{minipage}
\begin{minipage}[t]{0.49\textwidth}
\begin{figure}[H]
\includegraphics[width=\textwidth]{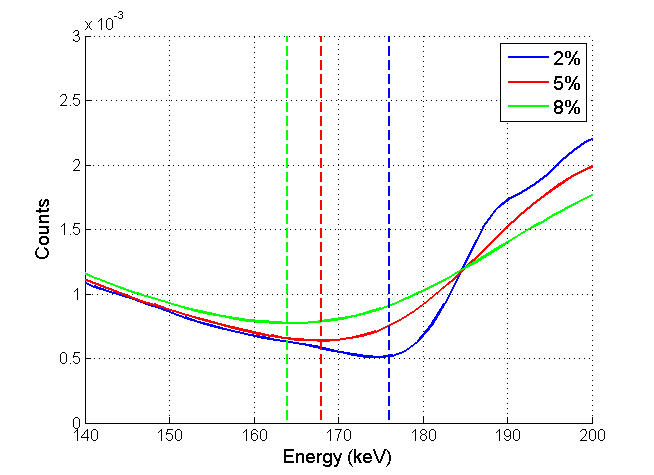}
\caption{Detail of the neutron valley. The dotted lines indicate the position of the neutron valley.}
\label{imm:valleyenzoom}
\end{figure}
\end{minipage}
\end{minipage}
\vspace{1mm}

The position of the neutron valley depends on the gamma flux (Figures \ref{imm:valley} and \ref{imm:valleyzoom}). The higher the gamma flux, the larger is the position of the neutron valley, i.e. it is shifted toward higher energies.

\begin{minipage}{\textwidth}
\centering
\begin{minipage}[t]{0.49\textwidth}
\begin{figure}[H]
\includegraphics[width=\textwidth]{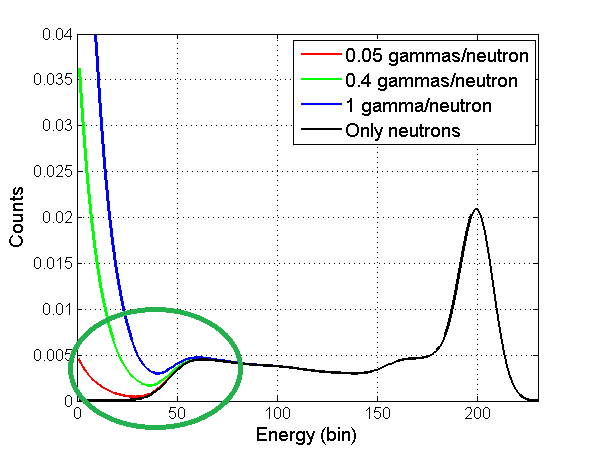}
\caption{PHS with different gamma flux intensity.}
\label{imm:valley}
\end{figure}
\end{minipage}
\begin{minipage}[t]{0.49\textwidth}
\begin{figure}[H]
\includegraphics[width=\textwidth]{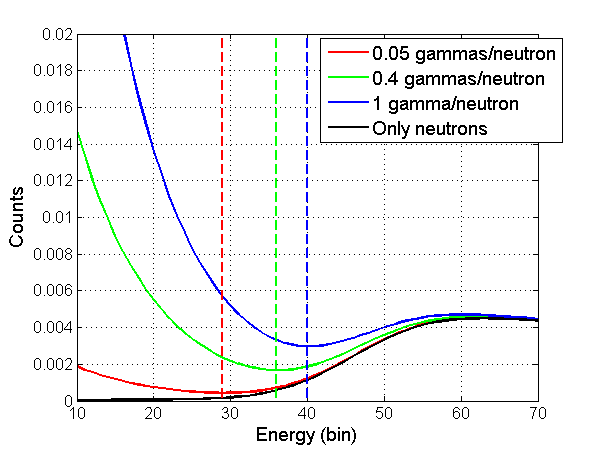}
\caption{Detail of the neutron valley. The dotted lines indicate the position of the neutron valley.}
\label{imm:valleyzoom}
\end{figure}
\end{minipage}
\end{minipage}
\vspace{1mm}

\begin{figure}[H]
\centering
\includegraphics[width=0.8\textwidth]{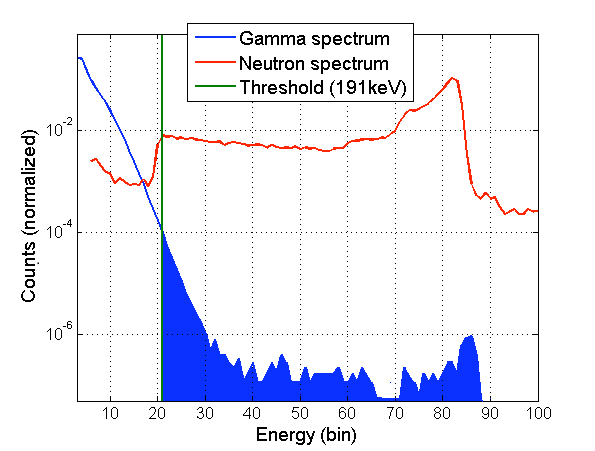}
\caption{Explanation of the calculation of gamma sensitivity. By setting the threshold on the proton-escape peak only the gamma-ray pulses over the 191 $keV$ are counted (blue area).}
\label{imm:sensexample}
\end{figure}

The position of the neutron valley depends on two external factors that are not under our direct control. The threshold we apply to reject the gamma pulses is not well defined and this could alter sensibly the gamma sensitivity results. Moreover, since the neutron valley often has a flat shape, its minimum is difficult to identify and this uncertainty can bring very different results for the gamma sensitivity. For these reasons the results we will show in Section \ref{sec:sensitivityvalley} should be considered semi-quantitative.
\\
\\
We decide then to use a different way to set the threshold for counting gamma-ray pulses. We use the peak that corresponds to the proton-escape at 191 $keV$ (Figure \ref{imm:sensexample}). This point has two advantages compared with the neutron valley: it does not depend on the gamma flux and it corresponds to a well-defined energy. This peak is easier to identify than the neutron valley. As a result, the uncertainty on the position of this peak does not affect dramatically the gamma sensitivity results.

Figure \ref{imm:sensexample} shows that although the detector is shielded with borated polyethilene and aluminium plates, some background neutrons are still recorded. This background does not alter significantly the gamma sensitivity results, in fact it gives a contribution approximately one order of magnitude smaller than the gamma-rays coming from the source we used.
\\
\begin{wrapfigure}{r}{0.4\textwidth}
\includegraphics[width=0.5\textwidth]{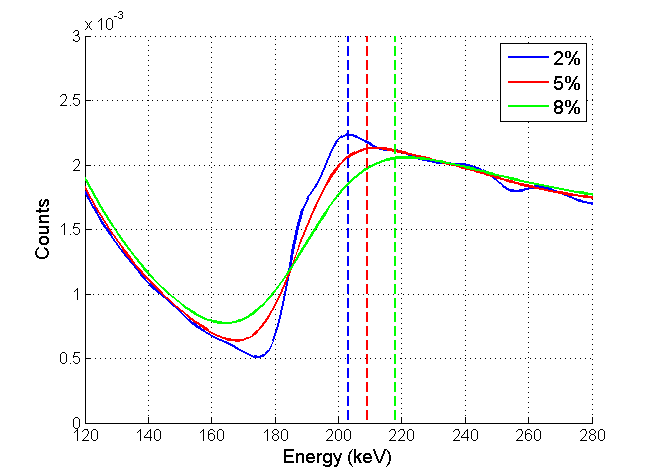}
\caption{Detail of the neutron valley. The dotted lines indicate the position of the peak at 191 $keV$.}
\label{imm:valley191}
\end{wrapfigure}

It must be pointed out that even the proton-escape peak depends on energy resolution (Figure \ref{imm:valley191}). In this case, however,  when the energy resolution gets worse the threshold is shifted toward higher energies and the gamma sensitivity decreases, exactly the opposite of what happens when the threshold is set in the neutron valley. As we will explain in the following Sections,  the measurements of the gamma sensitivity show that it increases with the applied voltage despite the fact that the position of the threshold tends to decrease it. This helps us to link the gamma sensitivity with the space charge effects. This would not be possible if we were using the gamma sensitivity defined by the neutron valley threshold, where the degradation of the energy resolution can be the only responsible for the increase of the gamma sensitivity.

\section{Conversion of Gamma-Rays}
\label{sec:conversion}

\begin{figure}[H]
\centering
\includegraphics[width=0.9\textwidth]{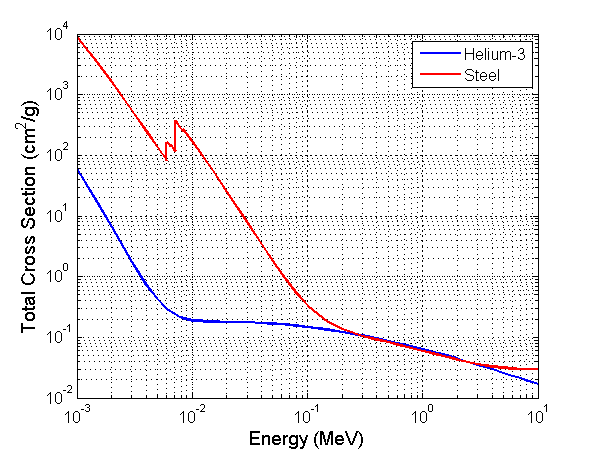}
\caption{Cross section of interaction of a gamma-ray in $\mathrm{^{3}He}$ and steel taken from \cite{Berger}}
\label{imm:gammacross}
\end{figure}

As explained in Section \ref{sec:interactiongamma} gamma-rays are converted into photo-electrons, which then generate the signals in the detector. Figure \ref{imm:crosssteel} shows that gamma-rays in steel in the energy region of interest for us ($\approx$100 $keV$-3 $MeV$, that correspond to the most common background gamma-rays with energy high enough to be recorded in the $\mathrm{^{3}He}$ detectors) interact mainly via Compton scattering. The conversion of a gamma-ray can occur either in the gas or in the detector walls. Figure \ref{imm:gammacross} shows the total cross sections (Photoelectric absorption + Compton scattering + pair production) for the interaction of a gamma-ray with $\mathrm{^{3}He}$ and steel.

In the energy range of interest for us the cross-sections have approximately the same magnitude. Because of its higher density, however, the probability of interaction is much higher in the steel than in$\mathrm{^{3}He}$. For example, if we consider a 10 $bar$ $\mathrm{^{3}He}$ tube with 1'' diameter and with walls of $\approx$0.5$mm$ (as the RS 25 $cm$ tube descibed in \ref{sec:detector}), we find for a 1 $MeV$ gamma-ray in $\mathrm{^{3}He}$ a probability of interaction of:

\begin{equation}
\label{eq:heliumphoton}
p_{\gamma interaction}^{^{3}He}= n \cdot \sigma \cdot d = 0.0012 \; \frac{g}{cm^{3}} \cdot 0.064 \; \frac{cm^{2}}{g} \cdot 2.54 \; cm = 1.9 \cdot 10^{-4} \ (0.019\%)
\end{equation}

On the other hand the probability of interaction in steel of the tube is:

\begin{equation}
\label{eq:steel}
p_{\gamma interaction}^{steel}= n \cdot \sigma \cdot d = 8 \; \frac{g}{cm^{3}} \cdot 0.055 \; \frac{cm^{2}}{g} \cdot 0.05 \; cm = 2.18 \cdot 10^{-2} \ (2.18\%)
\end{equation}

This probability is actually a lower limit for $p_{\gamma interaction}^{steel}$, because we are considering a gamma-ray to have a incidence orthogonal to the steel wall. In most cases a gamma-ray has an oblique incidence and travels more in the steel. The probability of interaction for a gamma-ray in steel is about 100 times larger than in $\mathrm{^{3}He}$. 
\\
\\
Let us consider the gamma-rays that interact with the walls of the tube. In order to create a signal the photo-electrons must travel in the steel until they reach the gas and consequently generate a detectable charge. During this process the photo-electrons lose part of their energy and most of them are absorbed. To create a pulse above the threshold that can consequently contribute to the gamma sensitivity, one photo-electron must release more than 191 $keV$ in the gas, the energy threshold that we set on the proton-escape peak.
\\
\\
If we can assume that the interaction in $\mathrm{^{3}He}$ is negligible with respect to the one in steel, we can divide the contribution to the gamma sensitivity arising from the gamma interacting with only the walls of the tube into 3 main contributions:

\begin{itemize}
\item Probability of interaction of a photon in the steel, with consequent emission of a photo-electron.
\item Probability that a photo-electron reaches the gas, i.e. is not absorbed in the wall.
\item Probability that the photo-electron releases in the gas more than 191 $keV$ (the energy threshold we set).
\end{itemize}

The gamma sensitivity, defined as the efficiency of recording a gamma-ray pulse corresponding to an energy >191 $keV$, is given by the multiplication of these 3 probabilities:

\begin{equation}
\label{eq:sensitivity}
GS=p_{\gamma interaction} \cdot p_{e^{-} transmission} \cdot p_{e^{-} release}
\end{equation}

We are interested in searching the order of magnitude of these 3 parameters. Equation \ref{eq:steel} shows that $p_{\gamma interaction}^{steel}\approx10^{-2}$.

\tocless\subsection{Simulations}
\label{sec:simulations}

\begin{figure}[H]
\centering
\includegraphics[width=0.9\textwidth]{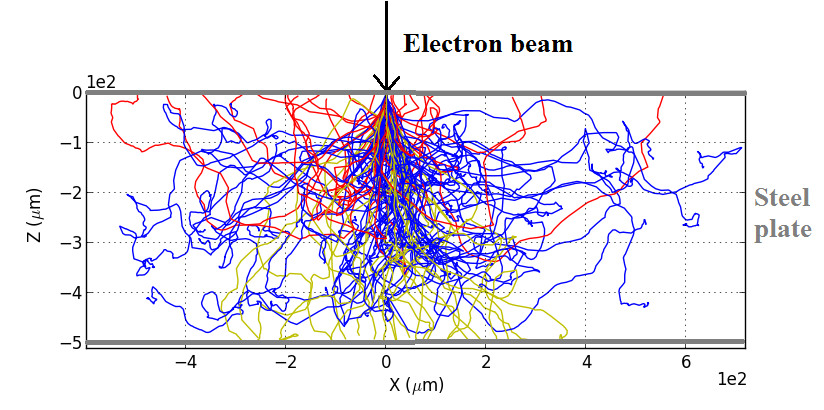}
\caption{The traces of an electron beam incident on a steel plate obtained with pyPENELOPE. In blue are indicated the electrons absorbed in the plate, in yellow the electrons transmitted to the the other side of the plate and in red the backscattered electrons.}
\label{imm:expenelope}
\end{figure}

For a better understanding of the generation and transport of the electrons in steel we perform simulations with pyPENELOPE \cite{Pinard}. This is an open-source software that implements the Monte Carlo code PENELOPE (\textit{Penetration and ENErgy LOss of Positrons and Electrons}) \cite{Salvat}, developed to simulate the transport of photons, electrons and positrons in matter. pyPENELOPE allows us to estimate the quantity $p_{e^{-} transmission}$. Figure \ref{imm:expenelope} shows an example of a simulation with pyPENELOPE involving an electron beam incident on a steel plate. There are indicated the traces of the electrons absorbed in the plate (blue), those transmitted to the other side of the plate (yellow) and those backscattered, i.e. transmitted to the same side of the plate where the beam enters (red).

\begin{figure}[H]
\centering
\includegraphics[width=0.7\textwidth]{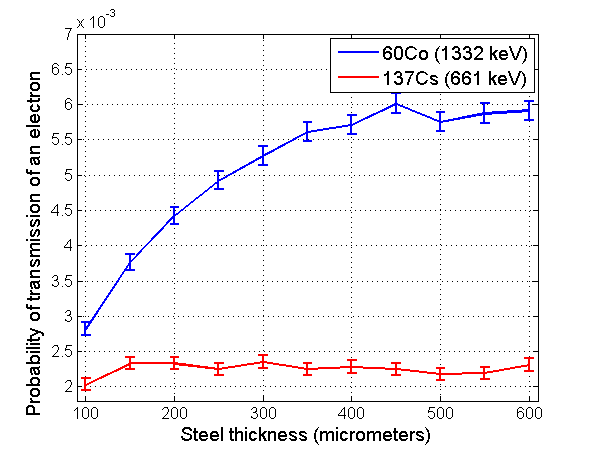}
\caption{The probability that a photon of a given energy interacts in a layer of steel and that it gives rise to a photo-electron that is transmitted to the other side of the plate as a function of the layer thickness.}
\label{imm:penelopethickness}
\end{figure}

pyPENELOPE allows to calculate the number of electrons that are transmitted into the gas starting from a given number of gamma-rays. Since we are determining the order of magnitude of $p_{e^{-}transmission}$, we neglect the curvature of the steel walls and we consider the case of a planar geometry. We simulate several thicknesses of the steel plate, in order to understand how the gamma sensitivity should change with it. We use photons of two energies: 1332 $keV$, as $\mathrm{^{60}Co}$-source photons, and 661 $keV$, as $\mathrm{^{137}Cs}$-source photons. The probability that a photon of such an energy interacts and the corresponding photo-electron is transmitted to the other side of the plate is shown in Figure \ref{imm:penelopethickness}. This quantity corresponds to the product $p_{\gamma interaction} \cdot p_{e^{-} transmission}$ and we refer to it as $p_{transmission}$.

\begin{figure}[H]
\centering
\includegraphics[width=0.8\textwidth]{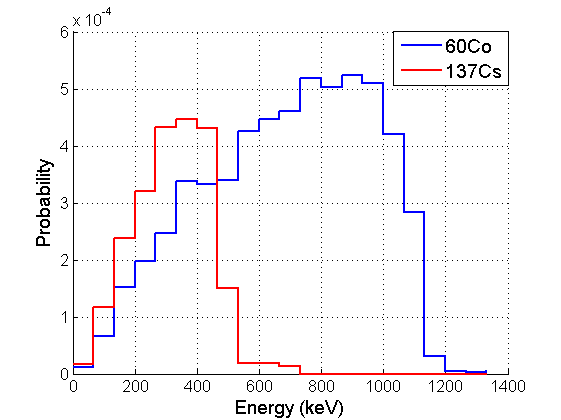}
\caption{Photo-electron PHS transmitted through a steel layer of thickness 500 $\mu m$.}
\label{imm:sensitivityspectrum500}
\end{figure}

Note that the probability for a photon to interact in the steel and transmit a photo-electron into the gas, i.e. $p_{transmission}$, is $\approx$10$^{-3}$. -This means that one photon over $\approx$1000 gives rise to a signal in the detector. Although not all the signals will be recorded, this quantity allows to estimate the gamma-ray signal rate (and pile-up) in the case of a measurement with a source of known activity. In the case of the thicknesses of our tubes (250 $\mu m$ and 500 $\mu m$), $p_{transmission}$ of a $\mathrm{^{137}Cs}$ photon is lower than that of $\mathrm{^{60}Co}$ by a factor between 2 and 3. We expect then the gamma sensitivity to follow a similar behavior. Note that for thicknesses larger than about 300 $\mu m$ for $\mathrm{^{60}Co}$-photons and about 100 $\mu m$ for $\mathrm{^{137}Cs}$ the gamma sensitivity does not increase further. This is due to the fact that only the photo-electrons generated in the last layers are transmitted into the gas. Given that $p_{e^{-} transmission}^{steel}\approx$10$^{-2}$ (from Equation \ref{eq:steel}) and $p_{transmission}= p_{\gamma interaction} \cdot p_{e^{-} transmission} \approx$10$^{-3}$ we find that $p_{e^{-} transmission}\approx$10$^{-1}$. From Equation \ref{eq:heliumphoton} we found that the probability for a photon to interact in the gas releasing a photo-electron is $\approx$10$^{-4}$, an order of magnitude lower than $p_{transmission}$. In this way we prove that, at a first approximation, we can neglect the interactions of photons in the $\mathrm{^{3}He}$ with respect of the ones in the wall steel. 
\\
\\
Moreover the probability that a photo-electron is backscattered is approximately 5$\cdot$10$^{-4}$ and 1.5$\cdot$10$^{-4}$ for $\mathrm{^{60}Co}$ and $\mathrm{^{137}Cs}$ photons respectively. These probabilities do not depend critically on the steel thickness. For this reason the contribution given by the wall facing the source is larger (about a factor 10) than that of the opposite wall.
\\
\\
Figure \ref{imm:sensitivityspectrum500} shows the PHS originated by the photo-electrons (for $\mathrm{^{60}Co}$ and $\mathrm{^{137}Cs}$) transmitted through a steel layer of 500 $\mu m$.

The photo-electrons that enter into the gas have lost about 1/4 of the gamma-ray initial energy. The PHS extends toward zero energy and almost no photo-electrons reach the gas with the full gamma-ray energy.

\section{Gamma Sensitivity Measurements}
\label{sec:gamm191}

We perform the gamma sensitivity measurements with 2 gamma-ray sources: $\mathrm{^{60}Co}$ and $\mathrm{^{137}Cs}$. These sources are, between common gamma-ray sources, the ones with gamma-rays of highest energy. Table \ref{tab:sources} shows their half-lives, the energy and the relative intensity of the gamma-rays emitted and the activity of our sources. Only the gamma-rays with energy >191 $keV$ are shown.

\begin{table}[H]
\centering
\begin{tabular}{|c|c|c|c|c|}
\hline
Isotope & Half life & $\gamma$ energy (keV) & Intensity (\%) & Activity ($MBq$)\\
\hline
\multirow{2}{*}{$\mathrm{^{60}Co}$} & \multirow{2}{*}{1925.28 d} & 1173.2 & 99.85 & \multirow{2}{*}{2.39}\\
& & 1332.5 & 99.98 & \\
\hline
$\mathrm{^{137}Cs}$ & 30.08 y & 661.7 & 85.1 & 7.11\\
\hline
\end{tabular}
\caption{The gamma-ray sources we used for measuring gamma sensitivity.}
\label{tab:sources}
\end{table}

\begin{figure}[H]
\centering
\includegraphics[width=0.8\textwidth]{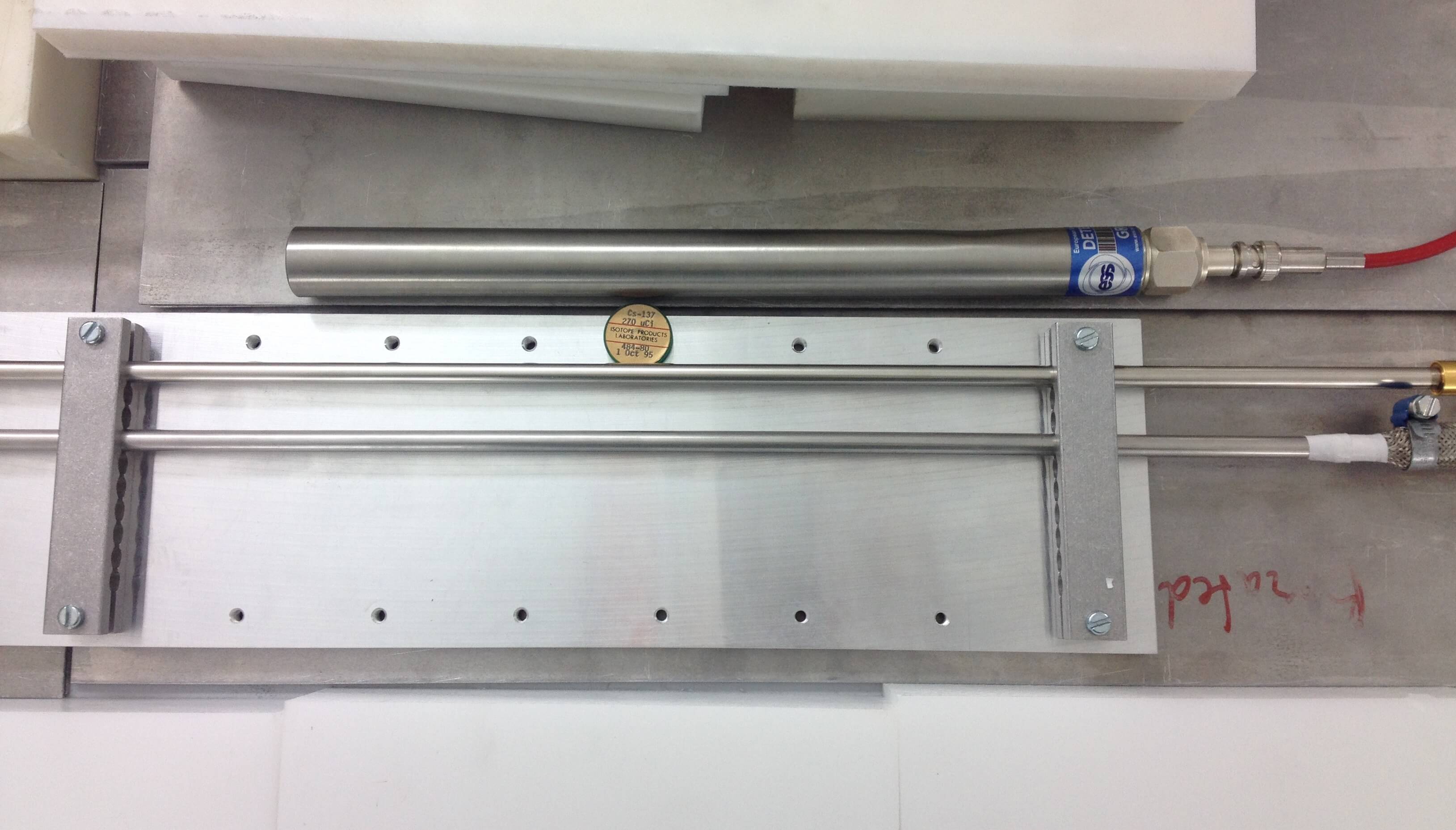}
\caption{A $\mathrm{^{137}Cs}$ source and the $\mathrm{^{3}He}$ tubes.}
\label{imm:expsetup1}
\end{figure}

The specifications of the detectors are discussed in Section \ref{sec:detector}. In order to decrease the neutron background, we move the neutron source in an other room of that of the experimental set-up. The signals of the tubes are amplified by the ORTEC 142PC preamplifier and then shaped by the ORTEC 855 Dual Spec Amplifier and finally are recorded with the Amptek MCA-8000D. Each measurement lasts between 6 hours and 3 days. In order to shield the tubes from the neutron background we cover them with some borated polyethylene and borated aluminium plates. Figure \ref{imm:expsetup1} shows the experimental set-up.

\begin{figure}[H]
\centering
\includegraphics[width=0.8\textwidth]{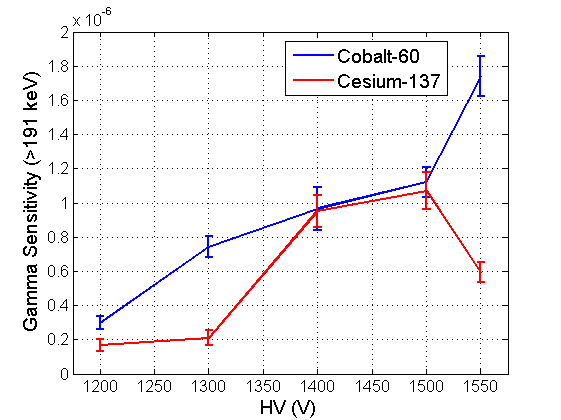}
\caption{Gamma sensitivity for the RS 25 $cm$ tube.}
\label{imm:sensitivityreuter}
\end{figure}

\begin{figure}[H]
\centering
\includegraphics[width=0.8\textwidth]{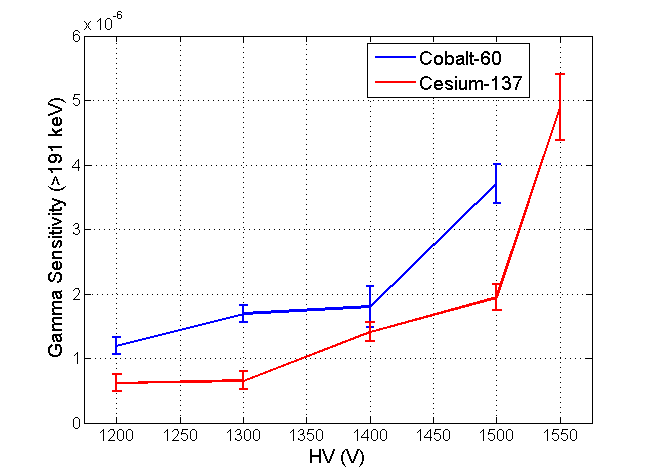}
\caption{Gamma sensitivity for the Toshiba 1 $m$ tube.}
\label{imm:sensitivitytoshiba}
\end{figure}

The distance between the gamma-ray source and the tubes is carefully measured. We assume an uncertainty of 3 $mm$ and 5 $mm$ for the 8 $mm$ and the 1 $inch$ tube respectively (the position of the smaller tubes is subject to a lower error). This distance must be carefully tuned because if the source is too close to the tubes, too many gamma-ray pulses will be generated and the pile-up alters significantly the measurements, on the other hand, the source must not be too far from the tube, in order to have enough counts and to decrease the effects of the background.
\\
We choose distances of a few $cm$ (1.5-8 $cm$), that correspond to a probability that a gamma-ray is directed toward the detector between 1\% and 15\% (the solid angle subtended by the detectors at the sources between 0.13 and 1.9 $sr$). 
\\
\\
Before the gamma sensitivity measurements, we record a neutron PHS in the same configuration. In order to calculate the gamma sensitivity only the gamma pulses that have an energy larger than the proton-escape peak at 191 $keV$ are counted (see Figure \ref{imm:sensexample}).
\\
\\
Figure \ref{imm:sensitivityreuter} shows the result for the RS 25 $cm$ tube, which has a diameter of 1 $inch$ and wall thickness of $\approx$500 $\mu m$ and Figure \ref{imm:sensitivitytoshiba} shows the result for the Toshiba 1 $m$ tube, which has a diameter of 8 $mm$ and wall thickness of $\approx$250 $\mu m$.

\begin{figure}[H]
\centering
\includegraphics[width=0.8\textwidth]{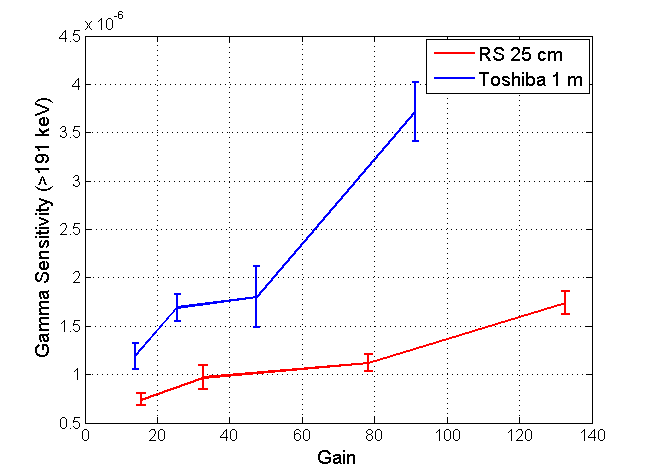}
\caption{Gamma sensitivity with the $\mathrm{^{60}Co}$ source.}
\label{imm:sensitivitycobalt}
\end{figure}

\begin{figure}[H]
\centering
\includegraphics[width=0.8\textwidth]{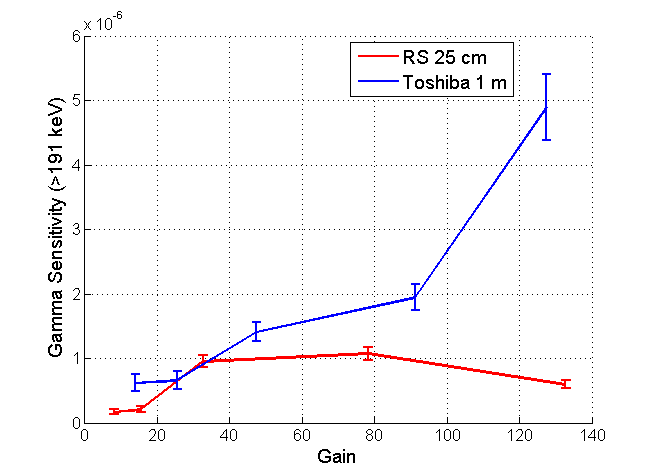}
\caption{Gamma sensitivity with the $\mathrm{^{137}Cs}$ source.}
\label{imm:sensitivitycesium}
\end{figure}

The gamma sensitivity is $\approx$10$^{-6}$-10$^{-7}$ and it increases with the HV. Referring to Section \ref{sec:conversion} we find that $p_{e^{-} release}$, the probability that an electron releases more than 191 $keV$ in the gas, is $\approx$10$^{-3}$-10$^{-4}$.
\\
\\
The gamma sensitivity increases with the high voltage applied to the tube with an approximately linear dependence. This is deeply related to the space charge effect (Section \ref{sec:gasdetectors}). Since the neutron fragments produce denser charge than the photo-electrons, they suffer more from the space charge effect as the HV applied to the tube is increased. This phenomenon is discussed in detail in Section \ref{sec:spacecharge}. 
\\
As expected, the gamma sensitivity for $\mathrm{^{60}Co}$ is larger than for $\mathrm{^{137}Cs}$, the latter in fact produces gamma-rays of lower energy. Usually the difference is a factor of 2-3, as obtained in the previous Section for $p_{e^{-} transmission}$.
\\
\\
Figures \ref{imm:sensitivitycobalt} and \ref{imm:sensitivitycesium} show a comparison of the two tubes. Note that in these Figures the gamma sensitivity is a function of the gain of the tubes (Section \ref{sec:gainn}).

The Toshiba 1 $m$ tube has a higher gamma sensitivity with respect to the RS 25  $cm$ tube, although it has a smaller diameter and thinner walls. This is a consequence of different gas compositions, in fact the first tube is a position sensitive detector and it contains a larger amount of Ar and $\mathrm{CO_2}$. The effects of the gas composition are discussed in Section \ref{sec:gascomposition}.

\tocless\subsection{Space Charge Effect}
\label{sec:spacecharge}

The gamma sensitivity increases with the applied voltage as a consequence of the space charge effect. The space charge effect occurs close to the wire, where a cloud of electrons and ions is generated during the Townsend avalanches. The electrons, thanks to their high mobility, are quickly collected on the wire, while the slow ions, at a first approximation, stay in place until all the electrons are collected. The dense cloud of positive ions decreases significantly the electric field reducing the growth of new avalanches. The net effect of the space charge effects is a partial loss of the gas gain. The denser the initial electron cloud, the more significant this loss is. Obviously, the space charge effects increase with the gas gain, i.e. the HV applied to the tube.
\\
\\
The neutron fragments have a stopping power about 1000 times higher than the photo-electrons in $\mathrm{^{3}He}$ and in the other fill gases (Figure \ref{imm:stoppingpower}) and therefore they generate a trace with a larger electron density. For this reason, the neutron fragments suffer more of the space charge effect. This explains why the gamma sensitivity increases with the HV. In fact, while the charge collected from neutron fragments is decreased due to the space charge effect, gamma-rays charge is still almost completely collected. As the HV increases, the neutron PHS experiences a lower gain than the gamma-ray exponential, hence more gamma-ray pulses overlap with the neutron PHS. For this reason the gamma sensitivity increases as the HV increases.

\tocless\subsection{Gas Composition}
\label{sec:gascomposition}

\begin{figure}[H]
\centering
\includegraphics[width=0.8\textwidth]{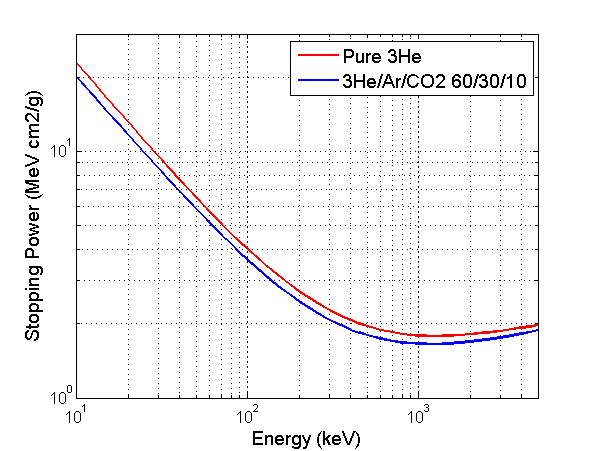}
\caption{Mass stopping power for electrons in two different gas mixtures.}
\label{imm:stoppingpowersensitivity}
\end{figure}

The gas composition strongly influences the gamma sensitivity. In fact, the amount of energy deposited by the electrons in the gas depends critically on their stopping power. 
\\
To analyze how different gas mixtures affect the stopping power without taking into account the gas density we use the mass stopping power. This is defined as the stopping power of a given particle in a gas, divided by the mass density of the gas. This quantity depends on the particle and on the gas composition, but not on its mass density. Figure \ref{imm:stoppingpowersensitivity} shows the mass stopping power for electrons in pure $\mathrm{^{3}He}$ and in a gas mixture of 60\%$\mathrm{^{3}He}$, 30\% Ar and 10\% $\mathrm{CO_2}$. 

Referring to Figure \ref{imm:stoppingpowersensitivity} the difference between the two is very small and it does not alter significantly the gamma sensitivity. Therefore almost only the mass density (proportional to the electron density) of the gas influences the gamma sensitivty and not its composition. It is common that a certain quantity (10-40\%) of stopping and quenching gases (such as Ar/$\mathrm{CO_2}$) is added to the $\mathrm{^{3}He}$ and this has a much stronger effect on the total mass density. I.e. 10 $bar$ of $\mathrm{^{3}He}$ count as about 0.75 $bar$ of Ar/$\mathrm{CO_2}$. 
\\
\\
We expect the position sensitive detectors to have a larger pressure of stopping gases than a neutron counter. This explains the results shown in Figures \ref{imm:sensitivitycobalt} and \ref{imm:sensitivitycesium}, where the Position Sensitive Toshiba 1 $m$ shows a larger or comparable gamma sensitivity than that of the neutron counter (the RS 25 $cm$ tube), although it has a smaller diameter and thinner walls. 

\section{Gamma Sensitivity calculated by using the Neutron Valley}
\label{sec:sensitivityvalley}

\begin{figure}[H]
\centering
\includegraphics[width=0.8\textwidth]{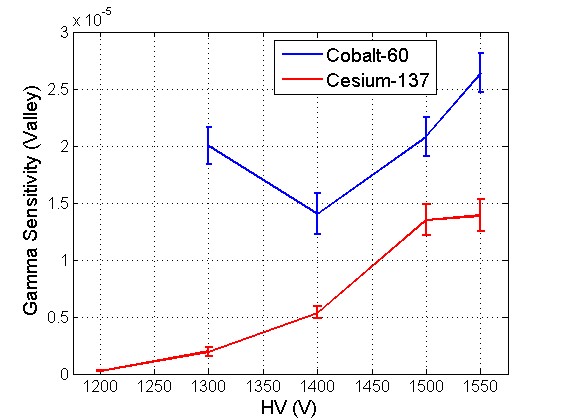}
\caption{Gamma sensitivity for the RS 25 $cm$ tube calculated setting the threshold in the neutron valley.}
\label{imm:gammalittlevalley}
\end{figure}

\begin{figure}[H]
\centering
\includegraphics[width=0.8\textwidth]{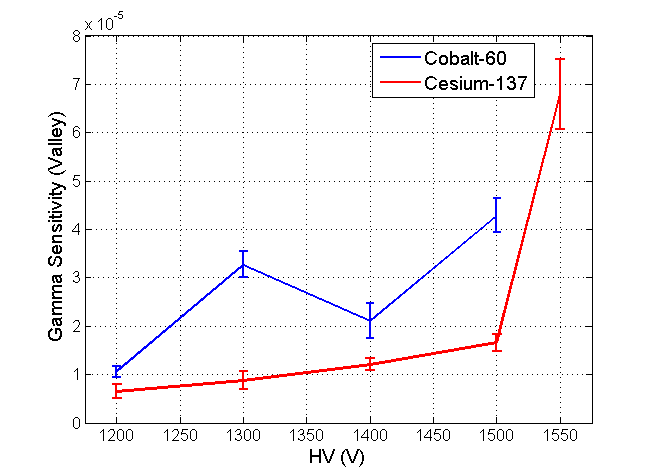}
\caption{Gamma sensitivity for the Toshiba 1 $m$ tube calculated setting the threshold in the neutron valley.}
\label{imm:gammatoshibavalley}
\end{figure}

Figures \ref{imm:gammalittlevalley} and \ref{imm:gammatoshibavalley} show the gamma sensitivity calculated setting the threshold in the neutron valley for the RS 25 $cm$ tube and the Toshiba 1 $m$ tube, respectively. As discussed in Section \ref{sec:definitiongamma}, these results should be considered semi-quantitative due to the uncertainty in the position of the neutron valley. For this reason, a comparison between such measurements performed with the two tubes is not satisfactory and thus it is not shown here.

The gamma sensitivity has an order of magnitude of 10$\mathrm{^{-5}}$, an order of magnitude larger than in the case with the threshold set on the peak at 191 $keV$. This result is consistent with an analogous measurement performed with the RS 1 $m$ tube. Moreover, the gamma sensitivity increases approximately linearly as the HV increases.

%% file: Chapters/Pulseshapediscrimination.tex

\chapter{Pulse Shape Discrimination}
\label{ch:PSD}

The usual information carried by a linear pulse is in its amplitude (and time of occurence). However, there are occasions in nuclear experiments when the shape of the pulse also assumes some importance. The Pulse Shape Discrimination (PSD) is a technique used to discriminate signals generated by different particles. This technique consists in analyzing the signal shape and extracting a parameter that assumes different values for different particles.
\\
\\
The PSD technique is widely used in particle and nuclear physics. In neutron physics it is commonly used in lithium-loaded scintillators\cite{Flaska2007, Zaitseva2012, Lintereur2012}, in order to discriminate neutrons from gamma-rays. A scintillator signal has a slow and a fast component and the ratio between these two depends on the linear stopping power $dE/dx$ of the particle. When a neutron interacts with lithium, fast ions are generated and these have a higher stopping power than the photo-electrons generated by a gamma-ray, thus the gamma discrimination via PSD is possible (see Figure \ref{imm:lithiumPSD}).
\\
\\
In the literature the feasibility of the PSD technique to discriminate neutron and $\mathrm{\alpha}$ signals\footnote{In detectors with Al walls the $\mathrm{\alpha}$-particles are a significant background. In fact, natural Al contains a relatively high concentration of $\mathrm{\alpha}$-emitter impurities \cite{Birch2015}. If walls are made of steel, the $\mathrm{\alpha}$-particle background is negligible.} in $\mathrm{^{3}He}$ is discussed \cite{Langford2013}, but there is not a comprehensive discussion about the gamma-ray rejection though.
\\
\\
In this Chapter we discuss the feasibility of the gamma-ray discrimination via PSD in $\mathrm{^{3}He}$ tubes. As discrimination parameter we choose the rise time of the signals, the time that the signal needs to go from a certain amplitude to another. We also examine the effects that the choice of the amplitude limits has on the discrimination.
\\
\\
The PSD technique reduces the number of gamma-rays miscounted as neutrons, enhancing the accuracy of neutron counting systems. This is particularly relevant in environments with a large gamma-ray background and for detectors with a low neutron efficiency, e.g. beam monitors\footnote{Note that as the $\mathrm{^{3}He}$ pressure decreases, the neutron efficiency decreases, but the same does not occur for the gamma sensitivity. In fact, gamma sensitivity depends primarily on the stopping and quenching gases pressure (Section \ref{sec:gascomposition}). In order to limit the range of the neutron fragments, the beam monitors have a pressure comparable to that of the neutron counters. For this reason the ratio (neutron efficiency)/(gamma sensivitity) is lower in beam monitors than in high efficiency neutron detectors.}. 
\\
\\
The information provided in this Chapter will be exploited in the future to implement the PSD technique on the $\mathrm{^{10}B}$ detectors. In those the signal is generated in the stopping gas (usually Ar/$\mathrm{CO_2}$) in the same way as in the $\mathrm{^{3}He}$ counters. The $\mathrm{^{10}B}$ detector PHS, unlike that of the $\mathrm{^{3}He}$, extends to low energies (where most of the gamma-ray signals occur), thus in this case the PSD is even more crucial.

\section{Introduction}

\begin{figure}[H]
\centering
\includegraphics[width=0.9\textwidth]{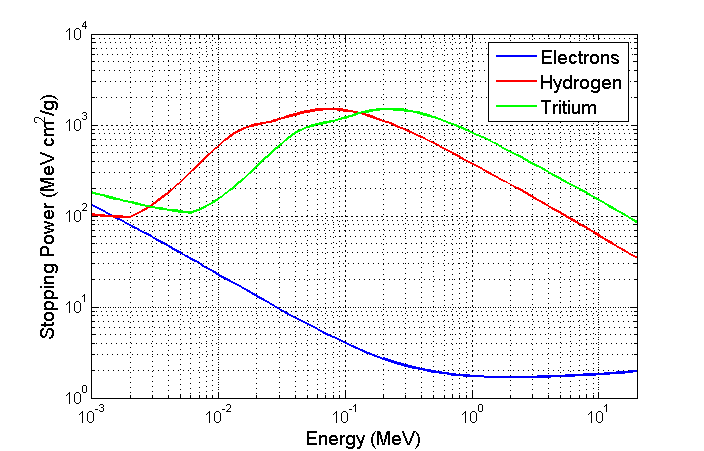}
\caption{The stopping power in $\mathrm{^{3}He}$ of the electrons, which was calculated with  ESTAR \cite{ESTARprogramwebpage}, and of the $\mathrm{^{1}H}$ and the $\mathrm{^{3}H}$ ions, which were calculated with the package SRIM \cite{Ziegler}.}
\label{imm:stoppingpower}
\end{figure}

In the energy range of interest for $\mathrm{^{3}He}$ detectors (100 $keV$-10 $MeV$), the neutron fragments have a larger mass stopping power than the photo-electrons ($\approx 10^{3}\cdot MeV\cdot cm^{2}/g$ and $\approx 10^{0}\cdot MeV\cdot cm^{2}/g$ respectively) (Figure \ref{imm:stoppingpower}. 

In order for a photo-electron to release the same energy in the gas as the neutron fragments, it must travel about 1000 longer. The neutron fragments have ranges of a few $mm$ in $\mathrm{^{3}He}$ (<5 $mm$ in 10 $bar$ of $\mathrm{^{3}He}$). The photo-electrons generally are not completely stopped in the gas, but they often reach the wall of the detector. Thus they release their energy on a larger distance up to the total active length of the detector. The secondary electrons, that is the electrons generated by ionization in the gas by the photo-electrons or by the neutron fragments, in this case are more dispersed, then their collection requires more time. The total collection time (the time span between the collection of the signal of the first electron and that of the last one) is longer for the gamma-rays than for the neutron fragments. We use as a discrimination parameter the rise time of the signal, which is proportional to the collection time of the secondary particles.

\section{Results}

\begin{figure}[H]
\centering
\includegraphics[width=0.8\textwidth]{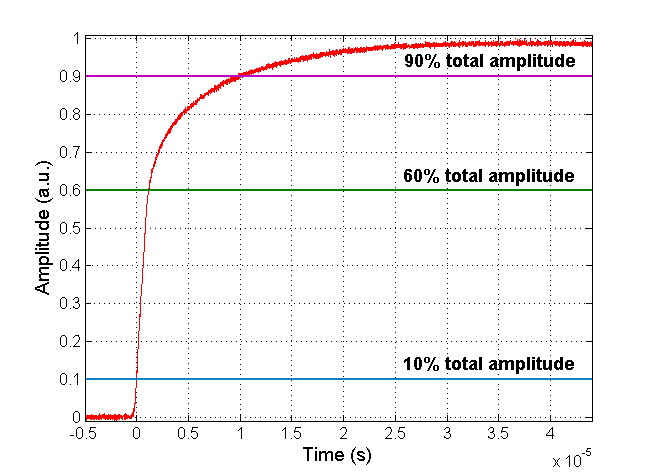}
\caption{A neutron signal from the RS 25 $cm$ tube. Three amplitude thresholds are also shown.}
\label{imm:neutronsignal}
\end{figure}

We perform the measurements with the RS 25 $cm$ tube and the RS 1 $m$ tube. The output from the connector of the tube is amplified by the FAST ComTec CSP10 preamplifier, which has a rise time of 7 $ns$. The waveforms are then recorded by using the oscilloscope LeCroy HDO4054. 
\\
Figure \ref{imm:neutronsignal} shows an example of a recorded neutron signal. This measurement is performed with the RS 25 $cm$ tube and HV 1600 $V$. Three amplitude thresholds are also shown.

\begin{figure}[H]
\centering
\includegraphics[width=0.8\textwidth]{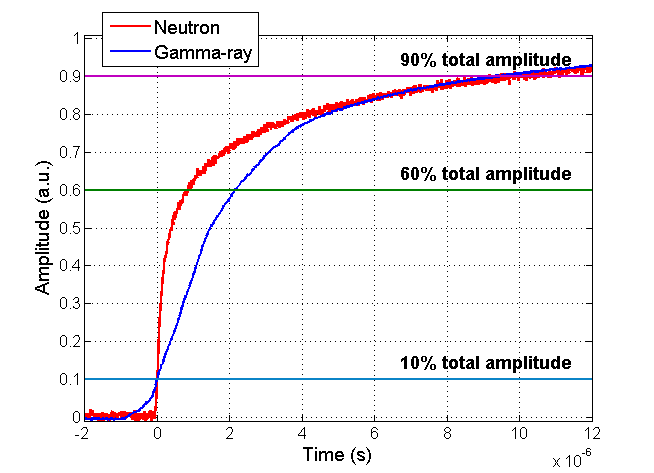}
\caption{Comparison between a neutron and a gamma-ray signal with the same amplitude. The neutron signal is noisier as a consequence of the different amplitude scales used in the oscilloscope during the two measurements.}
\label{imm:neutrongammasignal}
\end{figure}

Note that approximately 60\% of the total signal is quickly collected ($\approx$ 1 $\mu s$), while the remaining is collected in a much longer time ($\approx$ 10 $\mu s$). 
\\
\\
Figure \ref{imm:neutrongammasignal} shows a comparison between a neutron signal and a gamma-ray signal (from a $\mathrm{^{60}Co}$ source) with the same amplitude.

\begin{figure}[H]
\centering
\includegraphics[width=\textwidth]{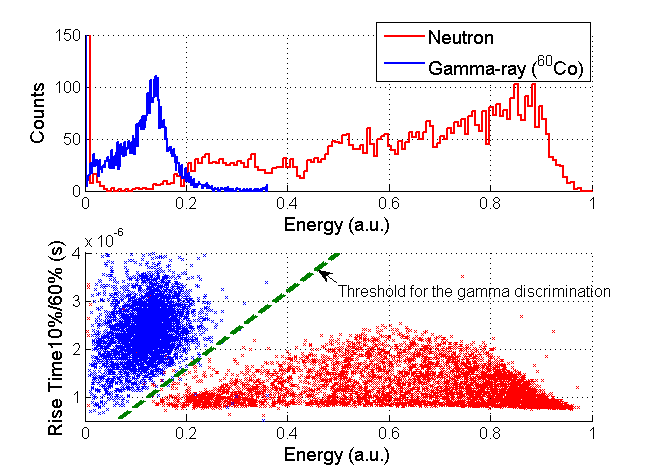}
\caption{PHS (upper plot) and rise time (lower plot) of the neutron and the gamma-ray source, with a possible choice for the threshold used to discriminate the gamma-ray signals (green dotted line). The measurements have been performed with the RS 25 $cm$ tube operated at 1600 $V$. The upper plot is the projection on the X-axis of the lower plot. The anomalous shape of the gamma-ray PHS at low energies is due to the threshold applied with the oscilloscope.}
\label{imm:psdrs25cm}
\end{figure}

Note that the difference in the rise time has a maximum at a fraction of 60\% of the total amplitude. Above 60\% the difference diminishes until no significant difference in the two rise time is noticeable (90\% of the total amplitude). For this reason we define the rise time as the time difference between the 10\% and the 60\% of the total amplitude. 
\\
\\
In order to obtain the neutron rise time distribution we use the neutron source and we shield the detector from the gamma-ray background with several steel plates. In order to obtain the gamma-ray rise time distribution we use a gamma-ray source and we shield the detector from the neutron background with several borated polyethylene and aluminium plates. We perform the measurement with the RS 25 $cm$ tube operated at 1600 $V$ and the $\mathrm{^{60}Co}$ as the gamma-ray source. We record about 5000 waveforms with both the sources. Figure \ref{imm:psdrs25cm} shows the neutron and gamma-ray PHS (upper plot) and their rise time distribution (lower plot). The upper plot is the projection on the X-axis of the lower plot. A possible choice for the threshold used to perform the neutron to gamma-ray discrimination is also shown (green dotted line).

\begin{figure}[H]
\centering
\begin{subfigure}[t]{0.49\textwidth}
\includegraphics[width=\textwidth]{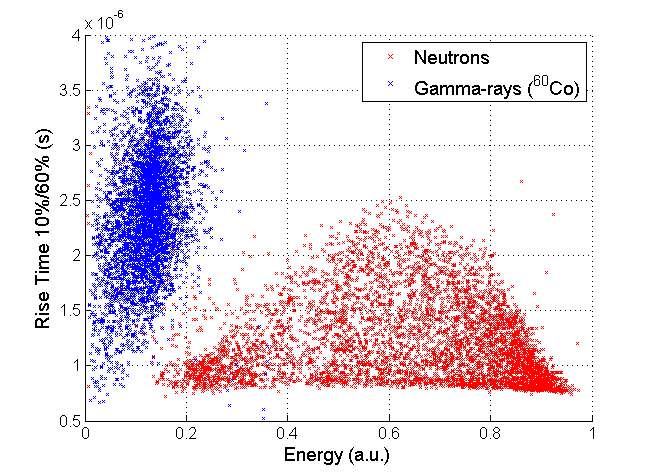}
\caption{Rise Time 10\%/60\%}
\end{subfigure}
\begin{subfigure}[t]{0.49\textwidth}
\includegraphics[width=\textwidth]{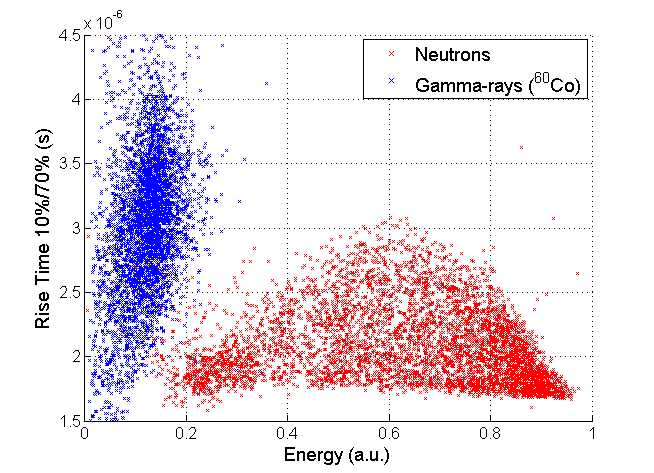}
\caption{Rise Time 10\%/70\%}
\end{subfigure}

\begin{subfigure}[t]{0.49\textwidth}
\includegraphics[width=\textwidth]{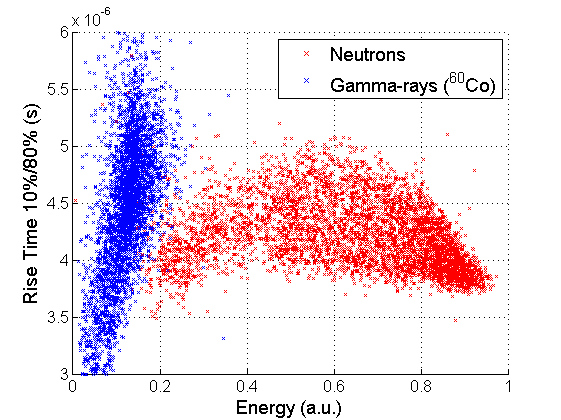}
\caption{Rise Time 10\%/80\%}
\end{subfigure}
\begin{subfigure}[t]{0.49\textwidth}
\includegraphics[width=\textwidth]{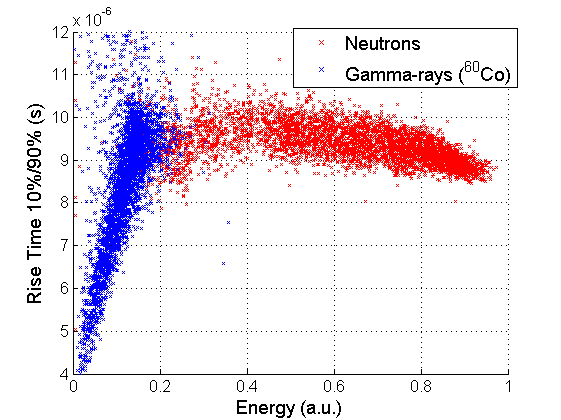}
\caption{Rise Time 10\%/90\%}
\end{subfigure}
\caption{Several amplitude limits for the rise time definition. The results obtained with the rise time 10\%/60\% exhibit the best behavior for the PSD technique. The discrimination deteriorates for higher amplitude thresholds.}
\label{imm:PSDrisetimes}
\end{figure}

Note that the gamma-ray signals have a larger rise time than the neutron signals of the same amplitude. By using the proposed threshold, we reject more than 97\% of the gamma-ray signals we would count using a simple amplitude threshold in the valley, while losing less than 0.6\% of the neutron pulses. The gamma sensitivity therefore decreases of almost two order of magnitude if the PSD technique is used.
\\
\\
Using other amplitudes limits for the rise time definition deteriorates the discrimination (Figure \ref{imm:PSDrisetimes}).

\begin{figure}[H]
\centering
\includegraphics[width=0.6\textwidth]{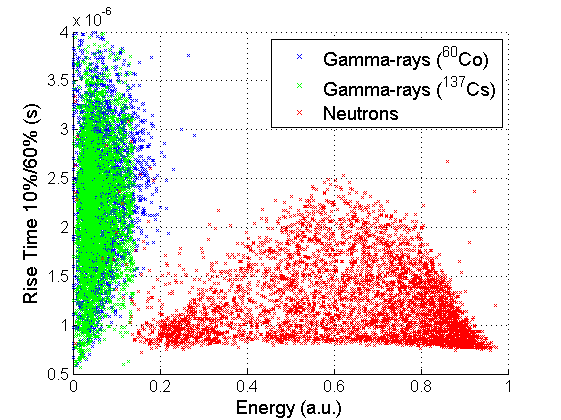}
\caption{Comparison between the rise time distribution of the pulses generated by the $\mathrm{^{60}Co}$ source and the $\mathrm{^{137}Cs}$ source.}
\label{imm:PSDgammasources}
\end{figure}

Figure \ref{imm:PSDgammasources} shows that no significant difference arises between the rise time distribution of the gamma-ray pulses from the $\mathrm{^{60}Co}$ and from the $\mathrm{^{137}Cs}$ sources.

\begin{figure}[H]
\centering
\includegraphics[width=\textwidth]{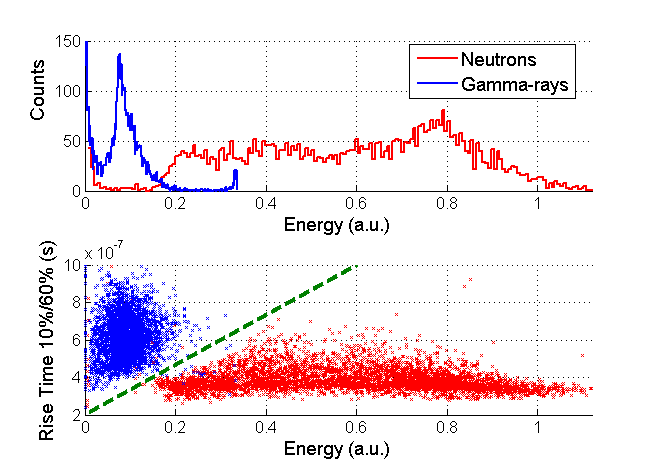}
\caption{PHS (upper figure) and rise time distribution (lower figure) of the neutron and a gamma-ray source, with a possible choice for the threshold used for the gamma-ray discrimination (green dotted line). The measurements have been performed with the RS 1 $m$ tube with applied 1500 $V$. }
\label{imm:PSDrs1m}
\end{figure}

Figure \ref{imm:PSDrs1m} shows the results obtained with the RS 1 $m$ tube operated at 1500 $V$. 

With the proposed threshold we discriminate more than 96\% of the gamma-ray signals, while losing only 0.3\% of the neutron signals.
\\
The PSD technique exhibits no significant differences between the results obtained with the RS 25 $cm$ tube and with the RS 1 $m$ tube, although they have a different gas composition and different mechanical characteristics.

%% file: Chapters/conclusions.tex

\chapter{Conclusions}

Helium-3 based detectors have been the main actor in thermal neutron detection in the last 70 years, thanks in particular to their high efficiency. Since 2001 the world is now experiencing a shortage of this rare isotope, thus its price is increasing over time and its availability is very limited. Moreover, the detector requirements in the new high-flux facilities, such as the European Spallation Source (ESS) that is under construction in Lund (Sweden), are becoming more and more challenging to fulfill. Because of the limited availability of helium-3 it is not realistic anymore to build large area detectors (approximately 50 square meters) with this technology-On the other hand, the main concern for small area detectors (approximately 1 square meter) is about the performance of the helium-3 detectors, in particular related to the spatial resolution and the counting rate capability. 
\\
The detector research in the entire world is now focused on the replacement of helium-3 detectors with detectors based on alternative neutron converters, such as boron-10, lithium-6 and gadolinium. The ESS will have a prominent role in this process as the first large scale facility based mainly on boron-10 detectors. In order to validate these new technologies, a detailed knowledge of the properties of helium-3 detectors is needed. Although helium-3 have been used for decades and this technology is well known, some of its properties can not be found in the literature.
\\
\\
In this thesis we presented the results of the measurements performed with several helium-3 tubes commonly used at worldwide neutron scattering facilities, with a special focus on the spatial resolution and gamma discrimination. We quantified the limit of this technology by exploring new procedures, which can be used as a reference to study and characterize the performance of the new alternative technologies.
\\
\\
The energy resolution is closely related to the gamma discrimination and to how the detectors operate. As the energy resolution deteriorates, the gamma discrimination decreases. 
\\
We quantified the energy resolution of 3 helium-3 tubes at several HV. We measured an energy resolution between 1\% and 10\%, but we expect a worse energy resolution at large HV, typically applied when position sensitive detectors are used. At small HV the main contributions to the energy resolution are the electronic noise and the poissonian uncertainty on the production of secondary electrons by the neutron reaction fragments. At large HV the main contribution is the space charge effect. As a consequence, at small HV the energy resolution improves as the HV increases, while the opposite happens at large HV.
\\
We found that a minimum of the energy resolution occurs in our helium-3 tubes with operative HV of 900 $V$ and 1300 $V$. These HV correspond to the best gamma discrimination and is suggested to operate a detector in such configuration, if no other requirements on the HV are present.
\\
\\
An important limit of the helium-3 detectors concerns the spatial resolution. In fact, the alternative technology detectors can achieve a better spatial resolution than them. Although it is well known the operative spatial resolution of the helium-3 detectors exploited in the main facilities in the world, the limits of this technology are not quantified in the literature.
\\
We investigated the spatial resolution of 2 helium-3 position sensitive tubes. They measure the position of a neutron absorption through the charge division readout. We performed these measurements at the neutron beamline R2D2, at the Institute for Energy Technology (IFE), located in Lillestr\o m (Norway), in order to take advantage of a collimated neutron beam. 
\\
The spatial resolution improves as the HV applied to the tube increases, as a consequence of a larger signal-to-noise ratio. We measured a maximum spatial resolution of (5.6 $\pm$ 0.6) $mm$ and (6.7 $\pm$ 0.5) $mm$ for our tubes.
\\
We also tested the linearity of the spatial resolution with respect to the position of the beam along and across the wire and we found that no significant difference arises.
\\
\\
In a neutron facility a detector is always exposed to other kind of radiation that we consider as a background to be suppressed. The detection of a background event can give rise to misaddressed events in a neutron detector. 
\\
The gamma-rays are the most significant background in neutron instruments. They are generated directly by the neutron source or by interaction of neutrons with any material in their path, e.g. neutron collimators, choppers or beam stops. Generally the gamma-ray background can be a few orders of magnitude more intense than the neutron signal. For this reason the gamma sensitivity of a neutron detector is a crucial property to evaluate the performance of the detectors. The gamma-ray sensitivity is defined as the efficiency for miscounting a gamma-ray signal as a neutron signal.
\\
Although in the last decade in the literature several articles described measurements performed with boron-10 and lithium-6 based detectors, none of them involves helium-3 detectors.
\\
We measured the gamma sensitivity for 3 helium-3 tubes by using a cobalt-60 and a cesium-137 gamma ray source. By setting the amplitude threshold in the neutron valley, as usually happens when these detectors are used in the scattering instruments, the gamma sensitivity is approximately 10$\mathrm{^{-5}}$. The measurements with the 3 detectors are consistent with each other. By setting the amplitude threshold at a fixed energy, i.e. the 191 $keV$ peak of the helium-3 Pulse Height Spectrum (PHS), we found that the gamma sensitivity increases as the HV applied to the tube increases. This is a consequence of the space charge effect and the gamma sensitivity increases of almost one order of magnitude if the HV is increased from 1200 $V$ to 1550 $V$. Moreover, we found that the gamma sensitivity depends critically on the stopping and quench gas pressure and not on the helium-3 pressure. In fact, 10 $bar$ of helium-3 count as about 0.75 $bar$ of argon and carbon dioxide.
\\
\\
To improve the gamma discrimination of the helium-3 detectors we investigated the implementation of the Pulse Shape Discrimination (PSD) technique. This technique consists in analyzing the signal shape and extracting a parameter that assumes different values for different particles. This technique is commonly used in lithium-6 scintillators, but in the literature there is not a comprehensive discussion about the gamma-ray rejection in helium-3 detectors. 
\\
We found that in helium-3 detectors the rise time of a gamma-ray signal is larger than that  of a neutron signal of the same amplitude. The rise time is the time that the signal needs to go from a certain amplitude to another. The PSD technique can be exploited with both the tubes we analyzed and no significant difference arises between the discrimination obtained with a cesium-137 and a cobalt-60 source.
\\
We propose a threshold that takes into account both the amplitude and the rise time of a signal. By using this threshold more than 96\% of the gamma-ray signals that would be counted using an amplitude threshold are rejected, while less than 1\% of the neutron signals are lost. 
\\
The definition of the amplitude limits for the rise time calculation is crucial. The best discrimination is obtained with amplitude limits of 10\% and 60\% of the total amplitude of the signal. By using a common choice of 10\% and 90\% the discrimination is not viable.
\\
All this information will be exploited in the future to implement the PSD technique on the boron-10 detectors. In fact, the boron-10 detector PHS, unlike that of the helium-3, is extended down to zero energy and there is not a clear energy separation between neutron and gamma-ray pulses. For his reason, the PSD technique is even more crucial for boron-10 counters.
\\
\\
In this thesis we measured and analyzed some important parameters of the helium-3 tubes in order to characterize their performance. These parameters set a standard to systematically compare the different detector technologies.
\\
We used new experimental procedures that will be exploited to analyze also the boron-10 detector performance. We gained much insight into how the helium-3 detectors operate and this knowledge will be very helpful to improve the alternative technologies detectors.